\documentclass[%
 reprint,
superscriptaddress,
 amsmath,amssymb,
 aps, physrev,
]{revtex4-2}

\usepackage{graphicx}% Include figure files
\usepackage{dcolumn}% Align table columns on decimal point
\usepackage{bm}% bold math
\usepackage{braket}
\usepackage{bbold}
\usepackage{subcaption}
\usepackage{ulem}
\usepackage{xcolor}
\usepackage{ragged2e}
\usepackage{caption}

\DeclareCaptionJustification{justified}{\justifying}
\begin{document}

\preprint{APS/123-QED}

\title{\textbf{Semiclassical molecular dynamics simulations of electromagnetically induced transparency cooling of 2D and 3D ion crystals in a Penning trap} 
}% 

\author{John Zaris}
 \email{Contact author: john.zaris@colorado.edu}
\affiliation{
 Department of Physics, University of Colorado, Boulder, Colorado 80309, USA
}%

\author{Wes Johnson}
\affiliation{
 Department of Physics, University of Colorado, Boulder, Colorado 80309, USA
}%
\affiliation{%
 Sandia National Laboratories, Albuquerque, New Mexico, 87185, USA
}

\author{Krishna Balaji Sogathur}
\affiliation{%
 Department of Physics, Indian Institute of Technology Madras, Chennai, India, 600036
}%

\author{Athreya Shankar}
\affiliation{%
 Department of Physics, Indian Institute of Technology Madras, Chennai, India, 600036
}%
\affiliation{%
 Center for Quantum Information, Communication and Computing, Indian Institute of Technology Madras, Chennai 600036, India
}%

\author{John J. Bollinger}
\affiliation{
 National Institute of Standards and Technology, Boulder, Colorado 80309, USA
}%

\author{Allison L. Carter}
\affiliation{
 National Institute of Standards and Technology, Boulder, Colorado 80309, USA
}%

\author{Murray J. Holland}
\affiliation{
 Department of Physics, University of Colorado, Boulder, Colorado 80309, USA
}%
\affiliation{
 JILA, University of Colorado, Boulder, Colorado 80309, USA
}%

\author{Scott E. Parker}
\affiliation{
 Department of Physics, University of Colorado, Boulder, Colorado 80309, USA
}
\affiliation{%
 Renewable and Sustainable Energy Institute, Boulder, Colorado 80309, USA
}

\date{\today}% It is always \today, today,

\begin{abstract}
Motional modes of trapped-ion crystals in Penning traps can now be cooled to near their ground states via electromagnetically induced transparency (EIT) cooling.  Numerical simulations are valuable tools for interpreting EIT cooling results and optimizing experimental protocols, but existing molecular dynamics codes are limited to the simulation of Doppler cooling. In this work, we fill this gap by implementing a semiclassical model of EIT cooling in a novel MD code.  By simultaneously evolving the ions' internal quantum states and classical motion, we are able to calculate the instantaneous EIT laser forces acting on the ions. We first validate our code by comparing the single-ion cooling rate to fully quantum simulation results. Next, we analyze cooling of the drumhead modes in planar ion crystals, quantifying the effects of planar ion motion and a finite laser beam width. Finally, we demonstrate efficient EIT cooling of the axial modes in 3D ellipsoidal crystals containing up to 500 ions. Such crystals are a promising platform for future quantum sensing experiments with enhanced sensitivity. We find that the mode spectra of 3D crystals can be engineered to enable sympathetic cooling of the low-frequency $\boldsymbol{E}\times\boldsymbol{B}$ modes. This numerical framework provides a powerful tool for modeling sub-Doppler cooling techniques and for optimizing future quantum sensing and quantum information experiments. 

\end{abstract}

\maketitle

%\tableofcontents

\section{\label{sec:introduction}Introduction}

Ultracold trapped-ion crystals are leading platforms for quantum science experiments.  Various trap architectures are currently being developed to achieve precise control over the internal states of such systems \cite{mcmahon2020,mcmahon2024,wolf2024,jain2020,gilmore2021,kiesenhofer2023,guo2024, burton2023,donofrio2021}.  These advances have enabled tremendous progress in research areas such as quantum simulation \cite{monroe2021,bohnet2016,britton2012,safavi-naini2018,shankar2022,qiao2022,hawaldar2024,pham2024}, quantum sensing \cite{gilmore2021,affolter2020,gilmore2017}, and searches for physics beyond the standard model \cite{carney2021,budker2022, ito2024}. 

Key experimental improvements promise to further enhance the sensitivity of ion crystal systems and extend their range of applications. For instance, the signal-to-noise of quantum sensing protocols may be enhanced by scaling up existing setups from planar crystals with a few hundred ions to 3D crystals with many thousands, or even millions, of ions \cite{itano1998,hornekaer2002}. In preparation for near-future experiments with 3D crystals, recent theoretical efforts have studied Doppler laser cooling \cite{zaris2024}, quantum simulation \cite{hawaldar2024}, and engineering of spin-motion coupling \cite{hawaldar2026} in such crystals.

In this work, we advance efforts toward quantum science experiments with 3D ion crystals by developing a semiclassical molecular-dynamics (MD) code to simulate electromagnetically-induced transparency (EIT) cooling \cite{morigi2000,morigi2003, roos2000,lechner2016}.  As the number of ions increases, it is vital that the crystals can still be cooled to temperatures near the ground state (sub-millikelvin). When excess energy remains in the crystal, its normal modes may not be sufficiently resolved \cite{shankar2020,johnson2024, johnson2025}. In addition, satisfying the Lamb-Dicke confinement criterion can become more challenging as the number of modes increases, unless all modes are well-cooled \cite{leibfried2003, king1998}.  Effective cooling of the axial modes is particularly important, as they are typically utilized in quantum sensing experiments \cite{gilmore2021}. Recent simulation results suggest that Doppler laser cooling of large 3D crystals in Penning ion traps can be highly effective \cite{zaris2026}. However, to obtain near ground-state cooling, it will be necessary to implement a sub-Doppler cooling scheme, such as EIT cooling, in 3D crystal setups. This will require building on past work with planar crystals, where EIT cooling has already been used to reach near ground-state energies \cite{jordan2019, kiesenhofer2023}.

We develop a numerical integration scheme that evolves the internal and motional states of the ions during EIT cooling.  This framework includes all relevant trap and laser forces, as well as Coulomb interactions between ions.  While theoretical treatments of multi-ion EIT cooling have been developed \cite{bartolotta2024,shankar2019,morigi2000}, they have not yet been applied to 3D crystals.  Furthermore, such studies generally ignore complexities such as nonlinear mode coupling and finite laser beam widths, which are inherently captured in our MD simulations. Therefore, our code should be particularly useful in assessing the viability of EIT cooling for 3D crystals as well as guiding the choice of experimental parameters that optimize cooling.  This work builds upon previous simulations of Doppler cooling in Penning trap ion crystals \cite{zaris2024, tang2019}.

In Section~\ref{sec:theoretical framework}, we review the physics of the Penning trap and EIT cooling before introducing our simulation framework. In Section~\ref{sec:single_ion}, we apply our code to study EIT cooling of a single trapped ion.  We compare simulated cooling rates to those predicted by a fully quantum model to validate our code. In Section~\ref{sec:planar_crystals}, we scale up to planar ion crystals and study the cooling of their axial normal modes.  We analyze the effects of planar motion and finite laser beam width on the cooling of the axial modes. Finally, in Section~\ref{sec:3D_crystals}, we extend our analysis to 3D crystals.  We analyze how the normal mode spectra of 3D crystals affect cooling dynamics, highlighting potential advantages and limitations.

\section{Theoretical Framework}
\label{sec:theoretical framework}
\subsection{Penning Trap}

\begin{figure}
    %\centering
    %\includegraphics[scale=0.5]{prolate_crystal_cooling_072825.pdf}
    \centering
        \includegraphics[scale=0.50]{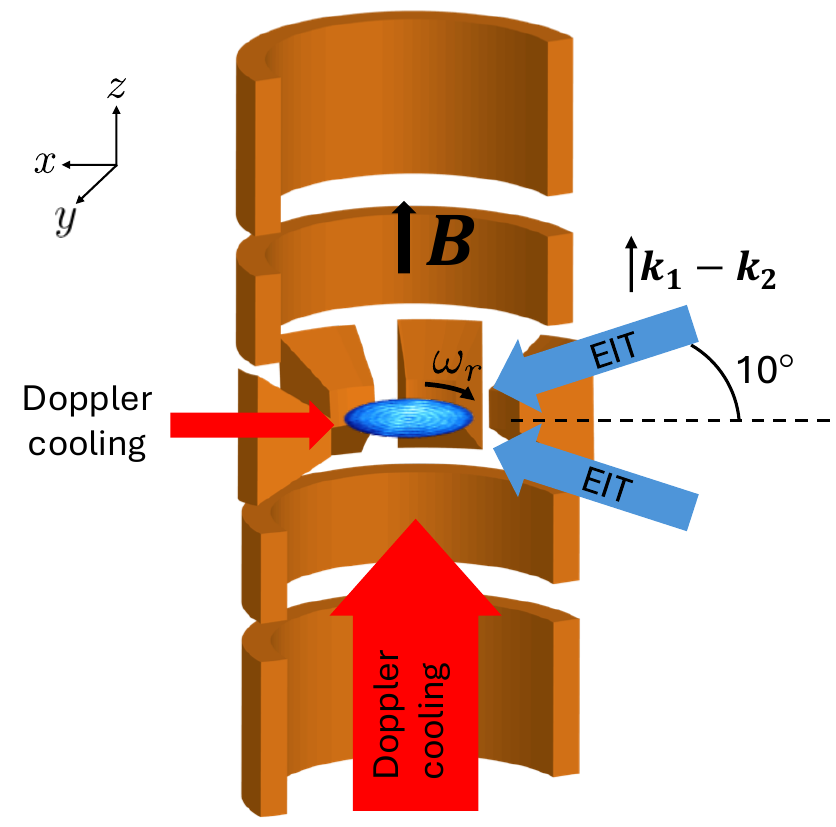}
        \label{fig:cry_shape}
    \caption{The NIST Penning trap is illustrated, with Doppler and EIT laser cooling beams shown by red and blue arrows, respectively.  Voltages applied to the four cylindrically symmetric electrodes produce axial confinement of $^9\text{Be}^+$ ions.  The axial frequency of the trap is $\omega_z = 2\pi\times 1.59$ MHz. Radial confinement is caused by the presence of a strong axial magnetic field $\boldsymbol{B} = B_z\hat{\boldsymbol{z}}$, where $B_z=4.4588$ T. A periodic time-dependent potential with frequency $\omega_r$ applied to the central electrodes produces a rotating-wall force on the ion crystal which stabilizes its rotation frequency.  The ions are first cooled using two Doppler beams, which provide planar and axial cooling, respectively.  Next, two EIT laser beams, whose difference wavevector points in the $\hat{\boldsymbol{z}}$ direction, are used to achieve sub-Doppler cooling.}
    \label{fig:penning_trap}
\end{figure}

A diagram of the NIST Penning trap is shown in Fig.~\ref{fig:penning_trap}.  Penning traps confine charged particles using static electric and magnetic fields \cite{dehmelt1968, brown1986, dubin1999}.  Voltages applied to the trap electrodes produce an electric potential that is confining in the axial direction.  Near the center of the trap, this potential is approximately harmonic and is given by

\begin{equation}
\label{pot_trap}
    \phi_{\text{trap}}(\boldsymbol{r}) = \frac{1}{4}k_z(2z^2-x^2-y^2),
\end{equation}
where $k_z$ is the strength of the potential, related to the trapping frequency by $\omega_z = \sqrt{qk_z/m}$, where $q$ and $m$ are the charge and mass of the trapped particle, respectively. It is necessary to introduce a static axial magnetic field because the force associated with $\phi_{\text{trap}}$ is radially repulsive.  In typical Penning traps, the magnetic field strength required for ion confinement is $B_z \gtrsim 1$ T ($B_z=4.4588$ T in the NIST Penning trap). The field $\boldsymbol{B}=B_z\hat{\boldsymbol{z}}$ produces radial confinement, but also leads to an $\boldsymbol{E}\times\boldsymbol{B}$ drift of the trapped particles about the symmetry axis of the trap.  To precisely control this global rotation of the crystal, a rotating-wall potential is often utilized \cite{huang1998}.  This is a time-varying potential that is periodic in the azimuthal direction, explicitly given by

\begin{equation}
\label{pot_wall}
    \phi_{\text{wall}}(\boldsymbol{r},t) = \frac{1}{2}k_z\delta(x^2+y^2)\cos[2(\varphi+\omega_r t)],
\end{equation}
where $\delta$ is the relative strength of the rotating-wall potential, $\varphi=\tan^{-1}(y/x)$ is the azimuthal coordinate, and $\omega_r$ is the rotating-wall frequency.

\subsection{EIT Cooling}

\begin{figure}
    %\centering
    %\includegraphics[scale=0.5]{prolate_crystal_cooling_072825.pdf}
    \centering
    \begin{subfigure}{0.5\textwidth}
        \subcaption{}
        \centering
        \includegraphics[scale=0.40]{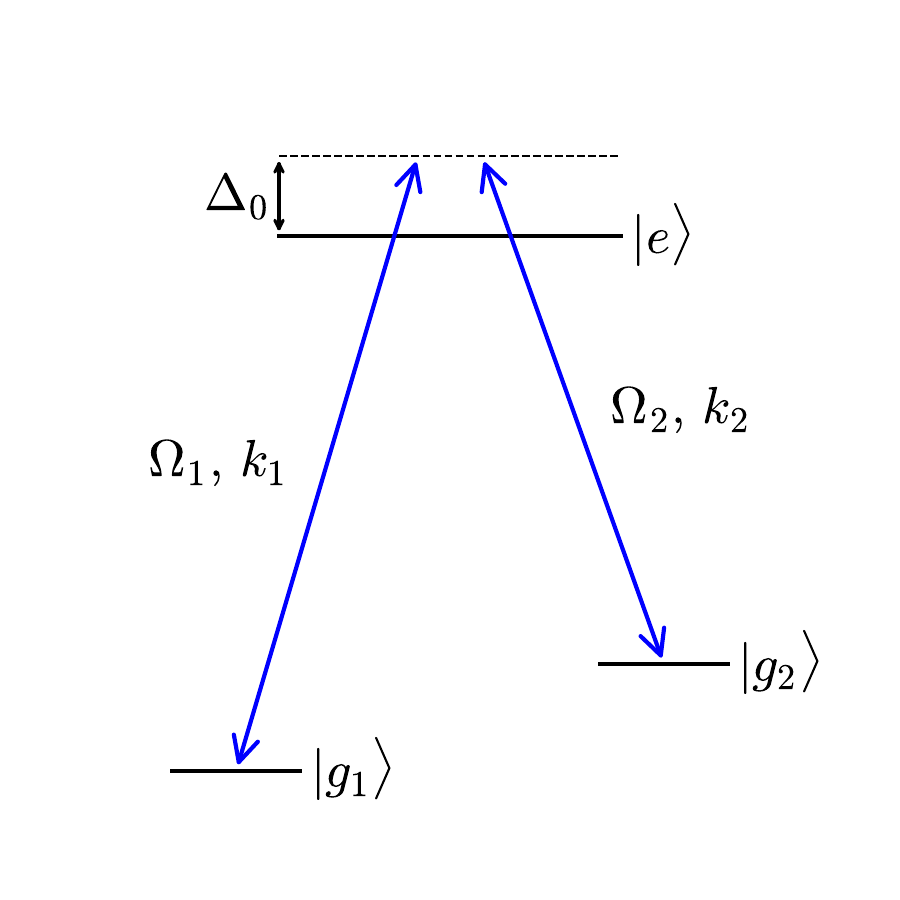}
        \label{fig:level_structure}
    \end{subfigure}
    ~ 
    \bigskip
    \begin{subfigure}{0.5\textwidth}
        \centering
        %\hspace*{0.15cm}  % shift image left
        \subcaption{}
        \includegraphics[scale=0.45]{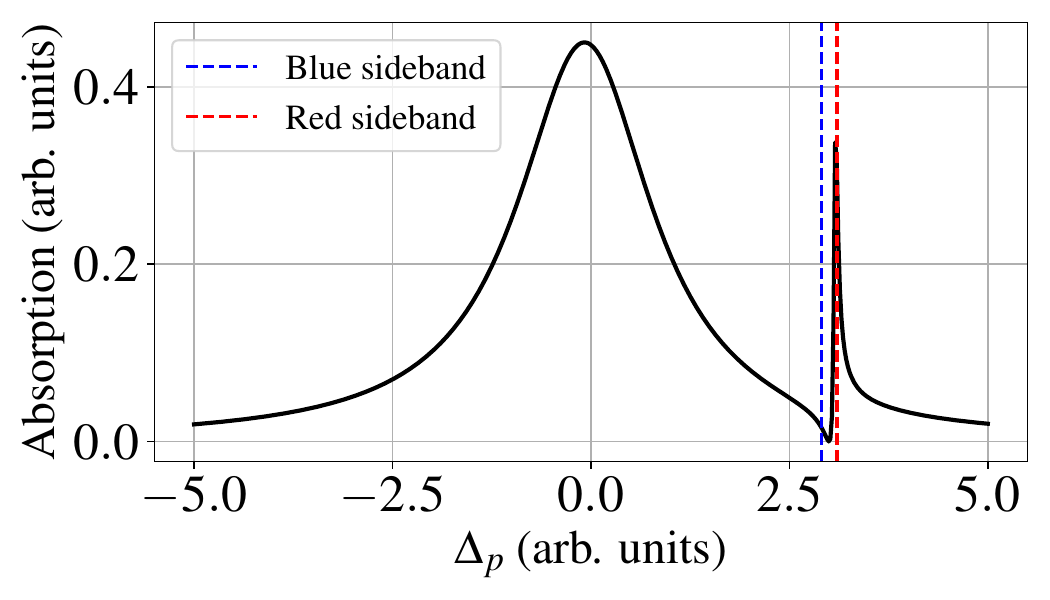}
        %\caption{}
        \label{fig:abs_spectrum}
    \end{subfigure}%
    ~
    \caption{(a) Level structure utilized in EIT cooling.  Two cooling lasers couple ground states $\ket{g_1}$ and $\ket{g_2}$ to an excited state $\ket{e}$.  The lasers are both blue-detuned from their respective transition frequencies, and are described by Rabi frequencies $\Omega_{1(2)}$ and wavevectors $\vec{k}_{1(2)}$. (b) The absorption spectrum of a weak probe is similar to a Fano profile, an example of which is plotted.  The spectrum includes a zero when the probe detuning is equal to the cooling beam detuning ($\Delta_p=\Delta_0$), followed by a narrow peak at positive $\Delta_p$.  When the motion-removing sideband transition associated with the crystal's normal modes (denoted by a red, dashed vertical line) is located within the bandwidth of this peak, photon scattering occurs frequently, removing energy from the crystal.}
    \label{prolate_crystal_cooling}
\end{figure}

Trapped ions must be cooled to ultracold temperatures for use in quantum information experiments \cite{leibfried2003}.  Cooling of the axial motional modes is particularly important, as they are often utilized in quantum simulation and sensing protocols \cite{gilmore2021, kiesenhofer2023}. Doppler cooling techniques can cool the axial motion to approximately $1$ mK. To achieve temperatures near the ground state, experiments with trapped-ion crystals now utilize sub-Doppler cooling techniques, most notably EIT cooling \cite{jordan2019,kiesenhofer2023,lechner2016}. In these systems, EIT cooling offers several advantages over resolved sideband cooling.  First, as we will describe, EIT cooling can be used to cool a large bandwidth of modes simultaneously, which is valuable in large ion crystals with many normal modes \cite{jordan2019}.  In contrast, sideband cooling can typically cool only a single mode, or possibly a few modes, at a time.  Furthermore, sideband cooling is typically inefficient outside of the Lamb-Dicke (LD) regime without the use of sophisticated protocols \cite{morigi1999,joshi2019,wu2023}. EIT cooling, however, is straightforward to implement and can remain efficient well outside of the LD regime \cite{bartolotta2024}. Therefore, it can be applied to Doppler-cooled crystals that may or may not satisfy the LD criterion.

We now briefly review the mechanism responsible for EIT cooling.  Further details can be found in \cite{morigi2000,morigi2003}.  The laser configuration (Fig.~\ref{fig:level_structure}) includes two beams that couple ground states $\ket{g_1}$ and $\ket{g_2}$, respectively, to the excited state $\ket{e}$ in a $\Lambda$ configuration.  In this work, we model the NIST Penning trap, which confines $^9\text{Be}^+$ ions.  In this case, $\ket{g_1}$ and $\ket{g_2}$ are the hyperfine levels $2s\;^2S_{1/2}(m_J=-1/2)$ and $2s\;^2S_{1/2}(m_J=+1/2)$, respectively, while $\ket{e}$ is the state $2p\;^2P_{3/2} (m_J=+1/2)$.  Assuming a magnetic field strength of $B_z = 4.4588$ T, the ground states are separated by 124 GHz, while $\ket{g_1}$ and $\ket{e}$ are separated by $\omega_{eg_1}=2\pi\times 957$ THz, corresponding to a laser beam wavelength of approximately 313 nm.  The beams coupling the ground states to the excited state are equally blue-detuned from their respective transitions ($\Delta_0\equiv\Delta_1=\Delta_2>0$).  In this setup, the absorption spectrum of a weak probe beam ($\Omega_p \ll \Omega_1,\Omega_2$) includes two distinct peaks, referred to as a ``broad" peak and a ``narrow" peak, as seen in Fig~\ref{fig:abs_spectrum}.  Ignoring ion motion, the maxima of these peaks occur at detunings $\Delta_p = (\Delta_0-\sqrt{\Delta_0^2+\Omega^2})/2$ and $\Delta_p = (\Delta_0+\sqrt{\Delta_0^2+\Omega^2})/2$, respectively, where $\Omega = \sqrt{\Omega_1^2+\Omega_2^2}$.  Furthermore, the absorption spectrum has a zero at $\Delta_p=\Delta_0$.  At this detuning, the system occupies a dark state, given by $\ket{\psi_d} = (\Omega_2\ket{g_1} - \Omega_1\ket{g_2})/\Omega$, in which the ground states are coherently superimposed and decoupled from $\ket{e}$ through destructive interference, preventing the absorption of photons from the cooling beams.  

By varying $\Delta_0$ and $\Omega$, one can tune the absorption profile to produce EIT cooling.  The carrier transition $\ket{g,n}\rightarrow\ket{e,n}$ is suppressed by placing the zero of the absorption spectrum near the carrier transition frequency. Then, $\Omega$ can be chosen such that the narrow peak in the spectrum coincides with the frequencies of the red-sideband transitions associated with the normal modes to be cooled.  In this case, transitions such as $\ket{g_{1,2},n}\rightarrow \ket{e,n-1}$ are enhanced while those like $\ket{g_{1,2},n}\rightarrow \ket{e,n+1}$ are suppressed, where $n$ represents the occupation number of a given normal mode.  When a normal mode with frequency $\omega$ is excited, the motion-removing (motion-adding) sideband experiences a probe detuning of $\Delta_0+\omega$ $(\Delta_0-\omega)$.  The probe detuning of the motion-removing sideband is near the narrow peak if

\begin{equation}
\label{EIT_condition}
\omega \approx \frac{-\Delta_0+\sqrt{\Delta_0^2+\Omega^2}}{2}.
\end{equation}

When this relation is satisfied, the red sideband transition occurs at a high rate while the blue sideband transition is suppressed, leading to overall cooling.  

\subsection{Simulation Code}

Our MD code evolves both the external and internal degrees of freedom of the ions, following the semiclassical model of EIT cooling presented in \cite{bartolotta2024}.  The internal state of each ion is evolved independently, according to the master equation

\begin{equation}
\label{dens_mat_eom}
\frac{d\hat{\rho}_j}{dt} = -\frac{i}{\hbar} \left[ \hat{H}_j(\boldsymbol{r}_j, t), \hat{\rho}_j \right] + \mathcal{D}_j \hat{\rho}_j.
\end{equation}
Here, $\hat{\rho}_j$ is the (3$\times$3) density matrix describing the internal state of the $j^{th}$ ion, $\hat{H}_j$ is the EIT laser-ion interaction Hamiltonian, and $\mathcal{D}_j \hat{\rho}_j$ is the Lindblad term describing spontaneous emission from the excited state to the ground states.  We begin by reviewing the ion-laser Hamiltonian, which is given by 

\begin{equation}
\label{beta}
\hat{H}^S = \sum_{j=1}^N (\hat{H}_{0,j}^S + \hat{H}_{1,j}^S(\boldsymbol{r}_j,t)).
\end{equation}
Here, $H_0^S$ describes the internal energy of the ion in the absence of the electric field and is expressed as 

\begin{equation}
\label{beta}
\hat{H}_{0,j}^S = -\hbar\sum_{\mu} (\omega_e-\omega_{g_\mu})\hat{\sigma}_{g_\mu g_\mu}^{(j)},
\end{equation}
where $\omega_e-\omega_{g_\mu}$ is the frequency difference between the excited state $\ket{e}$ and the ground state $\ket{g_\mu}$.  The Hamiltonian associated with the interaction of the EIT laser fields with ion $j$ is

\begin{equation}
\label{eq:H_1_schrodinger}
\hat{H}_{1,j}^S(\boldsymbol{r}_j, t) = \hbar\sum_{ \mu} \frac{\Omega^{(j)}_{\mu}(\boldsymbol{r}_j)}{2} \, \hat{\sigma}^{(j)}_{e g_\mu} \, e^{-i(\omega_{\mu}t - \boldsymbol{k}_{\mu}\cdot \boldsymbol{r}_j)} + \text{h.c.}
\end{equation}
In Eq.~\eqref{eq:H_1_schrodinger}, $\Omega^{(j)}_{\mu}(\boldsymbol{r}_j)$ is the Rabi frequency of the transition between $\ket{g_\mu}$ and $\ket{e}$ for ion $j$ and the operator $\hat{\sigma}^{(j)}_{e g_\mu} = \ket{e}_j\bra{g_\mu}_j$ describes internal-state transitions of ion $j$.  Since our simulation framework evolves ion motion classically, we take the position vector $\boldsymbol{r}_j$ to represent the classical position variable.  The spatial dependence of $\Omega^{(j)}_{\mu}$ allows for the possibility of non-uniform laser field intensities. The wavevector (frequency) of the laser beam which couples state $\ket{g_\mu}$ to $\ket{e}$ is $\boldsymbol{k}_{\mu}$ ($\omega_{\mu}$).  The EIT cooling lasers in the NIST Penning trap experiment, modeled in this work, propagate at an angle of $10^{\circ}$ from the horizontal, i.e $\boldsymbol{k}_1=k_1\cos(10^{\circ})\hat{\boldsymbol{x}}+k_1\sin(10^{\circ})\hat{\boldsymbol{z}}$ and $\boldsymbol{k}_2=k_2\cos(10^{\circ})\hat{\boldsymbol{x}}-k_2\sin(10^{\circ})\hat{\boldsymbol{z}}$.  The beams intersect at the center of the trap, as shown in Fig.~\ref{fig:penning_trap}.

To remove high-frequency oscillatory behavior from the Hamiltonian, it is convenient to perform the rotating-wave approximation (RWA) and move to the interaction picture.  For ion $j$, we use the free evolution Hamiltonian

\begin{equation}
\label{beta}
\hat{H}_{f,j}^S = -\hbar\sum_{\mu} \omega_\mu \hat{\sigma}^{(j)}_{g_\mu g_\mu}, 
\end{equation}
after which the total Hamiltonian for ion $j$ becomes

\begin{equation}
\begin{split}
\label{int_picture_Hamiltonian}
\hat{H}_j^I(\boldsymbol{r}_j, t) = &\hbar\sum_{\mu} \Delta_\mu \, \hat{\sigma}^{(j)}_{g_\mu g_\mu} \\
+ &\hbar\sum_{ \mu} \bigg(\frac{\Omega^{(j)}_{\mu}(\boldsymbol{r}_j)}{2} \, \hat{\sigma}^{(j)}_{e g_\mu} \, e^{i \boldsymbol{k}_{\mu}\cdot \boldsymbol{r}_j} + \text{h.c.}\bigg),
\end{split}
\end{equation}
where $\Delta_\mu = \omega_{\mu} - (\omega_e-\omega_{g_\mu})$ is the detuning of the EIT laser beam from the $\ket{g_\mu}$ to $\ket{e}$ transition.  This is the form of the Hamiltonian that we use in Eq.~\eqref{dens_mat_eom} and implement in our MD code.

The second term in Eq.~\eqref{dens_mat_eom} describes spontaneous emission from the excited state $\ket{e}$ and is given by

\begin{equation}
\label{dissipator}
\mathcal{D}_j \hat{\rho}_j = \sum_{\mu} \Gamma_{\mu}\big(\hat{\sigma}^{(j)}_{g_\mu  e}\hat{\rho}_j \hat{\sigma}^{(j)}_{e g_\mu}  - \frac{1}{2}\{\hat{\sigma}^{(j)}_{e  e},\hat{\rho}_j\}\big).
\end{equation}
Here, $\Gamma_{\mu}$ is the rate of spontaneous decay from $\ket{e}$ to $\ket{g_\mu}$.  In the NIST setup, $\Gamma_1 =2\pi\times 6$ MHz and $\Gamma_2 =2\pi\times 12$ MHz. Note that Eq.~\eqref{dissipator} does not include recoil from spontaneous emission. Our goal is to study cooling rates of large ion crystals, so we ignore recoil in this work.  This simplification is discussed further in the following section.

The previous discussion provides the necessary formalism to implement an EIT cooling model in our MD code.  Our numerical framework utilizes a cyclotronic integrator \cite{patacchini2009} to evolve the motional degrees of freedom of the ions and a Runge-Kutta (RK4) method to evolve the internal states. The ion motion is determined by the electric and magnetic fields of the Penning trap, Coulomb forces between ions, and EIT laser forces. Therefore, the total force on ion $j$ is

\begin{equation}
\begin{split}
\label{eq:total_force}
\boldsymbol{F}_j = &-q\nabla\phi_{\text{wall}} - q\nabla\phi_{\text{trap}} + q\boldsymbol{\dot{r}}_j\times B_z\boldsymbol{\hat{z}} \\ &+\boldsymbol{F}_{\text{laser}}^{(j)}
+\frac{q^2}{4\pi\epsilon_0}\sum_{i\neq j}\frac{\boldsymbol{r}_{j} - \boldsymbol{r}_{i}}{|\boldsymbol{r}_{j} - \boldsymbol{r}_{i}|^3},
\end{split}
\end{equation}
where the laser force is semiclassically approximated as 
\begin{equation}
\label{eq:laser_force}
\boldsymbol{F}_{\text{laser}}^{(j)} = -\hbar\sum_{\mu} \bigg(i\boldsymbol{k}_{\mu}\frac{\Omega_{\mu}^{(j)}}{2}+ \frac{\nabla\Omega_{\mu}^{(j)}}{2}\bigg)\langle \hat{\sigma}^{(j)}_{e g_\mu} \rangle e^{i \boldsymbol{k}_{\mu}\cdot \boldsymbol{r}_j}  +\text{h.c.}
\end{equation}
The expectation values of the coherence operators are computed using $\langle \hat{\sigma}_{eg_\mu}^{(j)} \rangle
=
\mathrm{Tr}\![\hat{\rho}_j \hat{\sigma}_{eg_\mu}^{(j)}]
=
\rho_{g_\mu e}^{(j)}.$ Further details of the numerical integration technique are provided in Appendix \ref{app:integrator}. Our simulation code is publicly available \cite{zaris_eit_code}.

\section{Single Ion Simulations}
\label{sec:single_ion}

Before applying our semiclassical model to ion crystals, we consider the case of a single trapped ion.  Since this system has only one axial mode, studying the cooling rate is especially straightforward. In this simple setting, we validate our code and illustrate key aspects of EIT cooling.  These results provide a foundation for later simulations with large ion crystals.

\subsection{Time-step convergence}

The Rabi frequency sets the smallest timescale that must be resolved in our simulations. We assume that the Rabi frequencies of the transitions $\ket{g_1}\rightarrow\ket{e}$ and $\ket{g_2}\rightarrow\ket{e}$ are equal ($\Omega_1=\Omega_2\equiv\Omega_0$). Optimal cooling is typically achieved by tuning the condition in Eq.~\eqref{EIT_condition} to the axial center-of-mass (COM) mode. Therefore, we use the Rabi frequency computed according to

\begin{equation}
\label{eq:rabi_COM_opt}
\Omega_0 = \sqrt{2\omega_{z}(\omega_{z}+\Delta_0)}.
\end{equation}

Using values for $\omega_z$ and $\Delta_0$ typical of models of the NIST Penning trap experiment, one finds $\Omega_0 \approx 2\pi\times 33.9$ MHz. A complete list of trap and laser parameters is provided in Table \ref{table1}. 

\begin{figure*}[t]
    \centering

    % Left column (a) and (b)
    \begin{minipage}{0.28\textwidth}
        \centering

        % --- (a) ---
        \begin{subfigure}{\textwidth}
            \centering
            \subcaption{}
            \includegraphics[width=\linewidth]{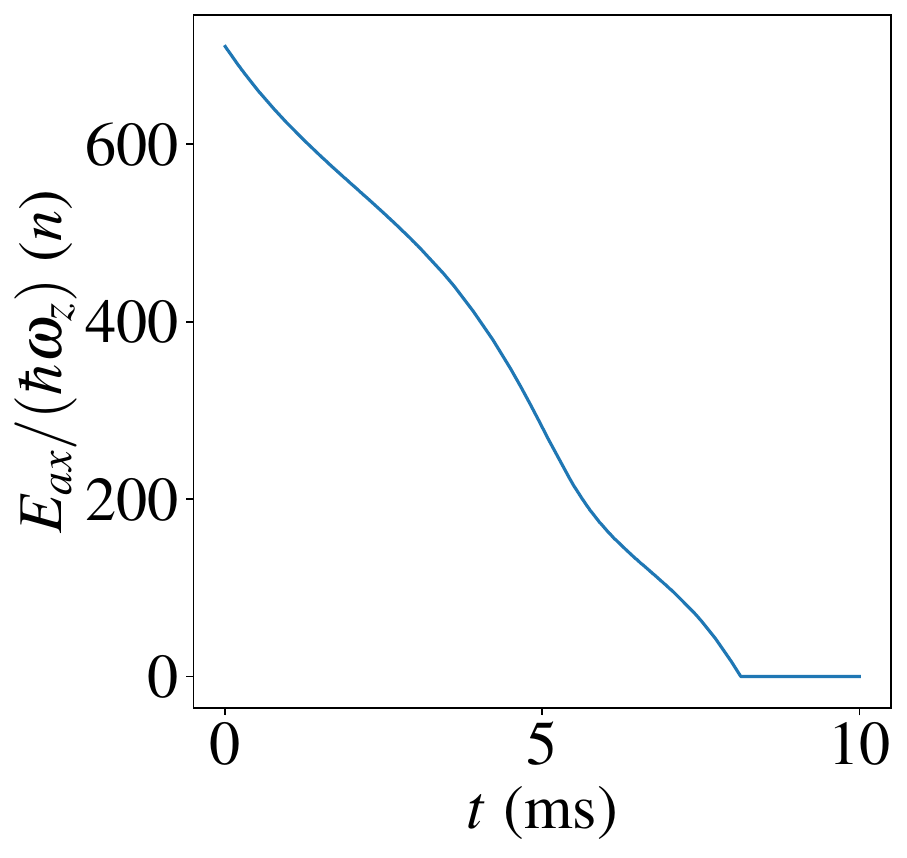}
            \label{fig:timestep_conv_a}
        \end{subfigure}

        \vspace{0.0cm}

        % --- (b) ---
        %\hspace*{0.4cm}
        \begin{subfigure}{\textwidth}
            \centering
            \subcaption{}
            \includegraphics[width=\linewidth]{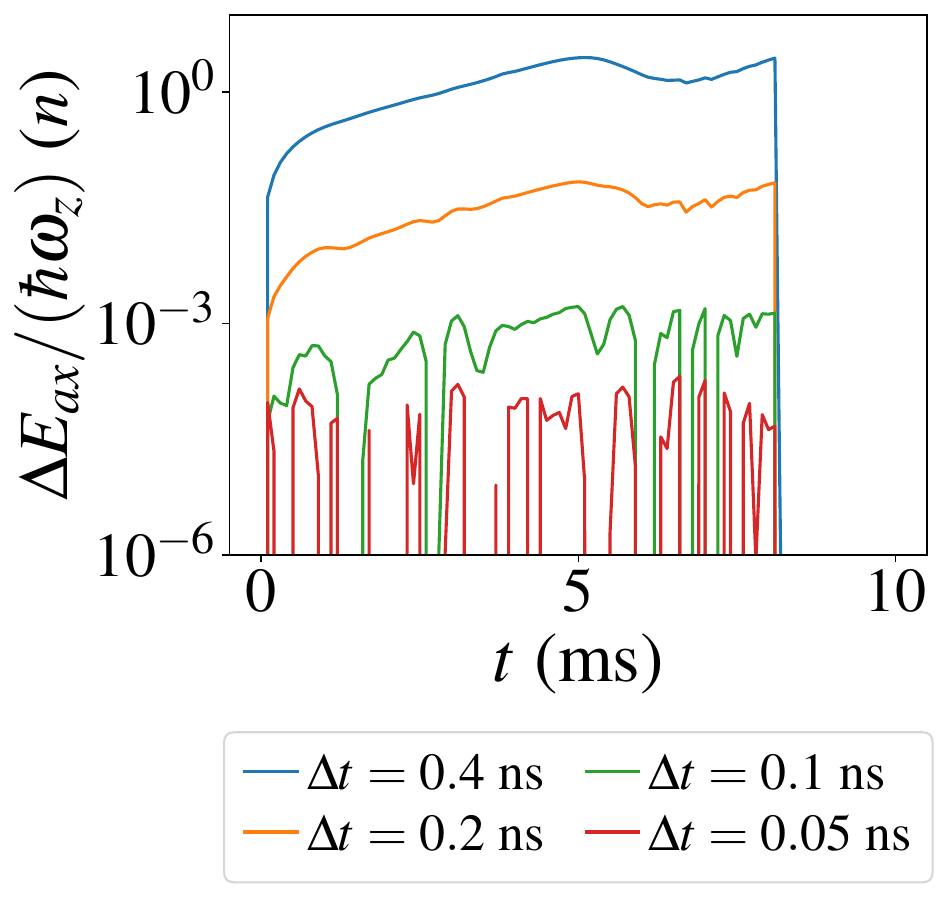}
            \label{fig:timestep_conv_b}
        \end{subfigure}

    \end{minipage}
    \hfill
    % Right column (c)
    \begin{minipage}{0.7\textwidth}
        \centering
        \begin{subfigure}{\textwidth}
            \centering
            \subcaption{}
            \includegraphics[width=\linewidth]{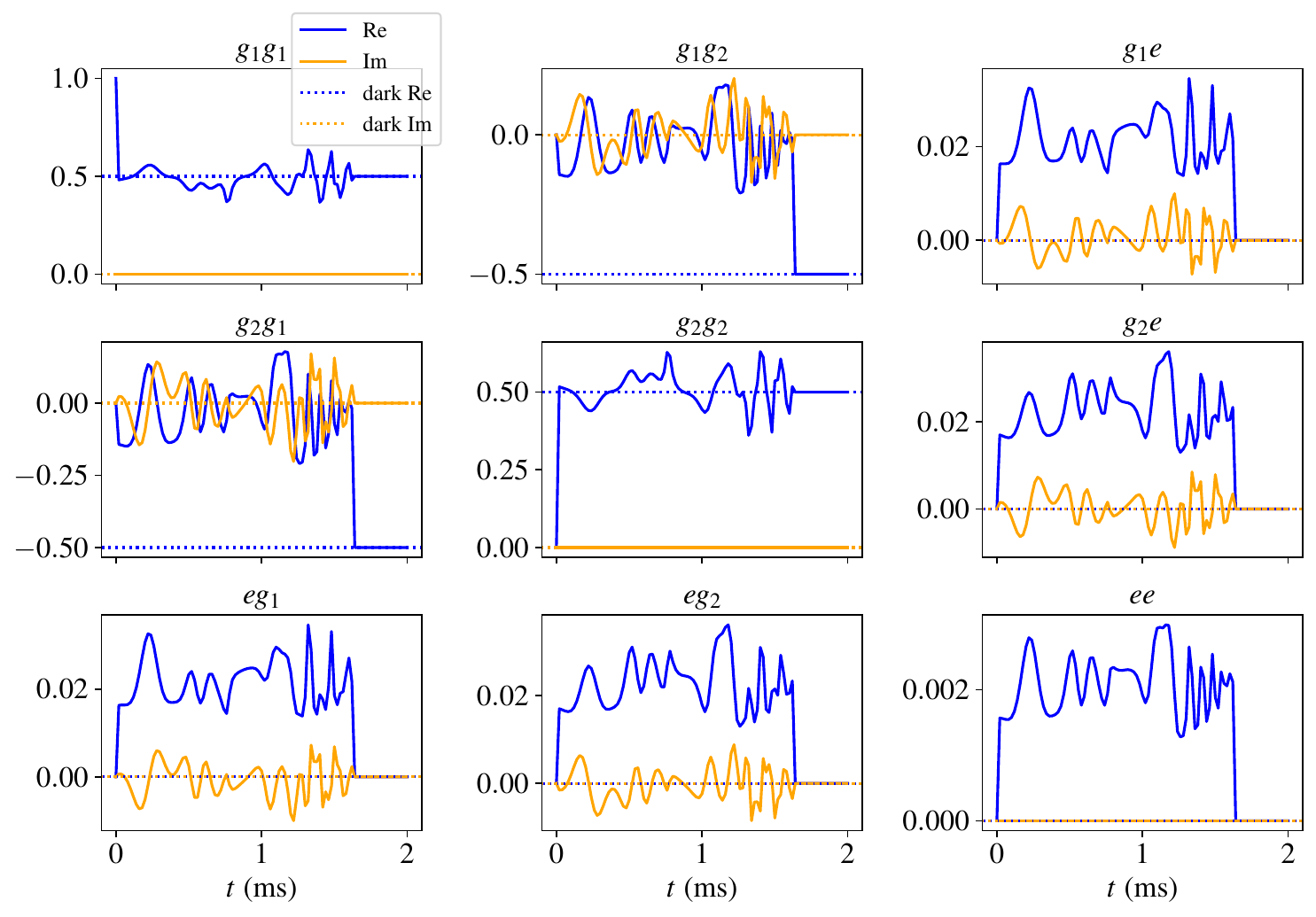}
            \label{fig:timestep_conv_c}
        \end{subfigure}
    \end{minipage}

    \caption{
    (a) The axial energy of a single trapped ion during cooling.  The ion is initialized at the center of the trap with an axial velocity of $10$ m/s (the thermal velocity corresponding to approximately 108 mK).  Since our semiclassical model does not include momentum diffusion, the energy approaches zero, which is not physical. Most trap and laser parameters can be found in Table~\ref{table1}.  This simulation uses no rotating wall (i.e., $\delta=0$) and the laser beam widths are assumed to be infinite. The ion begins in state $\ket{g_1}$ and the simulation time step is $\Delta t =0.1$ ns. (b) To test convergence with respect to the time step, we repeat the simulation with different values of $\Delta t$.  The energy difference $\Delta E_{\text{ax}}$ relative to the $\Delta t = 0.05$ ns simulation is plotted versus time.  Convergence is observed and the error at $\Delta t=0.1$ ns is small compared to the energy scale in (a).  (c)  Our code also integrates the equations of motion for the internal state of the ion.  Here, we plot the time-evolution of the value of each matrix element of the transformed density matrix $\tilde{\hat\rho}$.  The real and imaginary parts of each element are shown in blue and orange, respectively.  As cooling occurs, the ion eventually approaches the dark state, $\tilde{\hat\rho}_{D}$, whose matrix elements are shown using dotted lines.
    }
    
    \label{fig:timestep_conv}
\end{figure*}

We begin by initializing an ion at the center of the trap with an axial velocity of $10$ m/s (which is the thermal velocity associated with a temperature of approximately 108 mK).  The internal state of the ion is initialized to $\ket{g_1}$ by setting the density matrix, in the $\{\ket{g_1},\ket{g_2},\ket{e}\}$ basis, to
\begin{equation}
\label{init_int_state}
\hat{\rho}(t=0) = \begin{pmatrix}
    1&0&0\\ 0&0&0\\ 0&0&0
\end{pmatrix}.
\end{equation} 
EIT cooling is then simulated for 10 ms with a time step of $0.1$ ns.  The cooling curve is plotted in Fig.~\ref{fig:timestep_conv_a}.  Since the COM mode is the only axial mode in this system, we plot the total axial energy (in units of mode quanta) using
\begin{equation}
\label{state_vecotr_evolution}
\frac{E_{\text{ax}}}{\hbar\omega_z} = \frac{1}{\hbar\omega_z} \left(\frac{1}{2}mv_z^2 + \frac{1}{2}m\omega_z^2z^2 \right).
\end{equation}

In Fig.~\ref{fig:timestep_conv_a} the ion's energy approaches zero during cooling.  This is a consequence of ignoring spontaneous emission recoil and heating due to blue-sideband transitions in our model.  With these simplifications, the ion cooling rate remains large until $\bar n\ll 1$, which creates the appearance of a corner in the cooling curve near $\bar n=0$.  Later, we will compare the semiclassical cooling results to a quantum treatment to assess the validity of the semiclassical model. 

We test for time-step convergence by repeating the single ion simulation using various time steps. In Fig.~\ref{fig:timestep_conv_b}, we plot the difference in axial energy between the simulations as the time step is reduced.  Additional convergence tests using a large ion crystal are described in Appendix ~\ref{app:integrator}. In the remainder of this work, we use a time step of $\Delta t = 0.1$ ns.

Analysis of the density-matrix evolution shows that the internal state approaches the dark state of the Hamiltonian during cooling. For equal laser detunings and real Rabi frequencies, this state is given by

\begin{equation}
\label{state_vecotr_evolution}
\ket{D}(\boldsymbol{r}) = \frac{\Omega_2\ket{g_1} - \Omega_1e^{i\phi_D(\boldsymbol{r})}\ket{g_2}}{\sqrt{|\Omega_1|^2+|\Omega_2|^2}},
\end{equation}
where $\phi_{D}(\boldsymbol{r}) = (\boldsymbol{k}_1-\boldsymbol{k}_2)\cdot \boldsymbol{r}$.  When the Rabi frequencies are equal, this state corresponds to the density matrix

\begin{equation}
\label{state_vecotr_evolution}
\hat{\rho}_D = \frac{1}{2}\begin{pmatrix}
    1&-e^{-i\phi_D}&0\\ -e^{i\phi_D}&1&0\\ 0&0&0
\end{pmatrix}.
\end{equation} 
Note that the dark state is sensitive to ion motion due to its spatial dependence.  Therefore, the dark state is, in general, time-dependent. The dark state is approximately constant only when an ion is localized such that $(\boldsymbol{k}_1-\boldsymbol{k}_2)\cdot \boldsymbol{r}$ is approximately constant, known as the Lamb-Dicke regime.  To facilitate comparison with the dark state, we introduce a position-dependent unitary transformation of the density matrix of ion $j$, given by

\begin{equation}
\label{eq:transformed_rho}
    \tilde{\hat\rho}_j(t)= \hat U_j(t)\hat\rho_j(t)\hat U_j^\dagger(t),
\end{equation}
where $\hat{U}_j(t)$ is defined as

\begin{equation}
\label{eq:unitary_phase_transform}
    \hat U_j(t) =
    \begin{pmatrix}
        e^{i\boldsymbol{k}_1\cdot\boldsymbol{r}_j(t)} & 0 & 0 \\
        0 & e^{i\boldsymbol{k}_2\cdot\boldsymbol{r}_j(t)} & 0 \\
        0 & 0 & 1
    \end{pmatrix}.
\end{equation}
The transformation given in Eq.~\eqref{eq:transformed_rho} is useful because the transformed dark state is independent of position, given explicitly, when $\Omega_1=\Omega_2$, by
\begin{equation}
\label{state_vecotr_evolution}
\tilde{\hat{\rho}}_D = \frac{1}{2}\begin{pmatrix}
    1&-1&0\\ -1&1&0\\ 0&0&0
\end{pmatrix}.
\end{equation}

The evolution of the ion's transformed density matrix elements, computed using Eq.~\eqref{eq:transformed_rho}, is plotted in Fig.~\ref{fig:timestep_conv_c}.  During cooling, the matrix elements $\tilde{\rho}_{g_1g_1}$ and $\tilde{\rho}_{g_2g_2}$ quickly approach a value of $0.5$ and fluctuate about this value.  Toward the end of cooling, the ion's internal state approaches the dark state. In our simulations, the ion rapidly approaches the dark state only after the axial motion has been cooled to $E_{\text{ax}}/(\hbar\omega_z) < 1$. In this regime, spontaneous emission recoil is expected to significantly affect cooling dynamics, so the semiclassical model is inaccurate.  Nonetheless, experiments have shown that EIT cooling can cool ions to near the ground state ($E_{\text{ax}}/(\hbar\omega_z) < 1$) \cite{jordan2019}.  Therefore, it is plausible that the system would approach the dark state even if spontaneous emission recoil were accounted for. In later sections, we will comment on the convergence to the dark state for large ion crystals.

\begin{table}[b]%The best place to locate the table environment is directly after its first reference in text
\caption{\label{table1}%
The values of key Penning trap and EIT laser beam parameters.  For parameters that are varied, we list the range of values used.
}
\begin{ruledtabular}
\begin{tabular}{ll}
\textrm{Parameter}&
\textrm{Value(s)}\\
\colrule
$\omega_r$ & $2\pi\times 180$ kHz - $2\pi\times 240$ kHz\\
$B_z$ & 4.4588 T\\
$\omega_{z}$ & 2$\pi\times$ 1.59 MHz\\
$\Omega_0$ & 2$\pi\times$ 33.9 MHz\\
$\Delta_1=\Delta_2$ & 2$\pi\times$ 360 MHz\\
$\Gamma_1$ & 2$\pi\times$ 6 MHz\\
$\Gamma_2$ & 2$\pi\times$ 12 MHz\\
$k_1$ & 2$\pi\times$ (957 THz + 360 MHz)/c\\
$k_2$ & 2$\pi\times$ (957 THz - 124 GHz + 360 MHz)/c\\
$w_1=w_2$ & 10 $\mu$m-$\infty$\\
\end{tabular}
\end{ruledtabular}
\end{table}

\subsection{Capture range}

While EIT cooling is often studied in the Lamb-Dicke regime, it is evident from Fig.~\ref{fig:timestep_conv} that EIT can effectively cool systems that are excited well beyond the Lamb-Dicke limit.  The range of occupation numbers $n$ of a given mode that result in cooling is known as the ``capture range.''  Above this range, EIT cooling fails and can instead heat the trapped ion or ion crystal.  In this section, we characterize the capture range of a single trapped ion. This study provides insights into the experimental limitations of EIT cooling. Furthermore, by comparing the results to theoretical predictions, we further validate our MD code.

Before studying the capture range, it is useful to define the cooling rate of a single ion's axial mode. We define the axial energy of the ion as $E_{\text{ax}}$. The rate at which energy is removed from the ion's axial motion, in the semiclassical approximation, is then given by \cite{bartolotta2024} 

\begin{equation}
\label{eq:cooling_rate}
\frac{dE_{\text{ax}}}{dt} = \hbar\omega_z\frac{d\bar n}{dt}  = F_{\text{laser},z}\dot{z}.
\end{equation}

\begin{figure}
    %\centering
    %\includegraphics[scale=0.5]{prolate_crystal_cooling_072825.pdf}
    \centering
    \begin{subfigure}{0.5\textwidth}
        \centering
        %\hspace*{-0.45cm}  % shift image left
        \subcaption{}
        \includegraphics[scale=0.50]{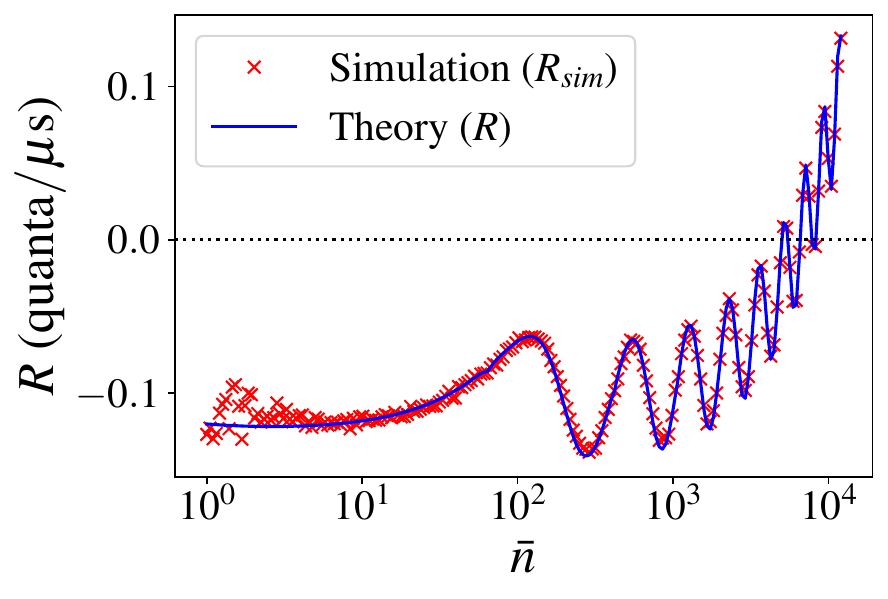}
        \label{fig:single_ion_R_a}
    \end{subfigure}
    \par\vspace{0.3cm}
    \begin{subfigure}{0.5\textwidth}
        \centering
        %\hspace*{-0.45cm}  % shift image left
        \subcaption{}
        \includegraphics[scale=0.50]{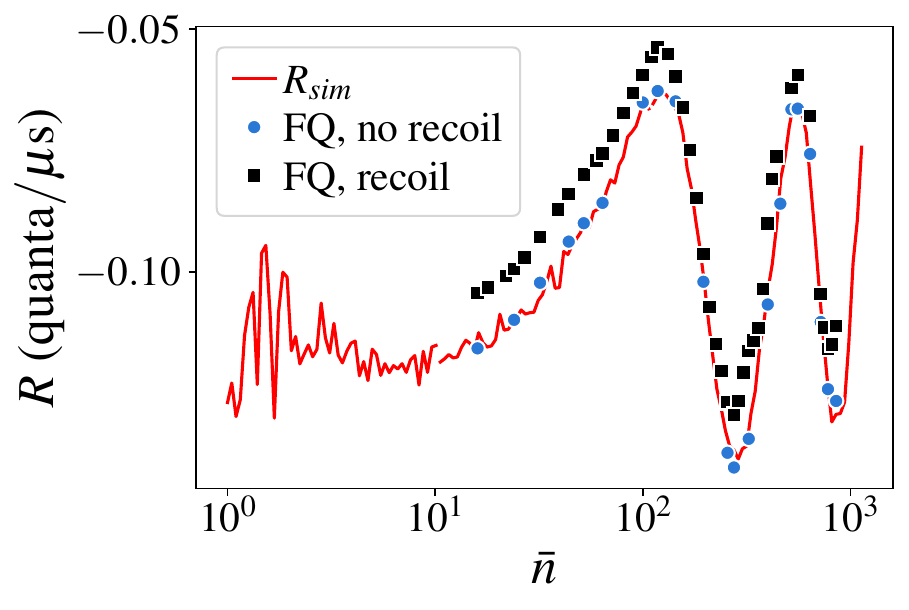}
        \label{fig:single_ion_R_b}
    \end{subfigure}
    \caption{(a) The cooling rate of the axial COM mode for a single trapped ion, in the semiclassical approximation, is plotted as a function of occupation number.  Most trap and laser beam parameters are given in Table~\ref{table1}.  In these simulations there is no rotating-wall potential and we use $w_1=w_2=\infty$.  The theoretical values are computed using the dynamic steady-state approach from \cite{bartolotta2024}. Good agreement is observed between the simulated and theoretical semiclassical cooling rates.  For $\bar n\leq 10$, the simulated cooling rate is harder to estimate since the ion's motion is cooled to near $\bar n=0$ very quickly.  The capture range of this cooling setup is several thousand motional quanta. (b)  The FQ cooling rates are plotted for a range of $\bar n$ values.  When recoil is omitted, they agree with the rates from the semiclassical simulations, although at sufficiently low $\bar n$ (not shown here), the results are expected to deviate.  Including recoil in the FQ simulations introduces a heating mechanism and, therefore, slightly reduces the cooling rate.}
    \label{fig:theory_sim_cooling_rates}
\end{figure}

Here, the mean occupation number $\bar n$ is defined according to $E_{\text{ax}} = \hbar\omega_z\bar n$ (ignoring the zero-point energy) and $F_{\text{laser},z}$ refers to the $z$-component of the laser force.  The time-averaged rate of change of the mean occupation number, $R$, is then computed according to 

\begin{equation}
\label{state_vecotr_evolution}
R = \frac{1}{T_{\text{av}}}\int_0^{T_{\text{av}}} \frac{d\bar n(t)}{dt}dt,
\end{equation}
where $T_{\text{av}}$ is an averaging time that is large compared to $1/\omega_z$. An algorithm to compute $R$ is presented in \cite{bartolotta2024}. Briefly, it finds the dynamic steady state of the internal-state density matrix by solving the optical Bloch equations.  Then, the laser force is computed according to Eq.~\eqref{eq:laser_force} and the cooling rate is found using Eq.~\eqref{eq:cooling_rate}.  We note that, to simplify the optical Bloch equations, they are solved on the reduced subspace of the ground states. This reduction requires adiabatically eliminating the excited state.  For more details, see \cite{bartolotta2024}.  In Fig.~\ref{fig:single_ion_R_a}, we plot the theoretical and simulated cooling rates as a function of $\bar n$. In the simulations, the ion is initialized with position $z=C\cos\alpha$ and velocity $\dot{z} = -C\omega_z\sin\alpha$, where $\alpha$ is a random angle between 0 and $2\pi$ and $C = \sqrt{2\hbar \bar n/(m\omega_z)}$.  For each value of $\bar n$, 10 independent realizations are simulated.  Since we are only concerned with axial motion in this case, we remove the planar components from the EIT laser beam wavevectors and set $\boldsymbol{k}_1 = -\boldsymbol{k}_2 = (2\pi\times (957$ THz + $360$ MHz$)/c)\sin(10^\circ)\hat{\boldsymbol{z}}$ in these simulations. The cooling rate $R_{\text{sim}}$ is approximated by averaging the cooling rate over a time $t_{\text{avg}}$.  As in \cite{bartolotta2024}, we wait a short time (2 $\mu$s) before averaging to allow the internal state of the ion to reach a dynamic steady state. For $\bar n\gtrsim 10$, the cooling rate is then computed by averaging the cooling rate over 50 $\mu$s.  For $\bar n<10$, a shorter duration is used since the ion is quickly cooled to near-zero occupation.  This leads to some deviation from the theoretical cooling predictions at low $\bar n$.

\subsection{Comparison to quantum model}
\label{sec:fq}

To validate the semiclassical model, we also compare our single-ion cooling results to those obtained using a `fully quantum' (FQ) simulation code capable of modeling EIT cooling of a single ion in one spatial dimension \cite{balaji_eit_code}.  The 1D Hamiltonian is explicitly given by

\begin{equation}
\label{H_FQ}
\begin{split}
\hat{H}_{\text{FQ}} = &\frac{\hat{p}^2}{2m}+\frac{1}{2}m\omega_z^2\hat{z}^2+\hbar\sum_{\mu} \Delta_\mu \, \hat{\sigma}_{g_\mu g_\mu} \\
+ &\hbar\sum_{ \mu} \bigg(\frac{\Omega_{\mu}}{2} \, \hat{\sigma}_{e g_\mu} \, e^{i k_{z,\mu}z} + \text{h.c.}\bigg),
\end{split}
\end{equation}
where $\hat{p}$ is the ion's axial momentum, $k_{z,\mu}$ is the axial wavevector component of laser beam $\mu$, and we have assumed a spatially uniform Rabi frequency. Dissipative dynamics are included by defining an effective non-Hermitian  Hamiltonian

\begin{equation}
\label{H_eff_FQ}
\hat{H}_{\text{eff}} = \hat{H}_{\text{FQ}}-\frac{i\hbar}{2}\sum_\mu L_\mu^\dagger L_\mu.
\end{equation}
Here, $L_\mu$ is the jump operator associated with spontaneous emission from $\ket{e}$ to $\ket{g_\mu}$. Denoting the angular distribution of photon emission from $\ket{e}$ to $\ket{g_{\mu}}$ as $f_\mu(\Omega_{\text{em}})$ and the associated direction as $\hat{\boldsymbol{n}}(\Omega_{\text{em}})$, the jump operator can be parameterized by $L_\mu = \sqrt{\Gamma_\mu f_\mu(\Omega_{\text{em}})}e^{-ik_{\mu}\hat{\boldsymbol{n}}(\Omega_{em})\cdot\hat{z}}\hat{\sigma}_{g_{\mu}e}$.  When recoil is ignored, this reduces to the form of the jump operators used in Eq.~\eqref{dissipator}.

The FQ code simulates individual pure-state trajectories by evolving $\ket{\psi(t)}$ according to

\begin{equation}
\label{eq:Heff_evolution}
\ket{\psi(t+\Delta t)} = e^{-i\hat{H}_{\text{eff}}\Delta t/\hbar}\ket{\psi(t)}.
\end{equation}
Note that $\ket{\psi(t)}$ belongs to the tensor-product Hilbert space of the motional and internal states of the ion. A description of the numerical integration technique used for Eq.~\eqref{eq:Heff_evolution} is provided in Appendix \ref{app:FQ}. The non-Hermitian term in Eq.~\eqref{H_eff_FQ} describes the conditional backaction associated with the absence of a spontaneous emission event, resulting in a reduction of the norm $|| \psi(t)||$ over time. A spontaneous emission event is deemed to occur when $|| \psi(t)||^2 \leq r$, where $r$ is a random number sampled uniformly from $U(0,1)$ \cite{daley2014}.  At the time step when an event occurs, the decay pathway is randomly chosen according to the relative decay rates.  Explicitly, the probability of decay to state $\ket{g_\mu}$ is given by $P_\mu = \Gamma_\mu / (\Gamma_1+\Gamma_2)$. The direction of the emitted photon, $\hat{\boldsymbol{n}}$, is then sampled from the distribution $f_\mu(\Omega_{\text{em}})$ and the state is renormalized according to  

\begin{equation}
\label{renorm}
\ket{\psi(t)}\to\ \frac{e^{-ik_\mu\hat{\boldsymbol{n}}\cdot \hat{z}}\ket{g_\mu}\braket{e|\psi(t)}}{||e^{-ik_\mu\hat{\boldsymbol{n}}\cdot \hat{z}}\ket{g_\mu}\braket{e|\psi(t)}||}.
\end{equation}
Finally, a new random number $r'\in U(0,1)$ is generated to set the norm threshold for the next spontaneous emission event.

The expectation value of an observable $\hat{O}$ can then be estimated by averaging over many ion trajectories.  Specifically, if $N_{t}$ is the number of simulated trajectories, then the expected value of $\hat{O}$ at time $t$ is given by \cite{daley2014}

\begin{equation}
\label{FQ_exp_value}
\braket{\hat O(t)} \approx \frac{1}{N_t} \sum_{k=1}^{N_t}
\frac{\bra{\psi_k(t)} \hat O \ket{\psi_k(t)}}{\braket{\psi_k(t)|\psi_k(t)}}
\end{equation}

Equation \eqref{FQ_exp_value} allows us to compute the time evolution of the occupation number during cooling in the FQ simulations.  The resulting cooling rates are plotted in Fig.~\ref{fig:theory_sim_cooling_rates}. We note that our FQ code initializes the motional component of $\ket{\psi}$ in a Fock state, $\ket{m}$, of the axial trap Hamiltonian. However, to compare the FQ cooling rates to the semiclassical results, it is useful to consider initialization in a phase-averaged coherent (PAC) state \cite{bartolotta2024}.  The PAC state with mean occupation number $\bar{n}$ can be written in terms of the Fock states $\ket{m}$ as

\begin{equation}
\label{eq:pac}
\hat\rho_{\text{PAC}}(\bar{n}) = e^{-\bar{n}}\sum_{m=0}^\infty \frac{\bar{n}^m}{m!}\ket{m}\bra{m}.
\end{equation}
Fortunately, the cooling rates of PAC states can be estimated from the Fock-state simulations, as detailed in Appendix \ref{app:FQ}.  In Fig.~\ref{fig:single_ion_R_b}, we plot the estimated PAC-state cooling rates for a variety of values of $\bar{n}$, with and without spontaneous emission recoil.  Without recoil, we observe good agreement between the semiclassical and FQ results.  Including recoil slightly decreases the cooling rate, leading to an upward shift of the cooling rate curve.

\section{Planar crystal EIT cooling}
\label{sec:planar_crystals}

\begin{figure}
    %\centering
    %\includegraphics[scale=0.5]{prolate_crystal_cooling_072825.pdf}
    \centering
    \begin{subfigure}{0.2\textwidth}
        \centering
        %\hspace*{0.15cm}  % shift image left
        \subcaption{}
        \includegraphics[scale=0.18]{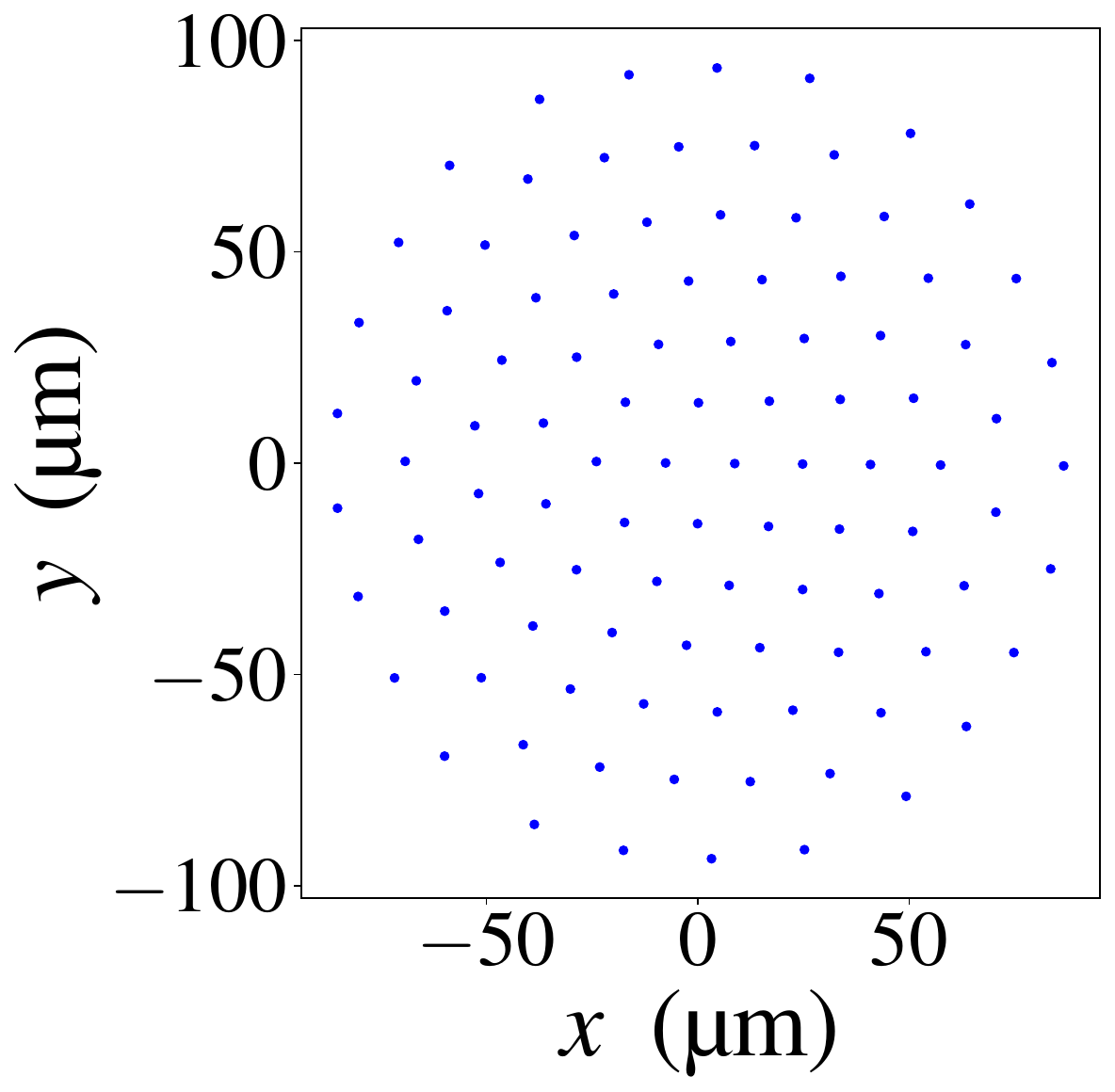}
        %\caption{}
        \label{fig:2D_crystal_eq}
    \end{subfigure}%
    \hspace*{0.25cm}
    \begin{subfigure}{0.2\textwidth}
        \centering
        %\hspace*{-0.45cm}  % shift image left
        \subcaption{}
        \includegraphics[scale=0.18]{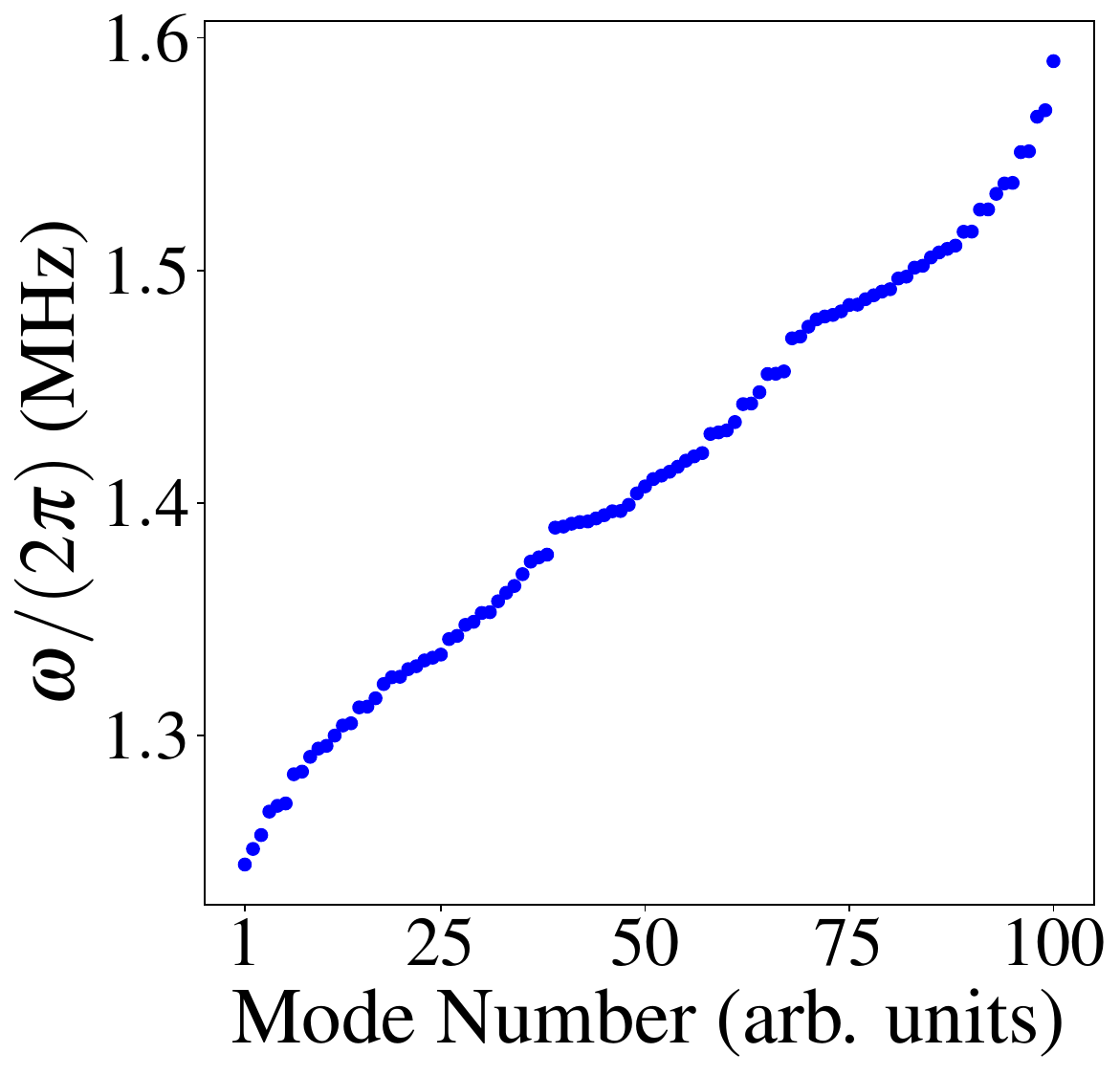}
        %\caption{}
        \label{fig:2D_crystal_mode_spec}
    \end{subfigure}
    ~ 
    \bigskip
    \vspace{-0.5cm}
    \begin{subfigure}{0.5\textwidth}
        \subcaption{}
        \centering
        \includegraphics[scale=0.36]{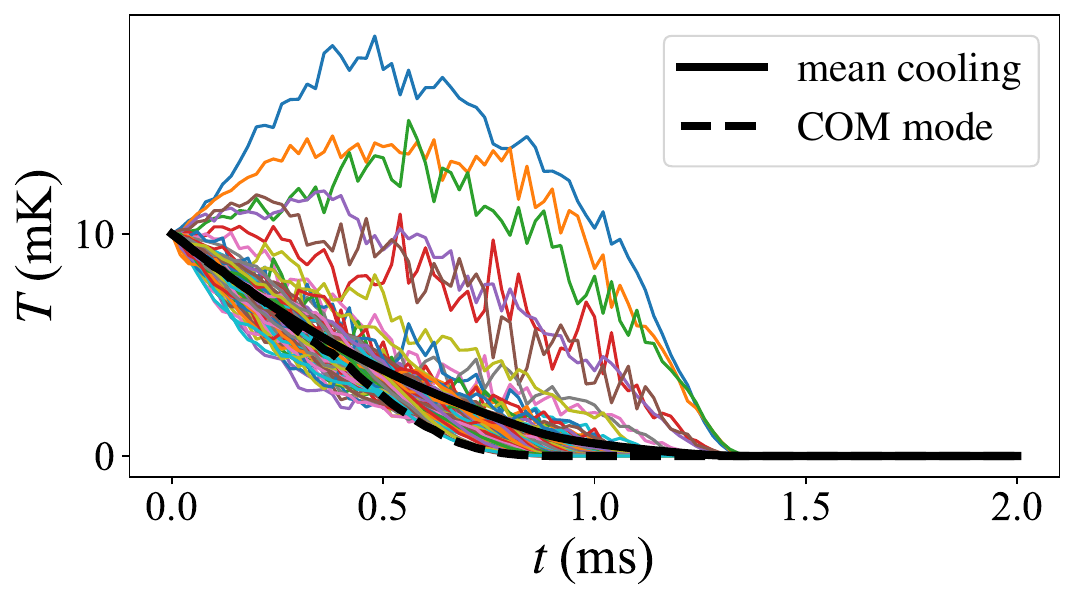}
        \label{fig:2D_crystal_axial_cooling_curves}
    \end{subfigure}
     ~ 
    \bigskip
    % Row 2
    \begin{subfigure}{0.2\textwidth}
        \centering
        %\hspace*{0.15cm}  % shift image left
        \subcaption{}
        \includegraphics[scale=0.18]{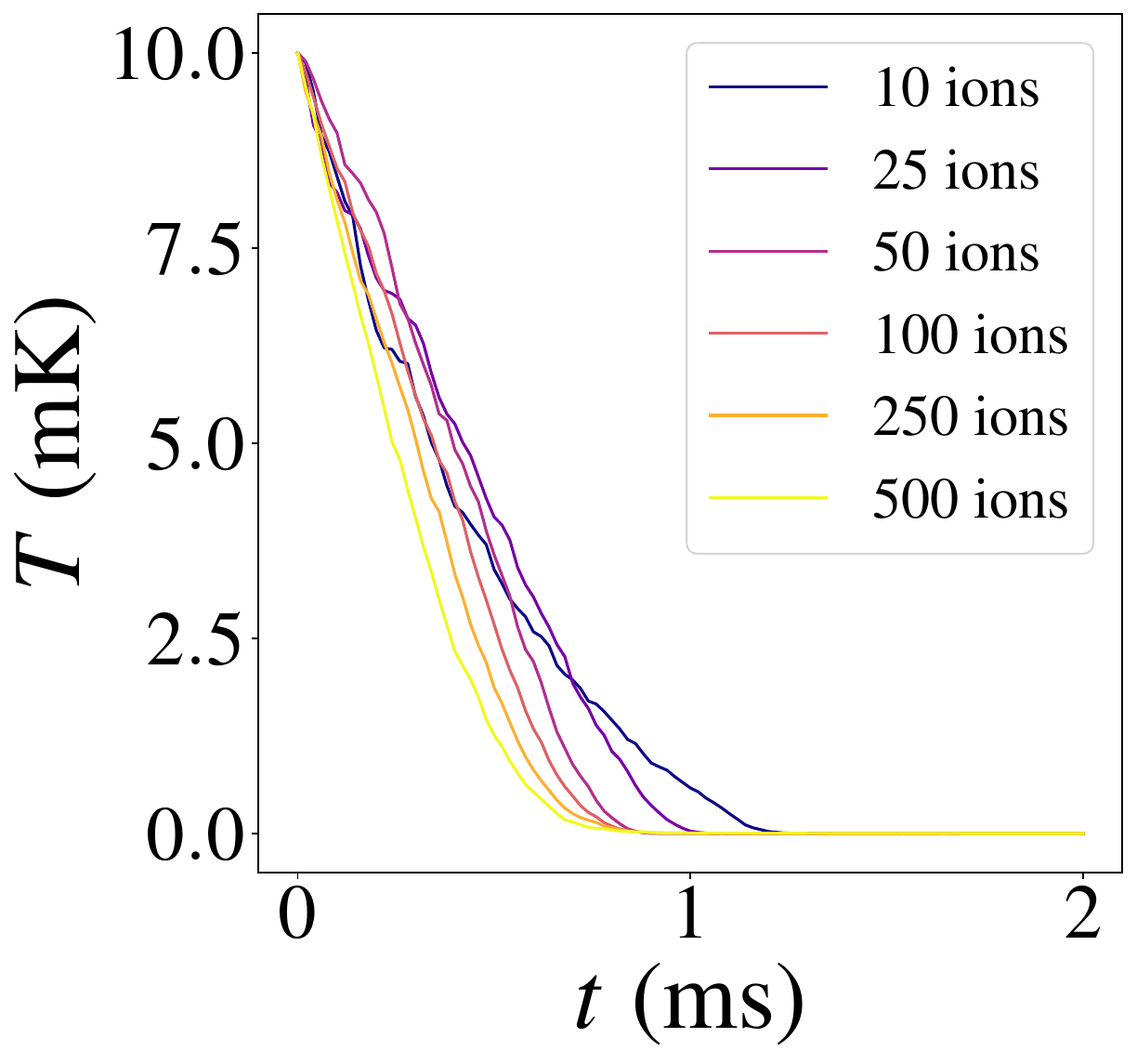}
        %\caption{}
        \label{fig:2D_crystal_COM_cooling}
    \end{subfigure}%
    \hspace*{0.25cm}
    \begin{subfigure}{0.2\textwidth}
        \centering
        %\hspace*{-0.45cm}  % shift image left
        \subcaption{}
        \includegraphics[scale=0.18]{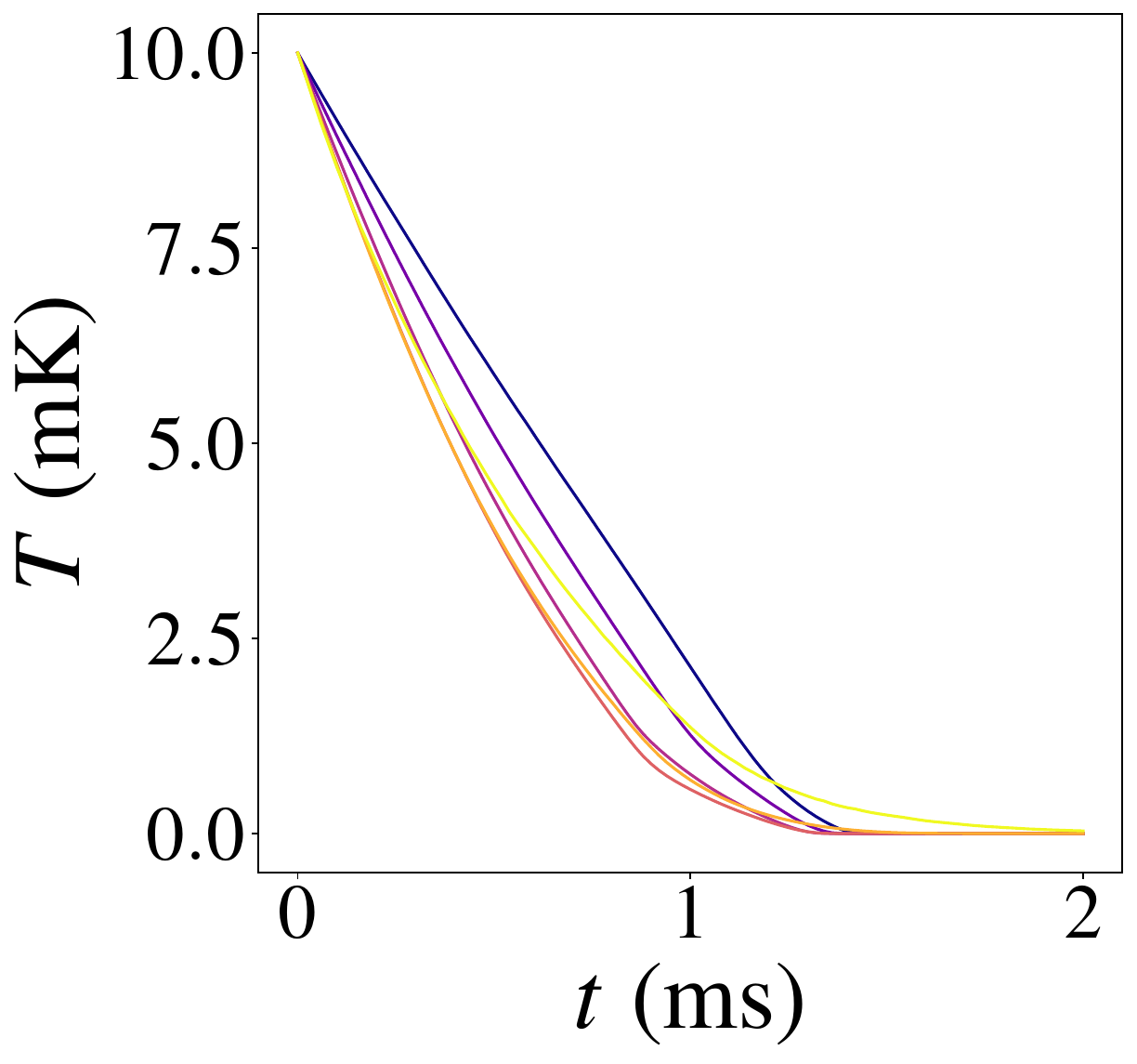}
        %\caption{}
        \label{fig:2D_crystal_mean_cooling}
    \end{subfigure}

    ~
    \caption{(a) Equilibrium configuration of a 2D crystal with $N=100$ ions.  Trap parameters include $\omega_r=2\pi\times 180$ kHz and $\delta = 1.7\times 10^{-3}$ (other parameter values are found in Table~\ref{table1}).  This crystal is excited by initializing the axial modes at $T_i$ = 10 mK. (b) Frequencies of the crystal's axial modes. The highest frequency mode is the center-of-mass (COM) mode with $\omega_z = 2\pi\times 1.59$ MHz.  (c) EIT cooling of the axial modes.  Each curve shows the temperature evolution of a single normal mode, averaged over 20 runs. In each run, the crystal is initialized with random mode phases.  The dashed black curve shows COM mode cooling and the solid black curve shows the mean cooling of all modes. The lowest frequency axial modes tend to be initially heated before cooling. (d)  Axial COM mode temperature (averaged over 20 runs) for planar crystals with different ion numbers, showing that the COM cooling rate increases as $N$ increases.  (e) Mean axial mode temperature as a function of $N$.  The mean cooling rate also increases with $N$. However, the $N=500$ crystal exhibits a reduced cooling rate at later times. This phenomenon is discussed in the text.}
    \label{fig:2D_crystal}
\end{figure}

\begin{figure}
    \centering
    \hspace*{-1.5cm}
    \begin{subfigure}{0.2\textwidth}
        \centering
        %\hspace*{0.15cm}  % shift image left
        \subcaption{}
        \includegraphics[scale=0.28]{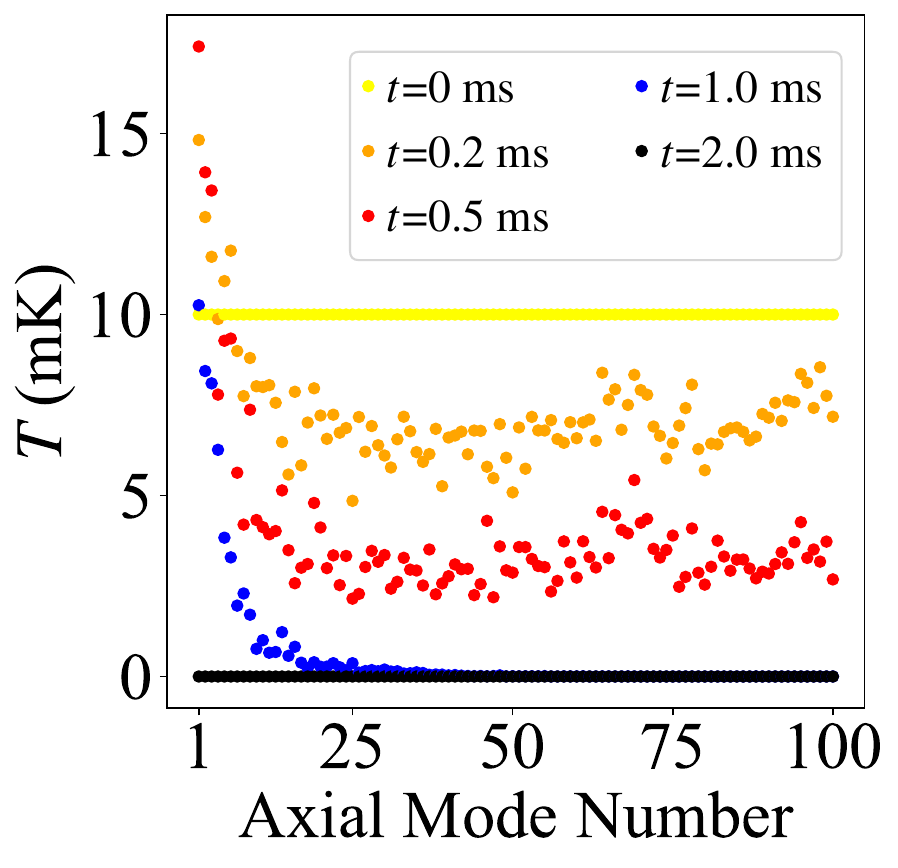}
        %\caption{}
        \label{fig:2D_100i_moed_resolved_cooling}
    \end{subfigure}%
    \hspace*{0.55cm}
    \begin{subfigure}{0.2\textwidth}
        \centering
        %\hspace*{-0.45cm}  % shift image left
        \subcaption{}
        \includegraphics[scale=0.28]{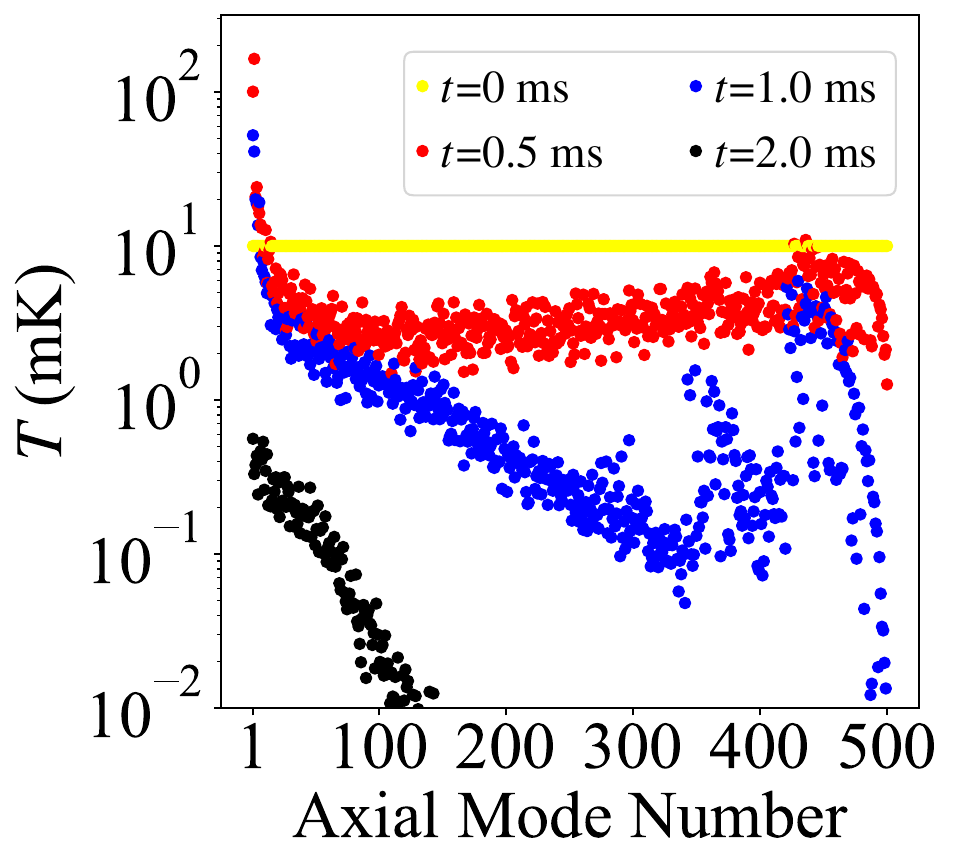}
        %\caption{}
        \label{fig:2D_500i_moed_resolved_cooling}
    \end{subfigure}
    ~ 
    \caption{(a) The temperature of each axial mode in an $N=100$ planar crystal is shown at several times during cooling.  The plotted mode temperatures are averaged over 20 realizations of the simulation. At $t=0$, all axial modes are initialized at $T_i^{\parallel}=10$ mK.  Resonant mode coupling asymmetrically transfers energy to the lowest-frequency modes, causing them to heat initially before cooling.  (b) The analogous plot for the $N=500$ ion crystal.  In this case, the initial heating of the low-frequency modes is more pronounced (note the logarithmic scale of the vertical axis).  For clarity, temperatures less than $0.1$ mK are not shown.}
    \label{fig:2D_mode_resolved_cooling}
\end{figure}

In this section, we extend our numerical study to the EIT cooling of trapped-ion crystals. Using our mode analysis code, we are able to initialize ion crystals with specified mode energies prior to running cooling simulations.  Of particular interest is the EIT cooling of the axial modes of crystals in the NIST setup.  These modes are utilized in quantum sensing experiments, and near-ground-state cooling provides increased sensitivity to small perturbations \cite{gilmore2021}. Here, we consider planar crystals; 3D crystals are studied in the following section. While theoretical EIT cooling curves have been obtained for planar trapped-ion crystals \cite{shankar2019}, our simulations provide valuable new insights. Unlike previous treatments, our MD code includes nonlinear mode-coupling effects. Furthermore, the simulated cooling dynamics are valid outside of the LD regime, providing information that is complementary to previous LD studies.

\subsection{Axial mode cooling without planar mode initialization}

Before simulating EIT cooling, an appropriate crystal must be initialized.  First, the equilibrium configuration of the crystal is found using energy minimization techniques \cite{wang2013,dubin1999}.  Next, the crystal's axial modes are initialized at a temperature $T_i^{\parallel}$.  We briefly explain this procedure here.  Further details are provided in Appendix \ref{app:normal_modes}. The axial equations of motion for the state vector $\ket{q_{\parallel}}=\ket{z_1, ...,z_N,\dot{z}_1, ...,\dot{z}_N} \equiv \ket{z_{1-N},\dot{z}_{1-N}}$ are given by

\begin{equation}
\begin{split}
\label{beta}
\frac{d\ket{q_{\parallel}}}{dt} &= \begin{bmatrix}
        \mathbb{0}_{N}& \mathbb{I}_{N}\\
        -\mathbb{K}_{\parallel}/m & \mathbb{0}_{N} \\
        \end{bmatrix}\ket{q_\parallel} \equiv \mathbb{D}\ket{q_\parallel},
\end{split}
\end{equation}
where $\mathbb{K}_{\parallel}$ is the axial stiffness matrix and $\mathbb{I}_N$ is the identity matrix of size $N$. The axial eigenmodes are found by diagonalizing the matrix $\mathbb{D}$.  This produces $N$ axial eigenmodes $\ket{u_j^{\parallel}}$, with associated frequencies $\omega_j^{\parallel}$, whose linear combinations can be used to describe any motional state of the crystal.  Specifically, the state vector can be expressed as 

\begin{equation}
\label{state_vecotr_evolution}
\ket{q_{\parallel}} = \sum_{j=1}^{N}\big[A_j^{\parallel}e^{-i(\omega_j^{\parallel}t+\varphi_j)}\ket{u_j^{\parallel}}+A_j^{\parallel }e^{i(\omega_j^{\parallel}t+\varphi_j)}\ket{u_j^{\parallel*}}\big],
\end{equation}
where $A_j^{\parallel}$ is the real amplitude of axial mode $j$.  To calculate the initial axial positions and velocities of the ions, we compute

\begin{equation}
\begin{split}
\label{state_vecotr_evolution}
\ket{q_{\parallel}}_{t=0} &= \sum_{j=1}^{N}\big[A_j^{\parallel}e^{-i\varphi_j}\ket{u_j^{\parallel}}+A_j^{\parallel}e^{i\varphi_j}\ket{u_j^{\parallel*}}\big] \\
&= 2\sum_{j=1}^{N}A_j^{\parallel}\text{Re}(e^{-i\varphi_j}\ket{u_j^{\parallel}}),
\end{split}
\end{equation}
and then set $A_j^{\parallel} = \sqrt{k_BT_i^{\parallel}}$ for all axial modes, where $k_B$ is the Boltzmann constant.  The phase $\varphi_j$ is useful, as it can be randomized to provide an ensemble of initial motional states.  For now, we assume that the planar modes are initialized at $0$ mK, but we relax this condition later. After initializing the crystal, its motion and internal state are evolved under EIT cooling in our MD simulation. 

In our planar crystal simulations, the rotating-wall frequency is set to $\omega_r=2\pi\times 180$ kHz and its strength is $\delta=1.7\times10^{-3}$.  The internal-state density matrix of each ion is initialized according to Eq.~\eqref{init_int_state}.  The EIT cooling beams are assumed to have uniform intensity in space.  Additional trap and laser parameters are provided in Table~\ref{table1}. Prior to cooling, all axial modes are initialized at $T_i^{\parallel}=10$ mK. The cooling results for crystals with various values of $N$ are shown in Fig.~\ref{fig:2D_crystal}.  Figure~\ref{fig:2D_crystal_eq} illustrates the equilibrium configuration of an $N=100$ ion crystal.  Figure~\ref{fig:2D_crystal_mode_spec} shows the frequencies of the crystal's axial modes.  The highest-frequency mode corresponds to the axial COM mode with $\omega_z=2\pi\times 1.59$ MHz. In Fig.~\ref{fig:2D_crystal_axial_cooling_curves}, we plot the time evolution of the axial mode temperatures of this crystal.  The plotted temperatures are obtained by averaging over 20 runs of the cooling simulation.  In each run, the crystal's normal modes are initialized with random phases $\varphi_j$.  All modes are cooled to a temperature below the ground state energy within a fraction of a millisecond.  The final cooling limits seen in Fig.~\ref{fig:2D_crystal_axial_cooling_curves} are not physical, but result from using a recoil-less semiclassical model that neglects diffusive heating effects. The cooling results at early times illustrate an interesting phenomenon in which certain modes (which are found to be the lowest-frequency axial modes) are heated before being cooled. This effect becomes more dramatic as $N$ increases, with the two lowest frequency modes of the $N=500$ ion crystal reaching a maximum temperature of approximately 150 mK before being cooled (not shown). This heating is not caused by the EIT lasers, but rather by energy exchange between the axial modes and the $\boldsymbol{E}\times\boldsymbol{B}$ modes.  Free-evolution simulations of the crystal (without EIT cooling) confirm this behavior.  They show that an increase in the planar potential energy, which is associated with the $\boldsymbol{E}\times\boldsymbol{B}$ modes, accompanies the transfer of energy into the lowest-frequency axial modes.  This phenomenon is discussed further in Appendix \ref{app:mode_coupling}.

The cooling of the axial COM mode is uniquely important, as this mode is often utilized in quantum sensing protocols.  As seen in Fig.~\ref{fig:2D_crystal_axial_cooling_curves}, the COM mode is among the most rapidly cooled modes as a result of optimizing the EIT cooling condition for this mode according to Eq.~\eqref{eq:rabi_COM_opt}.  In Fig.~\ref{fig:2D_crystal_COM_cooling}, we plot the COM mode temperature over time, averaged over 20 realizations, for simulations with different values of $N$.  As theoretically predicted \cite{shankar2019, khan2026}, we observe that the cooling rate tends to increase with ion number.  In Fig.~\ref{fig:2D_crystal_mean_cooling}, we plot the mean cooling rate of all axial modes. The mean cooling rate also improves as $N$ increases, up to $N=500$, where the cooling rate is initially high, but seems to degrade at later times.  To understand this reduced cooling rate at large $N$, it is useful to further analyze the mode-resolved cooling.  In Fig.~\ref{fig:2D_mode_resolved_cooling}, we plot the temperature of each mode at several snapshots during cooling for the $N=100$ and $N=500$ ion crystals. In the $N=100$ ion crystal, the lowest frequency modes are initially heated, as described previously. However, after 2 ms of cooling, all modes are cooled to well below their ground state temperatures. In contrast, in the $N=500$ ion crystal, approximately 15 low-frequency modes have temperatures greater than $0.3$ mK after 2 ms of cooling.  This means that, at later times, the energy present in the lower frequency modes accounts for the higher mean temperature for the $N=500$ crystal seen in Fig.~\ref{fig:2D_crystal_mean_cooling}.  The decreased cooling rate in the $N=500$ ion crystal may also be related to the cooling of the higher frequency modes seen in Fig.~\ref{fig:2D_mode_resolved_cooling}.  Several modes between mode numbers 400 and 500 appear to experience especially weak cooling between 0.5 ms and 1 ms.  This approximately coincides with the time at which the mean cooling rate begins to degrade in Fig.~\ref{fig:2D_crystal_mean_cooling}.

\subsection{Mode cooling with planar mode initialization}
\label{sec:2D_excited_planar_modes}

\begin{figure}
    %\centering
    %\includegraphics[scale=0.5]{prolate_crystal_cooling_072825.pdf}
    \centering
    \begin{subfigure}{0.2\textwidth}
        \centering
        %\hspace*{0.15cm}  % shift image left
        \subcaption{}
        \includegraphics[scale=0.25]{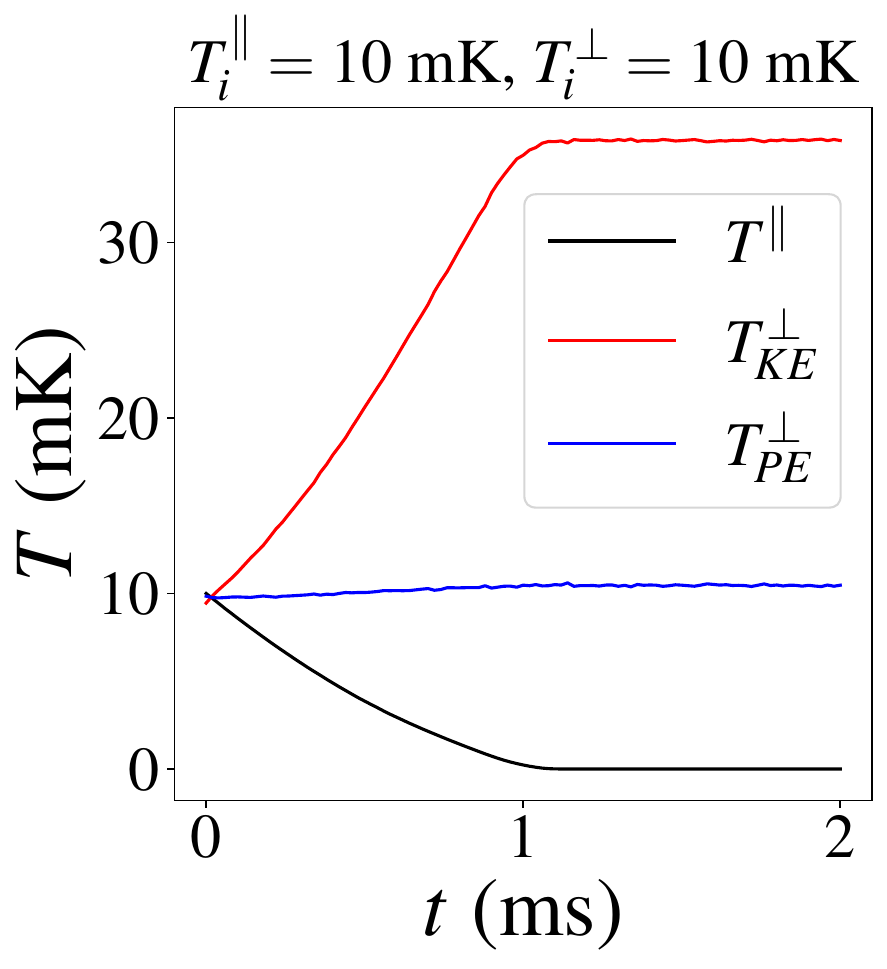}
        %\caption{}
        \label{fig:2D_T10mK_Tplanar10mK}
    \end{subfigure}%
    \hspace*{0.25cm}
    \begin{subfigure}{0.2\textwidth}
        \centering
        %\hspace*{-0.45cm}  % shift image left
        \subcaption{}
        \includegraphics[scale=0.25]{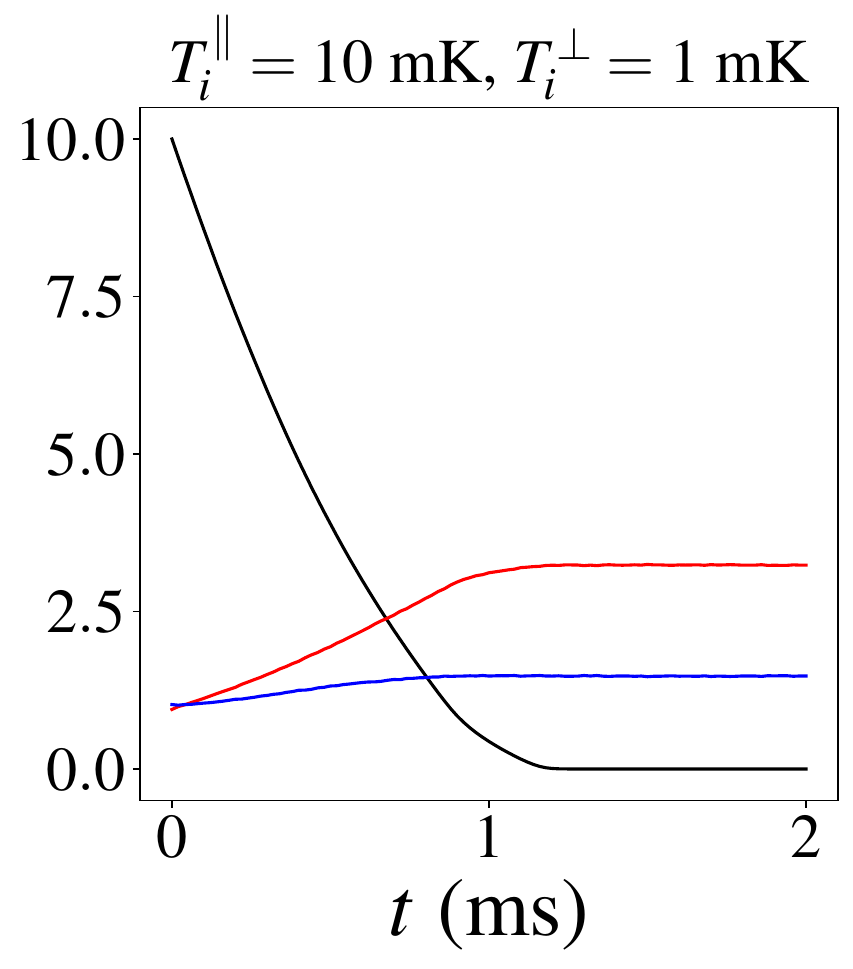}
        %\caption{}
        \label{fig:2D_T10mK_Tplanar1mK}
    \end{subfigure}
    ~ 
    % Row 2
    \begin{subfigure}{0.2\textwidth}
        \centering
        %\hspace*{0.15cm}  % shift image left
        \subcaption{}
        \includegraphics[scale=0.25]{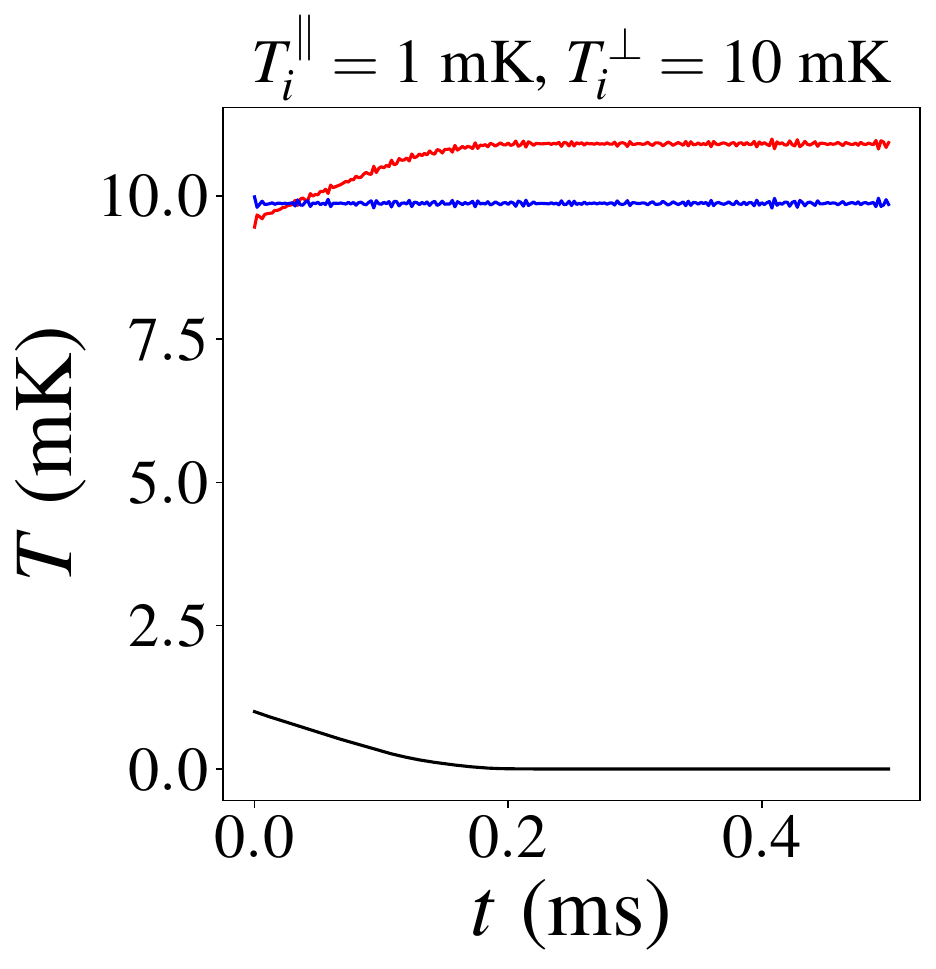}
        %\caption{}
        \label{fig:2D_T1mK_Tplanar10mK}
    \end{subfigure}%
    \hspace*{0.25cm}
    \begin{subfigure}{0.2\textwidth}
        \centering
        %\hspace*{-0.45cm}  % shift image left
        \subcaption{}
        \includegraphics[scale=0.25]{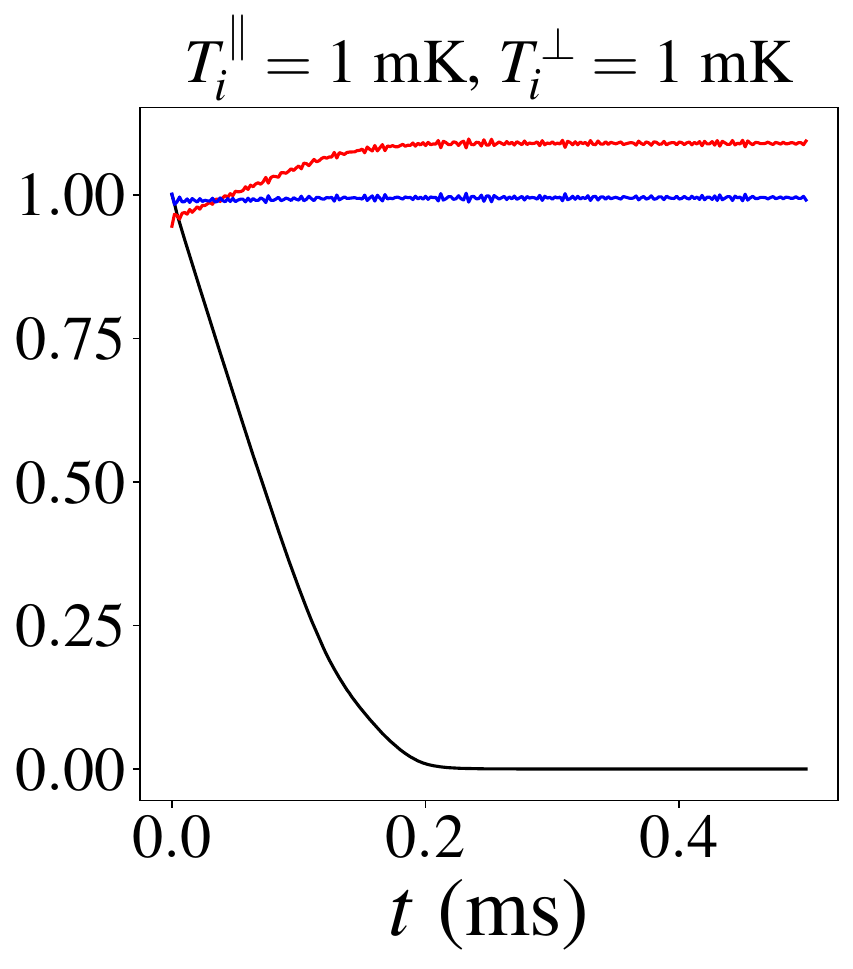}
        %\caption{}
        \label{fig:2D_T1mK_Tplanar1mK}
    \end{subfigure}
    ~
    \caption{The axial and planar temperatures of an $N=100$ ion crystal  ($\omega_r=2\pi\times180$ kHz, $\delta = 1.7\times 10^{-3}$) are plotted over time (averaged over 20 runs).  Panels (a), (b), (c), and (d) correspond to different combinations of the initial axial and planar temperatures, $T_i^\parallel$ and $T_i^\perp$.  The respective values are listed above each plot.  The temperatures $T_{\text{KE}}^\perp$ and $T_{\text{PE}}^\perp$ denote the temperatures associated with the planar kinetic energy and potential energy, respectively, and are computed using Eqs.~\eqref{eq:T_KE_perp} and ~\eqref{eq:T_PE_perp}. In (a), while the axial motion is still cooled very effectively, significant heating of the planar kinetic energy occurs. Lowering the initial planar temperature, as in (b), or the initial axial temperature, as in (c), reduces this heating. The lowest planar heating rate is seen in (d), where $T_i^\parallel$ and $T_i^\perp$ are both small.}
    \label{fig:2D_excited_planar}
\end{figure}

Up to this point, we have assumed that the motion of the crystal in the plane perpendicular to the magnetic field (which we will refer to as ``planar" or ``perpendicular" motion) is not initially excited.  In experiments, the planar modes are left with nonzero temperatures after Doppler laser cooling.  In particular, the potential energy-dominated $\boldsymbol{E} \times \boldsymbol{B}$ modes are typically difficult to cool using Doppler cooling alone. Therefore, simulating EIT cooling when the planar modes are excited would more closely reflect experimental conditions.  Such simulations would be useful for determining whether EIT cooling can also cool the $\boldsymbol{E} \times \boldsymbol{B}$ and cyclotron mode branches.  This is an especially interesting question in the context of 3D crystals, which will be considered in the following section.  

The planar motion of the ions near equilibrium is governed by the linearized equations of motion

\begin{equation}
\begin{split}
\label{beta}
\frac{d\ket{q_{\perp}}}{dt} &= \begin{bmatrix}
        \mathbb{0}_{2N}& \mathbb{I}_{2N}\\
        -\mathbb{K}_{\perp}/m & \mathbb{W} \\
        \end{bmatrix}\ket{q_\perp} ,
\end{split}
\end{equation}
where $\ket{q_{\perp}}=\ket{x_{1-N}, y_{1-N},\dot{x}_{1-N},\dot{y}_{1-N}}$. The matrix $\mathbb{W}$ accounts for the Lorentz force in the directions orthogonal to the magnetic field. It is anti-symmetric and block diagonal.  The form of  $\mathbb{W}$ can be found in Appendix \ref{app:normal_modes}. Diagonalizing this system produces $2N$ planar modes that separate into two branches that occupy distinct frequency bandwidths. The low-frequency modes constitute the $\boldsymbol{E} \times \boldsymbol{B}$ branch, while the high-frequency modes make up the cyclotron branch.  The branches are further differentiated by the ratio of potential to kinetic energy within each mode.  $\boldsymbol{E} \times \boldsymbol{B}$ modes are dominated by potential energy, while cyclotron modes contain mostly kinetic energy.  In fact, this is the reason why $\boldsymbol{E} \times \boldsymbol{B}$ modes are often difficult to laser cool \cite{johnson2024}.

To initialize the planar modes at temperature $T^{\perp}_i$, we use two methods. The cyclotron modes are initialized using the same method used to excite the axial modes in the previous section.  The $\boldsymbol{E} \times \boldsymbol{B}$ modes are excited using a Metropolis-Hastings (MH) algorithm that systematically perturbs the ion positions to initialize the temperature associated with the potential energy at $T_i^{\perp}$ \cite{shankar2020}.  Since the $\boldsymbol{E} \times \boldsymbol{B}$ modes contain nearly all of the planar potential energy, the MH algorithm primarily excites these modes and does not significantly affect the cyclotron mode energies.  Explicitly, the MH algorithm proceeds as follows. Beginning with the equilibrium ion configuration, we propose perturbing a given ion's position by independently displacing it in the $x$ and $y$ directions by values sampled from the uniform distribution $U(-\delta x/2,\delta x/2)$, with $\delta x=1.0\;\mu$m. The energy before the perturbation is denoted $E_{0}$ and the ion coordinates are denoted by $\{\boldsymbol{r}^n_0\}$.  After the proposed position change, these quantities are denoted by $E_{\text{new}}$ and $\{\boldsymbol{r}^n_{\text{new}}\}$, respectively.  Next, a random number $u$ is sampled from the distribution $U(0,1)$. The proposal is accepted if $E_{\text{new}} < E_0$ or if $u < \exp[-(E_{\text{new}}-E_{0})/k_BT_i^{\perp}]$.  Then we update $\{\boldsymbol{r}^n_0\} = \{\boldsymbol{r}^n_{\text{new}}\}$ and $E_{0}=E_{\text{new}}$. Otherwise, the proposal is rejected. This process is repeated for each ion in the crystal, after which one scan of the MH algorithm has been completed. We find that performing 500 scans is sufficient to initialize the potential energy of the crystals studied in this section at $T_i^\perp$.

In Fig.~\ref{fig:2D_excited_planar}, we plot the simulated cooling results for an $N=100$ ion crystal with different combinations of $T_i^{\parallel}$ and $T_i^{\perp}$.  In these plots, $T^{\parallel}$ is the mean axial mode temperature averaged over 20 runs.  It is more difficult to extract the mode-resolved cooling rates for the cyclotron and $\boldsymbol{E} \times \boldsymbol{B}$ mode branches. Instead, we plot the temperatures associated with the planar kinetic and potential energies, denoted $T_{\text{KE}}^{\perp}$ and $T_{\text{PE}}^{\perp}$, respectively. As explained earlier, the temperature of the $\boldsymbol{E} \times \boldsymbol{B}$ (cyclotron) mode branch is approximately $T_{\text{PE}}^{\perp}$ ($T_{\text{KE}}^{\perp}$). The planar kinetic energy is computed according to 

\begin{equation}
\label{eq:T_KE_perp}
T_{\text{KE}}^{\perp}= \frac{m\sum_{i=1}^N(\dot{x}_i^2+\dot{y}_i^2)}{2Nk_B},
\end{equation}
where $\dot{x}$ and $\dot{y}$ represent velocities in the rotating frame of the crystal.  To compute $T_{\text{PE}}^{\perp}$ at a given time, we first artificially set the $z$ coordinates of all ions to zero and calculate the potential energy of this modified ion configuration, $E_{\text{pot}}(\{\boldsymbol{r}_i^{z=0}\ | 1\leq i \leq N\})$.  We then subtract the equilibrium potential energy $E_{\text{pot}}^{\text{eq}}$ and convert the energy to a temperature using

\begin{equation}
\label{eq:T_PE_perp}
    T_{\text{PE}}^\perp (t)= \frac{E_{\text{pot}}(\{\boldsymbol{r}_i^{z=0}\ | 1\leq i \leq N\})-E_{\text{pot}}^{\text{eq}}}{Nk_B}.
\end{equation}

Figure~\ref{fig:2D_T10mK_Tplanar10mK} illustrates cooling when $T_i^{\parallel}=T_i^{\perp} = 10$ mK.  The axial modes are cooled at approximately the same rate as when the planar modes were not excited (see Fig.~\ref{fig:2D_crystal}).  However, the perpendicular kinetic energy is heated significantly. Since the EIT laser parameters are optimized to cool the axial COM mode, which has a much lower frequency than the cyclotron modes, it is not surprising that the cyclotron modes are heated in this setup.  Since the wavevectors of the EIT laser beams have large planar components, they can excite the planar motion as they cool the axial motion.  In contrast, the perpendicular potential energy increases only slightly.  In Fig.~\ref{fig:2D_T10mK_Tplanar1mK}, we plot the analogous cooling curves when the perpendicular motion is initialized at $T_i^{\perp}=1$ mK.  In this case, $T_{\text{KE}}^\perp$ still increases, but only by about 2 mK. This suggests that the initial Doppler laser cooling stage can dramatically affect the EIT cooling stage.  Specifically, more effective Doppler cooling of the cyclotron modes would prepare the crystal with a lower $T_{\text{KE}}^{\perp}$, which in turn results in less heating during the EIT stage.  We note that heating of the cyclotron modes is often less detrimental than heating of the $\boldsymbol{E} \times \boldsymbol{B}$ modes to quantum sensing protocols that utilize the axial modes. Excitation of the $\boldsymbol{E} \times \boldsymbol{B}$ modes results in the displacement of ions from their equilibrium positions, which has been shown to broaden the axial-mode spectrum \cite{shankar2020}.

Figures~\ref{fig:2D_T1mK_Tplanar10mK} and \ref{fig:2D_T1mK_Tplanar1mK} show cooling results when the axial modes are initialized at $T_i^\parallel = 1$ mK and the planar temperature is initialized at $T_i^\perp = 10$ mK and $T_i^\perp = 1$ mK, respectively.  At this reduced axial temperature, the perpendicular kinetic energy heating is significantly reduced.  This, once again, highlights the importance of preparing the crystal in a low-energy state during the Doppler cooling stage.  Fortunately, Doppler cooling typically cools the axial temperature to roughly $1$ mK, suggesting that EIT cooling will not dramatically heat the planar modes.

\subsection{Finite beam waist considerations}

\begin{figure}
    \centering
    \includegraphics[scale=0.3]{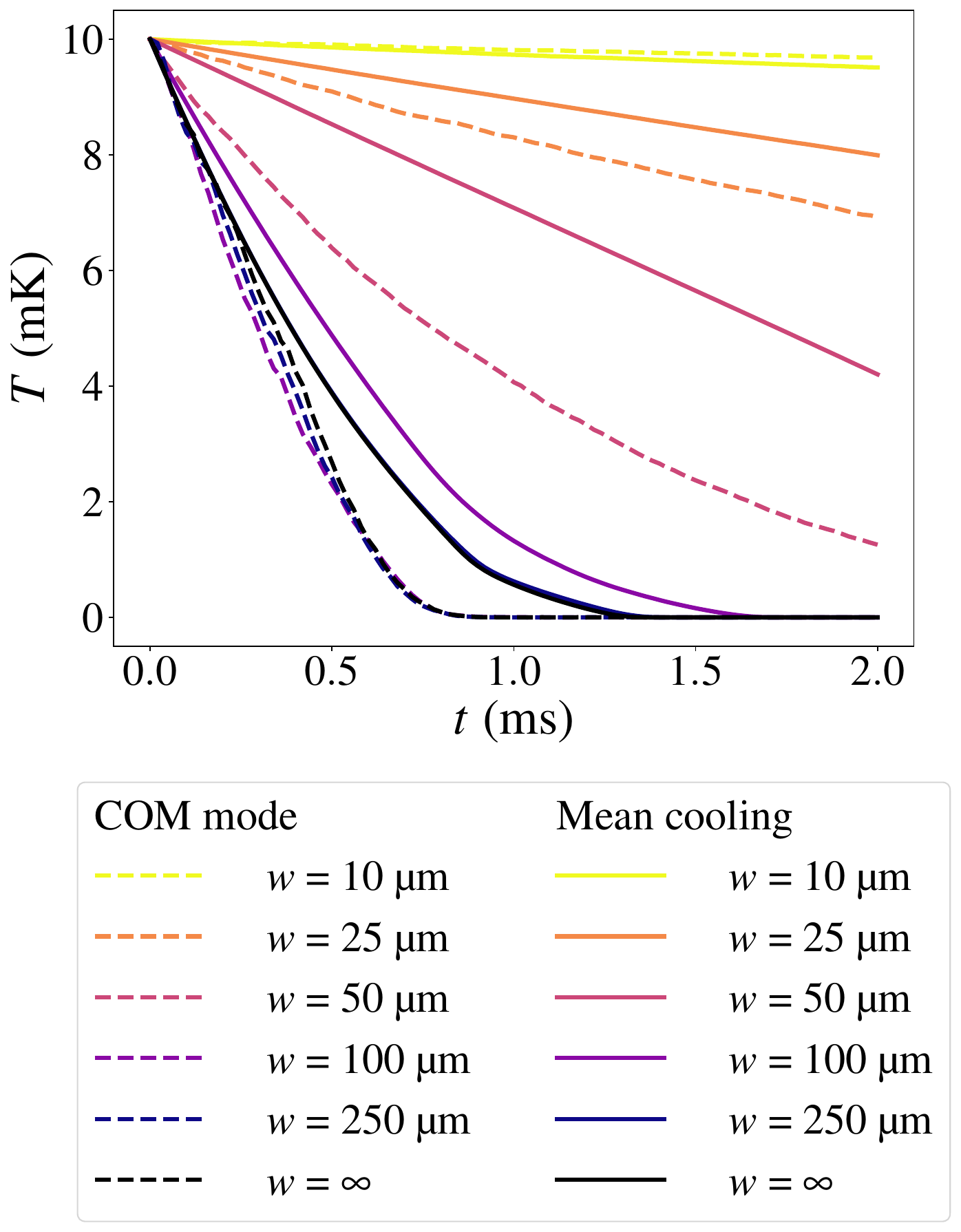}
    ~ 
    \caption{The COM and mean axial cooling rates are plotted for a range of laser beam waists.  The cooling rates are averaged over 20 runs for each waist value.  Using a small beam waist can significantly degrade cooling.  As the waist increases, the cooling rate increases and approaches that of the infinite-waist simulations.  For $w=250\;\mu$m, the mean cooling rate is visually indistinguishable from the infinite-waist result.}
    \label{fig:finite_beam_waist}
\end{figure}

While we have so far approximated the EIT cooling beam waist to be infinite, this is a simplification and does not accurately describe the laser beams used in experiments. While large-waist beams can be produced, deviations from the infinite waist approximation will likely affect cooling rates in ways that are not analytically tractable due to the complicated structure of crystal normal modes. Our MD code, however, can readily characterize changes in cooling rates when the beam waist is taken to be finite. Finite beam waists lead to spatially inhomogeneous Rabi frequencies.  Assuming that the EIT laser beams are focused at the trap center, ions closer to the trap center experience a larger electric field and, therefore, a larger Rabi frequency than ions at the edge of the crystal.  Mathematically, if $\Omega_\mu^0$ is the Rabi frequency at the maximum beam intensity of beam $\mu$, then the Rabi frequency at position $\boldsymbol{r}$ is given by

\begin{equation}
\label{state_vecotr_evolution}
\Omega_{\mu}(\boldsymbol{r}) = \Omega^0_{\mu} e^{-r_{\perp,\mu}^2/w_{\mu}^2},
\end{equation}
where $r_{\perp,\mu}^2 = |\boldsymbol{r}-(\boldsymbol{r}\cdot\hat{\boldsymbol{k}}_\mu)\hat{\boldsymbol{k}}_\mu|^2$. Here, we assume $\Omega_1^0 = \Omega_2^0 = 2\pi\times 33.9$ MHz. 

To study the effect of using lasers with finite beam waists, we run EIT cooling simulations with several beam-waist values between 10 $\mu$m and 250 $\mu$m.  We assume that $w_1 = w_2\equiv w$.  Axial mode cooling results are presented in Fig.~\ref{fig:finite_beam_waist}. Additional parameters used in these simulations are $N=100$, $\omega_r=2\pi\times 180$ kHz, $T_i^{\parallel}=10$ mK, and $T_i^{\perp} = 0$ mK.  The value of the beam waist strongly influences the cooling rate of the axial motion.  However, when $w\gtrsim 100 \;\mu$m, the cooling rate is nearly the same as the infinite-waist cooling rate.  This suggests that reasonably large waists should be sufficient to produce nearly optimal cooling in practice.  

\begin{figure}
    %\centering
    %\includegraphics[scale=0.5]{prolate_crystal_cooling_072825.pdf}
    \centering
    \hspace*{-1.5cm}
    \begin{subfigure}{0.2\textwidth}
        \centering
        %\hspace*{0.15cm}  % shift image left
        \subcaption{}
        \includegraphics[scale=0.28]{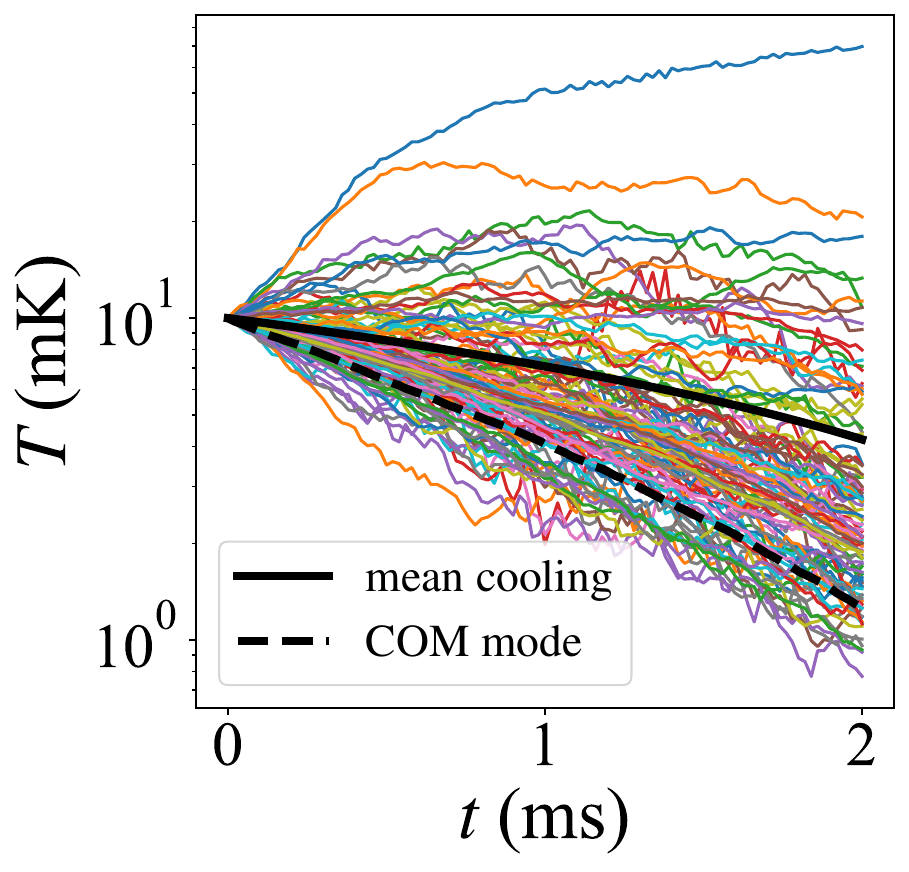}
        %\caption{}
        \label{fig:finite_beam_waist_all_axial_moeds}
    \end{subfigure}%
    \hspace*{0.55cm}
    \begin{subfigure}{0.2\textwidth}
        \centering
        %\hspace*{-0.45cm}  % shift image left
        \subcaption{}
        \includegraphics[scale=0.28]{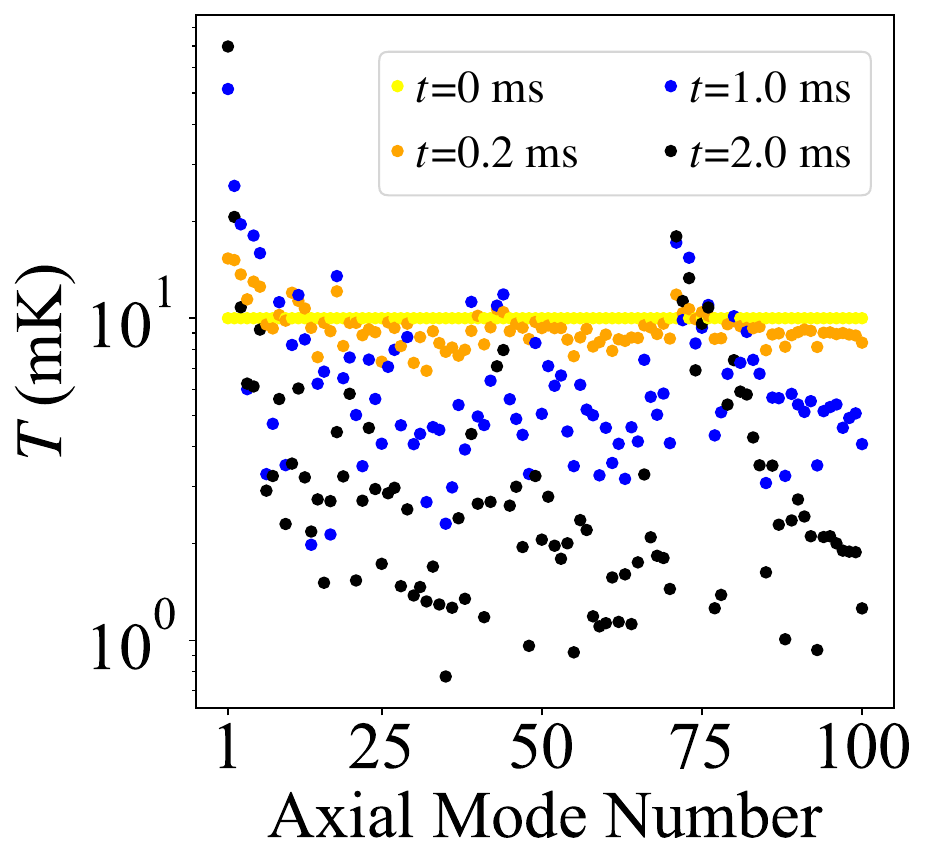}
        %\caption{}
        \label{fig:finite_beam_waist_mode_resolved}
    \end{subfigure}
    
    \caption{(a) Each colored curve shows the cooling, averaged over 20 realizations, of an axial mode in an $N=100$ ion 2D crystal when the EIT beams have a beam waist of $w_1=w_2=50\;\mu$m.  The dotted and solid black lines show the COM mode temperature and mean axial mode temperature, respectively. (b) Snapshots of the mode temperatures during cooling reveal that the lowest-frequency modes are heated.  Qualitatively similar behavior was seen in the infinite-waist simulations shown in Fig.~\ref{fig:2D_mode_resolved_cooling}.}
    \label{fig:finite_beam_wasit_mode_cooling}
\end{figure}

To better understand the details of the cooling dynamics with a finite beam waist, we provide mode-resolved cooling plots in Fig.~\ref{fig:finite_beam_wasit_mode_cooling}.  Figure~\ref{fig:finite_beam_waist_all_axial_moeds} shows the time evolution of each mode's temperature for $w=50\;\mu$m.  Compared to the infinite-waist simulations (Fig.~\ref{fig:2D_crystal_axial_cooling_curves}), certain low-frequency modes are now heated to significantly higher temperatures.  Figure~\ref{fig:finite_beam_waist_mode_resolved}, which shows the mode temperatures at different snapshots, clearly illustrates the heating of low-frequency modes. Once again, this heating is not caused by the EIT laser beams, but by thermalization with the initially unexcited $\boldsymbol{E} \times \boldsymbol{B}$ mode branch.  Since the heating rates caused by thermalization decrease over time, these modes can eventually be cooled, as seen in Fig.~\ref{fig:2D_crystal_axial_cooling_curves}.  However, with a finite beam waist of $w=50\;\mu$m, the cooling is significantly degraded.

\section{EIT Cooling Results: 3D crystals}
\label{sec:3D_crystals}

\begin{figure}
    %\centering
    %\includegraphics[scale=0.5]{prolate_crystal_cooling_072825.pdf}
    \centering
    \hspace*{-1.0cm}
    \begin{subfigure}{0.2\textwidth}
        \centering
        %\hspace*{0.15cm}  % shift image left
        \subcaption{}
        \includegraphics[scale=0.18]{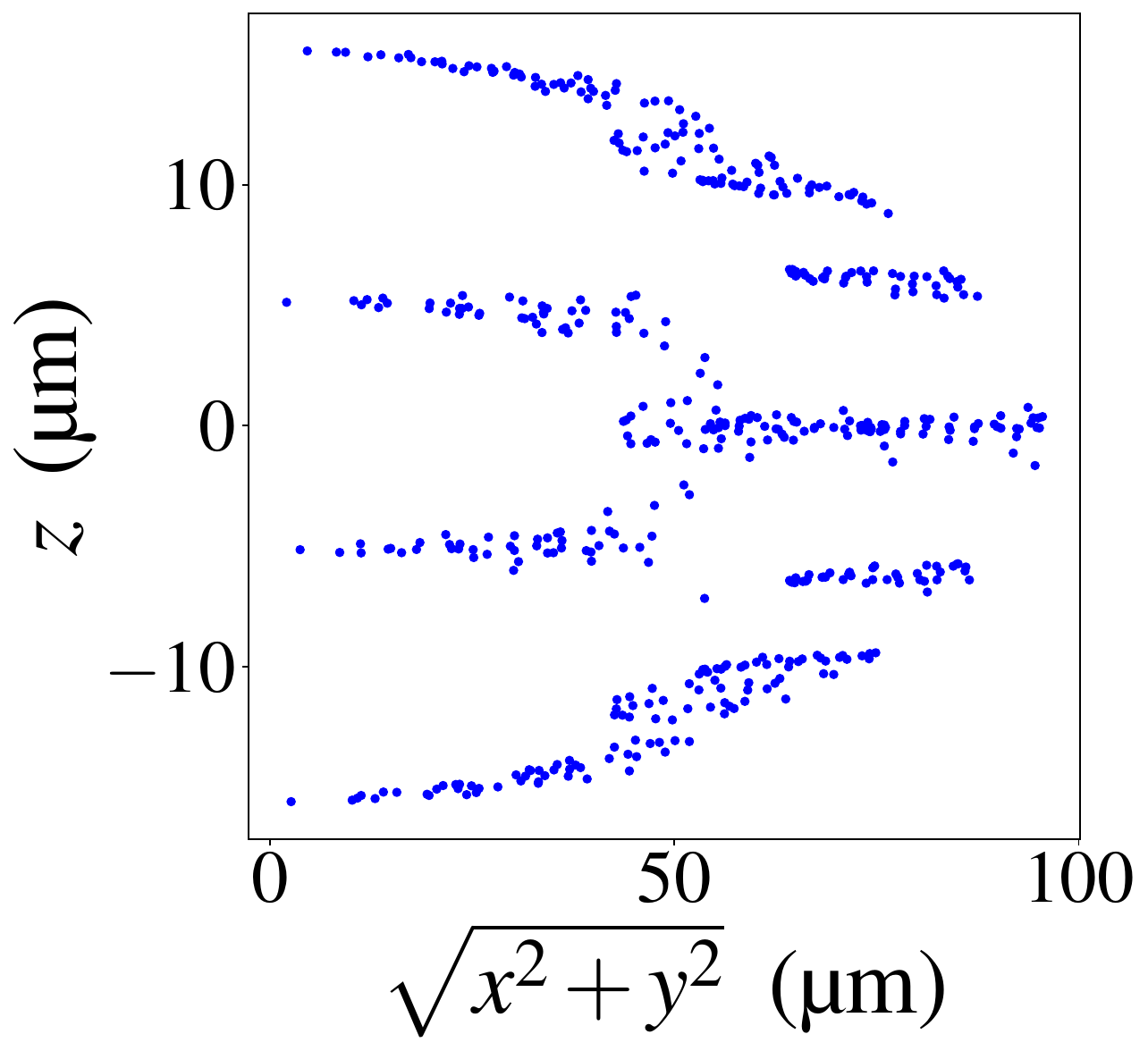}
        %\caption{}
        \label{fig:3D_eq}
    \end{subfigure}%
    \hspace*{0.55cm}
    \begin{subfigure}{0.2\textwidth}
        \centering
        %\hspace*{-0.45cm}  % shift image left
        \subcaption{}
        \includegraphics[scale=0.18]{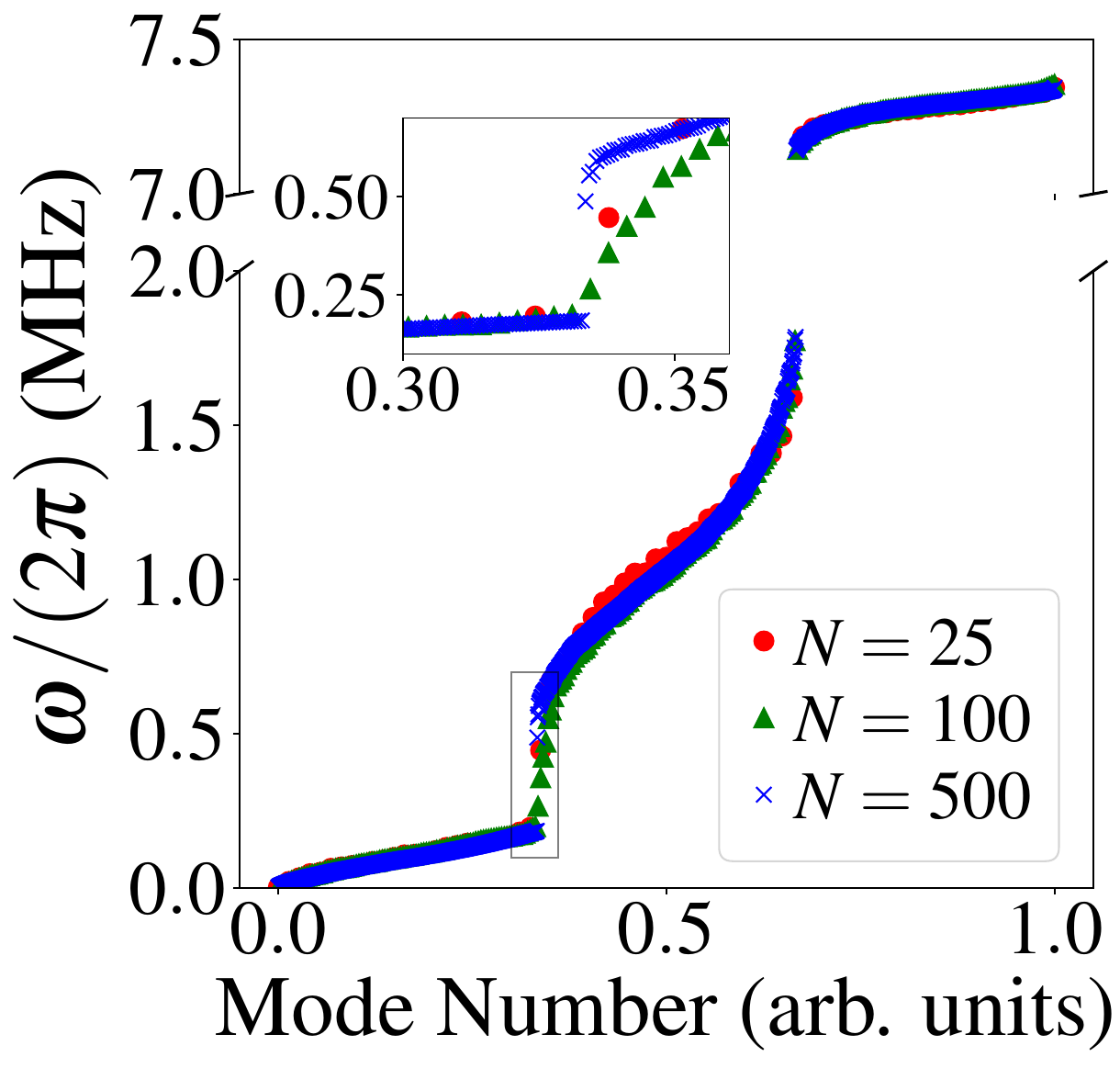}
        %\caption{}
        \label{fig:3D_spectrum}
    \end{subfigure}
    ~ 
    \vspace{-0.5cm}
    % Row 2
    \hspace*{-1.0cm}
    \begin{subfigure}{0.2\textwidth}
        \centering
        %\hspace*{0.15cm}  % shift image left
        \subcaption{}
        \includegraphics[scale=0.26]{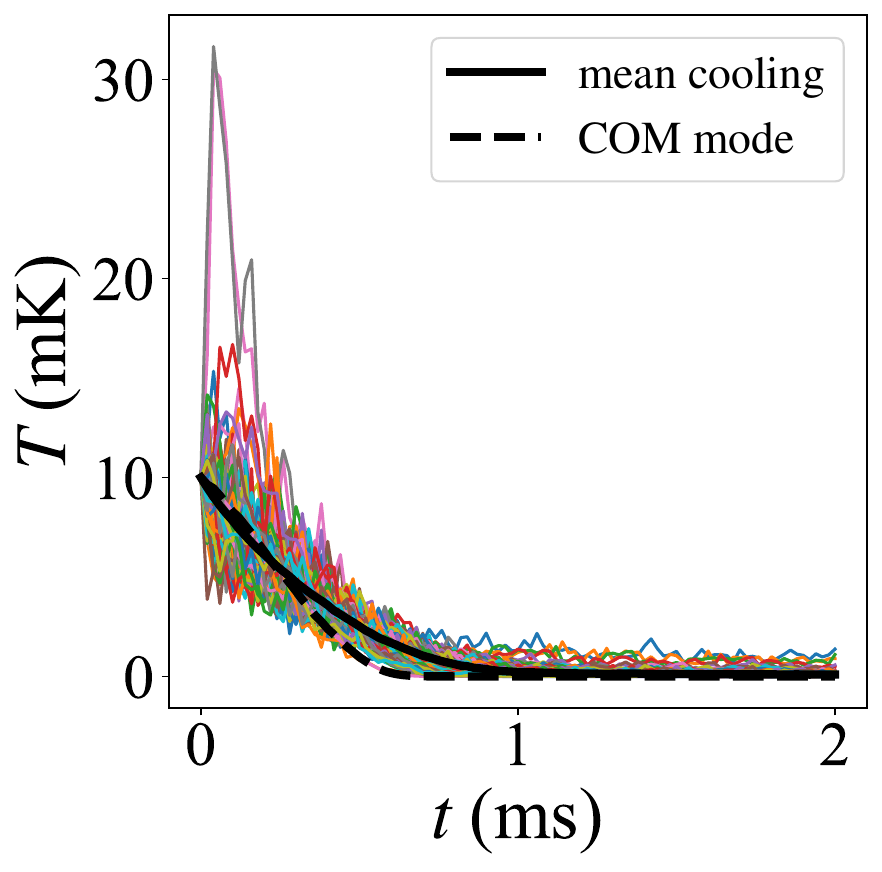}
        %\caption{}
        \label{fig:3D_all_axial_modes}
    \end{subfigure}%
    \hspace*{0.3cm}
    \begin{subfigure}{0.2\textwidth}
        \centering
        %\hspace*{-0.45cm}  % shift image left
        \subcaption{}
        \includegraphics[scale=0.26]{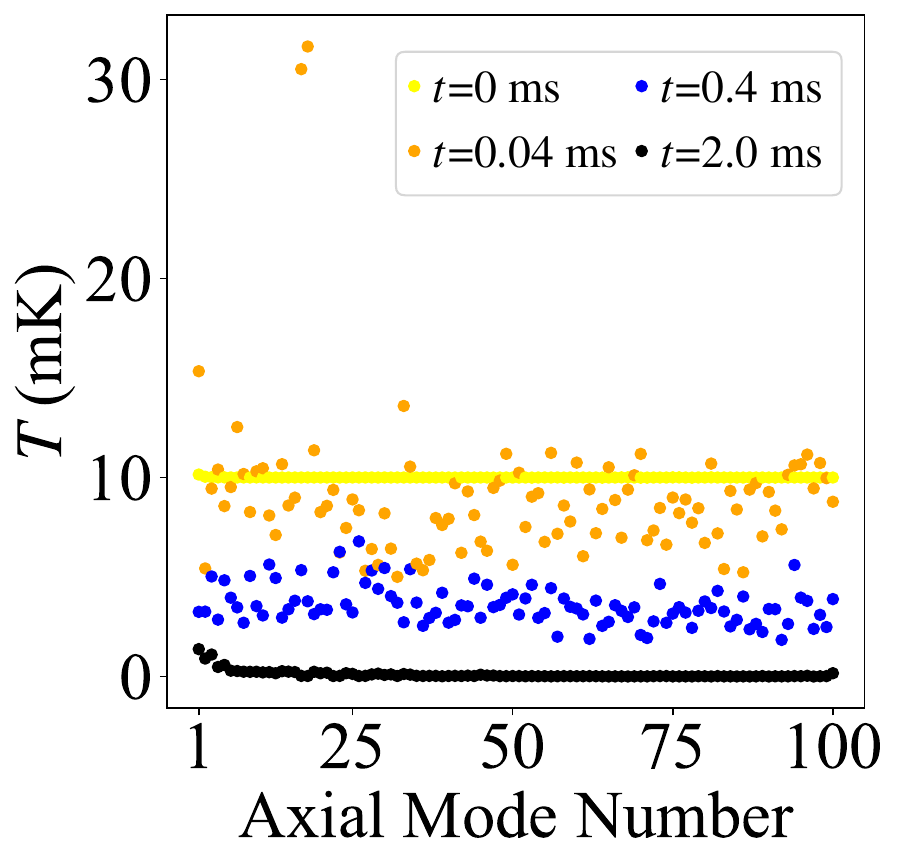}
        %\caption{}
        \label{fig:3D_mode_resolved}
    \end{subfigure}
    ~ 
    \vspace{-0.5cm}
    \hspace*{-0.5cm}
    % Row 2
    \begin{subfigure}{0.45\textwidth}
        \centering
        %\hspace*{0.15cm}  % shift image left
        \subcaption{}
        \includegraphics[scale=0.23]{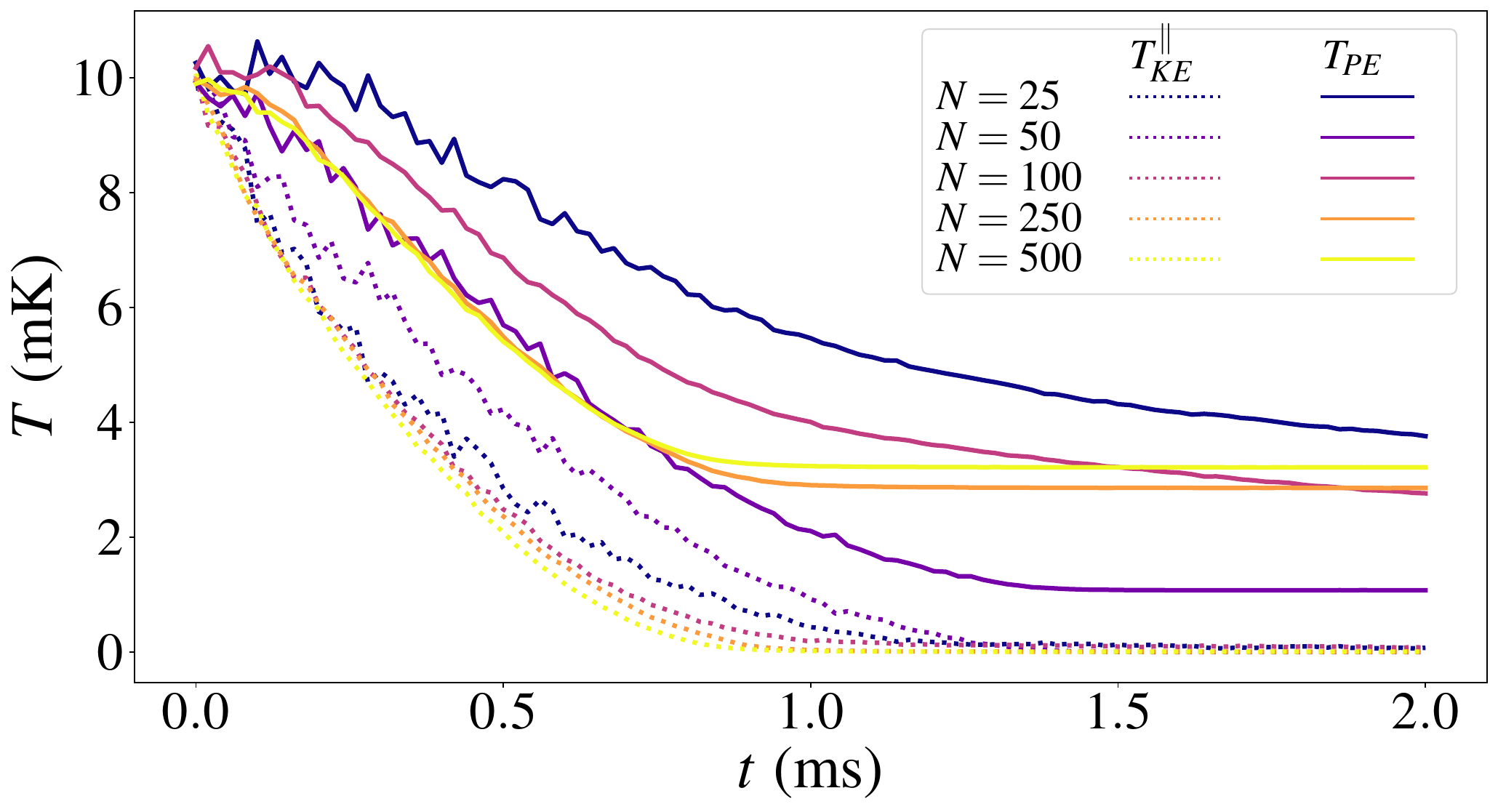}
        %\caption{}
        \label{fig:3D_mean_cooling}
    \end{subfigure}
    \caption{(a) Local equilibrium configuration of an oblate $N=500$ ion crystal with $\omega_r=2\pi\times240$ kHz and $\delta=0.035$. (b) The mode spectrum is shown for the crystal in (a), and also for two other crystals which differ only by ion number. The inset shows that the ion number affects the frequency gap between the $\boldsymbol{E} \times \boldsymbol{B}$ and axial mode branches. The gap vanishes in the $N=100$ ion crystal. (c) Cooling of axial modes of the $N=100$ ion crystal, averaged over 20 runs, after they are initialized at $T=10$ mK. (d) Axial mode temperatures at several snapshots show that mode numbers 17 and 18 are often heated at early times before being cooled. (e) The axial kinetic energy and total potential energy are plotted for crystals with different ion numbers. While the kinetic energy is efficiently cooled, the potential energy decreases at a slower rate.  This behavior likely indicates energy exchange between the $\boldsymbol{E} \times \boldsymbol{B}$ and axial mode branches, as described in the text.}
    \label{fig:3D_cooling}
\end{figure}

\subsection{Background}

While EIT cooling has been applied to 2D crystals in Penning traps \cite{jordan2019} and radio-frequency traps \cite{kiesenhofer2023}, there is motivation for extending this technique to large 3D crystals, which offer improved sensitivity in quantum sensing experiments.  Recent numerical simulations suggest that 3D crystals can be Doppler cooled to millikelvin temperatures, thereby preparing them for EIT cooling \cite{zaris2026}. However, the effectiveness of EIT cooling in 3D crystals has not been characterized.  The mode spectra of 3D crystals exhibit qualitative differences from those of 2D crystals, which could plausibly degrade or enhance cooling.  In general, the axial modes of 3D crystals occupy a larger frequency bandwidth, which could hinder cooling of the lower frequency axial modes, assuming the EIT parameters are still optimized to cool the COM mode. In contrast, the frequency gap between the axial and $\boldsymbol{E} \times \boldsymbol{B}$ mode branches can be made to vanish in certain 3D crystals, which may allow for sympathetic cooling of the $\boldsymbol{E} \times \boldsymbol{B}$ modes.  Our MD simulation includes these effects and can therefore be used to predict cooling rates of 3D crystals.

Experimentally, 3D crystals are typically produced by increasing the rotating-wall frequency $\omega_r$.  The shape of a trapped-ion crystal is parameterized by the ratio of confining forces in the planar and axial directions, given by 

\begin{equation}
\label{beta}
\beta = \frac{\omega_r(\omega_c-\omega_r)}{\omega_z^2}-\frac{1}{2},
\end{equation}
where $\omega_c = qB_z/m$ is the cyclotron frequency of the trapped species. Increasing $\beta$ beyond a given value will cause the minimum energy configuration to change from a single plane to a 3D crystal. Setting the initial temperature of the axial modes of a 3D crystal is slightly more complicated because these modes now have a small but nonnegligible planar component, due to the Coulomb interaction between ions. Therefore, it is necessary to consider the full $6N\times 6N$ matrix system that describes the 3D motion of the crystal.  Explicitly, the linearized 3D equations of motion are given by 

\begin{equation}
\label{beta}
\frac{d\ket{q}}{dt} = \begin{bmatrix}
        \mathbb{0}_{3N}& \mathbb{I}_{3N}\\
        -\mathbb{K}/m & \mathbb{W} \\
        \end{bmatrix}\ket{q} ,
\end{equation}
where $\ket{q}=\ket{x_{1-N}, y_{1-N},z_{1-N},\dot{x}_{1-N},\dot{y}_{1-N},\dot{z}_{1-N}}$ and $\mathbb{K}$ is the 3D stiffness matrix, defined in Appendix \ref{app:normal_modes}.

\subsection{Axial mode cooling in an oblate crystal}
\label{sec:3D_crystal_axial_only}

In this section, we simulate EIT cooling of oblate 3D crystals obtained by setting $\omega_r=2\pi\times240$ kHz and $\delta = 0.035$, while keeping other trap parameters fixed at their previous values. A local equilibrium configuration of such a crystal with $N=500$ ions is illustrated in Fig.~\ref{fig:3D_eq}.  We note that it is challenging to compute global equilibria of 3D crystals, and do not attempt to do so in this work.  Unlike in the 2D case, where crystals near the global energy minimum can be found using a hexagonal lattice seed configuration, no analogous seed exists for 3D crystals.  Nevertheless, we find that the energy differences among numerically obtained local minima are relatively small, typically on the order of a millikelvin.   Near the symmetry axis of the trap, the crystal in Fig.~\ref{fig:3D_eq} approximately consists of stacked planes of ions. The boundary of the crystal is an ellipsoid with relatively small axial extent (note the different scales on the horizontal and vertical axes in Fig.~\ref{fig:3D_eq}). We initialize the axial modes, defined as the $N$ middle-frequency modes, as previously described, again emphasizing that these modes are no longer purely axial. As a consequence, small-amplitude planar motion also contributes. We note that for sufficiently large $\beta$, the $N$ modes with the largest axial component do not necessarily correspond exactly to the $N$ middle-frequency modes.  However, this is not a concern at $\omega_r =2\pi\times 240$ kHz. 

Figure~\ref{fig:3D_spectrum} shows the mode spectrum for several crystals that differ only by ion number.  Three mode branches are evident, although the frequency gap between the low-frequency $\boldsymbol{E} \times \boldsymbol{B}$ branch and middle-frequency axial branch vanishes in the $N=100$ ion crystal. The inset clearly demonstrates this phenomenon. In recent studies, it has been shown that the size of this frequency gap can strongly affect cooling dynamics \cite{johnson2024, zaris2024}. We will discuss this further in the following analysis. Note that a break is used on the vertical axis in Fig.~\ref{fig:3D_spectrum} due to the large frequency separation between the axial branch and the high-frequency cyclotron branch. Finally, we note that the axial COM mode, with frequency $\omega_z = 2\pi\times 1.59$ MHz, is not necessarily the highest frequency axial mode in 3D crystals, as seen in Fig.~\ref{fig:3D_spectrum}.  For each crystal, the COM mode frequency is $\omega_z = 2\pi\times 1.59$ MHz, but this is the highest-frequency axial mode only in the $N=25$ ion crystal.

Axial mode cooling results are shown in Fig.~\ref{fig:3D_all_axial_modes}. Here, we plot the cooling of each mode, averaged over 20 runs, for the $N=100$ ion crystal.  While the axial modes are rapidly cooled, the final temperatures are higher than in the 2D crystal simulations shown in Fig.~\ref{fig:2D_crystal}. This behavior is further analyzed in Fig.~\ref{fig:3D_mode_resolved}, where we plot the axial mode temperatures at several times during cooling.  Several of the lowest-frequency modes retain significant energy after 2 ms of cooling. We propose that coupling between the $\boldsymbol{E} \times \boldsymbol{B}$ and axial modes may be responsible for the elevated temperatures of these axial modes.  When the frequency gap between the branches is small, energy can be transferred from the initially excited axial modes to the $\boldsymbol{E} \times \boldsymbol{B}$ modes \cite{johnson2024}. If such coupling were to occur, elevated axial mode temperatures would be expected, since the $\boldsymbol{E} \times \boldsymbol{B}$ modes are unlikely to be efficiently cooled by the EIT laser beams. To test for this mechanism, we performed free-evolution simulations, without EIT cooling, for the $N=100$ ion crystal.  As shown in Appendix \ref{app:mode_coupling}, we observe that the mean axial kinetic energy decreases significantly and that the total potential energy increases on the timescale of EIT cooling, suggesting that energy is transferred from the axial branch to the $\boldsymbol{E} \times \boldsymbol{B}$ branch. 

In Fig.~\ref{fig:3D_mean_cooling}, we plot cooling curves for crystals with different ion numbers, averaged over 20 runs.  Instead of plotting normal-mode temperatures, we show the axial kinetic energy and the total potential energy of the crystal. In the oblate crystals studied here, the axial modes remain dominated by motion parallel to $\hat{\boldsymbol{z}}$, so the axial kinetic energy provides a useful measure of the temperature of the axial mode branch. We observe rapid cooling of the axial kinetic energy. For an ion configuration $\{\boldsymbol{r}_i(t)\ | 1\leq i \leq N\}$ at time $t$, we define the temperature associated with the potential energy as

\begin{equation}
\label{PE_1D}
    T_{\text{PE}} (t)= \frac{2[E_{\text{pot}}(\{\boldsymbol{r}_i(t)\ | 1\leq i \leq N\})-E_{\text{pot}}^{\text{eq}}]}{Nk_B}.
\end{equation}
The factor of $2$ follows from the approximately equal partition of kinetic and potential energy in the $N$ axial modes. When only these modes are excited, the potential energy is therefore approximately one-half of the total energy. The potential energy also decreases, but at a significantly slower rate.  We attribute this behavior to the resonant coupling discussed previously. If energy is transferred to the $\boldsymbol{E} \times \boldsymbol{B}$ modes, which are not efficiently cooled in the EIT setup, slower potential-energy cooling is expected.

\begin{figure*}
    \centering
    \includegraphics[scale=0.58]{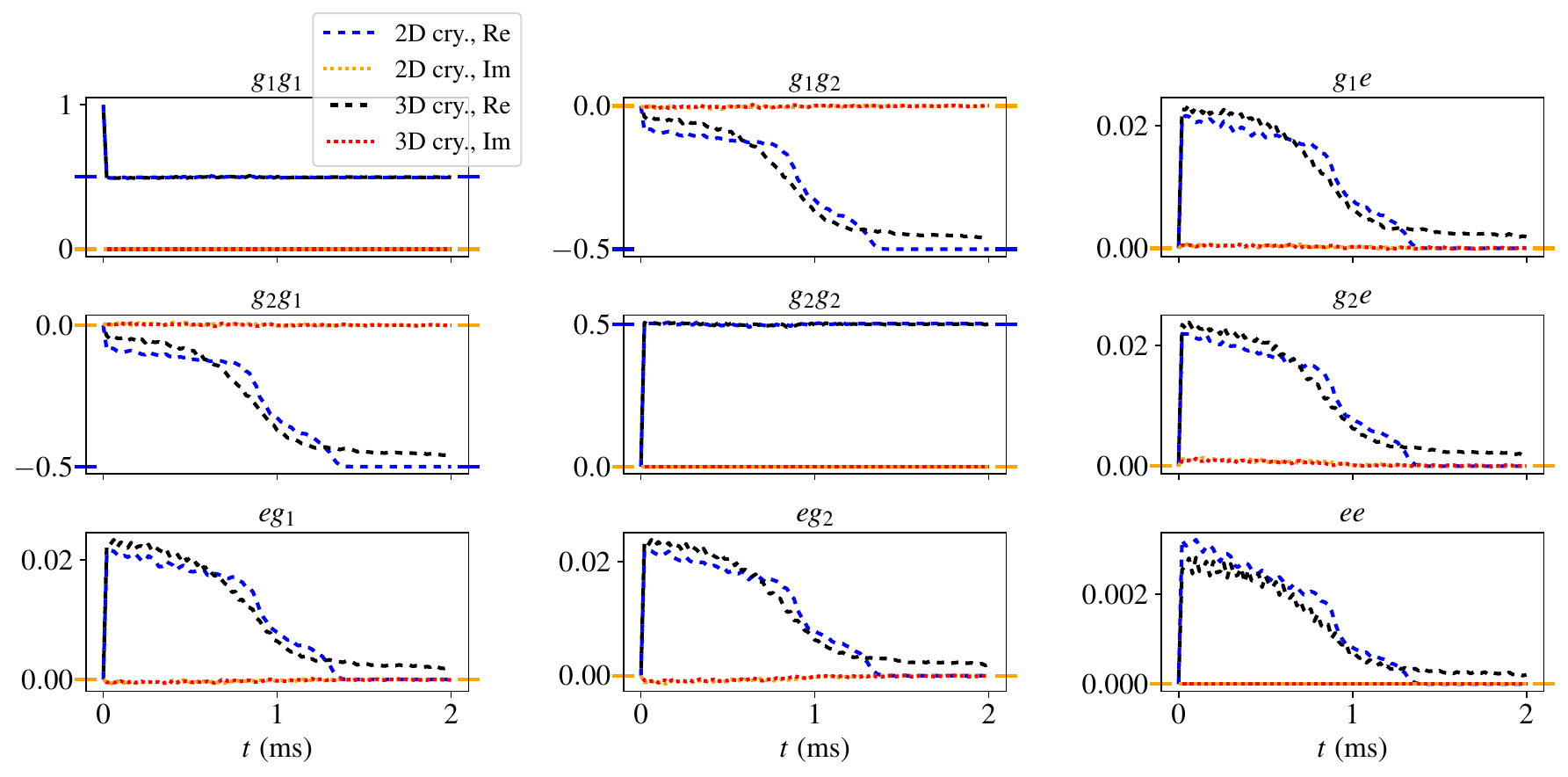}
    % Left column (a) and (b)

    \caption{The evolution of the elements of the transformed density matrix, averaged over all ions and independent simulation runs according to Eq.~\eqref{eq:mean_transformed_dens_mat}, for the 2D $N=100$ ion crystal from Fig.~\ref{fig:2D_crystal} and the 3D $N=100$ ion crystal from Fig.~\ref{fig:3D_cooling}.  The real and imaginary parts of the corresponding transformed dark-state density matrix elements are denoted by blue and orange dashes on the vertical axes, respectively.  When only one dash is visible, the real and imaginary parts of the dark-state matrix element are both equal to zero. The internal states of the ions in both crystals approach the dark state, although the convergence is slightly weaker for the 3D crystal at late times, possibly due to residual axial motion.
    }
    
    \label{fig:2D_3D_internal_state}
\end{figure*}

It is also informative to analyze how the internal state dynamics are modified in the presence of mode coupling.  In Fig.~\ref{fig:2D_3D_internal_state}, we plot the time evolution of the transformed density matrix elements for the $N=100$ ion crystal from Fig.~\ref{fig:3D_cooling} (see Eq.~\eqref{eq:unitary_phase_transform}).  To compare to 2D crystal results, we also plot the internal-state evolution for the 2D $N=100$ ion crystal from Fig.~\ref{fig:2D_crystal}.  In both cases, the plotted curves are obtained by averaging each density matrix element over all ions and all runs. Explicitly, the plotted values are found using

\begin{equation}
\label{eq:mean_transformed_dens_mat}
    \langle \tilde{\hat{\rho}}(t)\rangle =  \frac{1}{N_rN}\sum_{j=1}^N\sum_{i=1}^{N_r}\tilde{\hat\rho}_{j,i}(t),
\end{equation}
where $N_r=20$ is the number of independent runs and $\tilde{\hat\rho}_{j,i}(t)$ is the transformed density matrix of ion $j$ at time $t$ in run $i$.  Note that the transformation given in Eq.\eqref{eq:transformed_rho} is applied independently to each ion at every recorded timestep before averaging over ions and runs. We find that the ions in both crystals approach the dark state, but that the convergence to the dark state is slightly weaker in the case of the 3D crystal. This is likely due to the residual small-amplitude axial motion which persists even after 2 ms of cooling and may be sustained by coupling between the axial and planar degrees of freedom. 

\subsection{Cooling dynamics with excited $\boldsymbol{E} \times \boldsymbol{B}$ and cyclotron modes}

\begin{figure}
    \centering
    \begin{subfigure}{0.2\textwidth}
        \centering
        %\hspace*{0.15cm}  % shift image left
        \subcaption{}
        \includegraphics[scale=0.26]{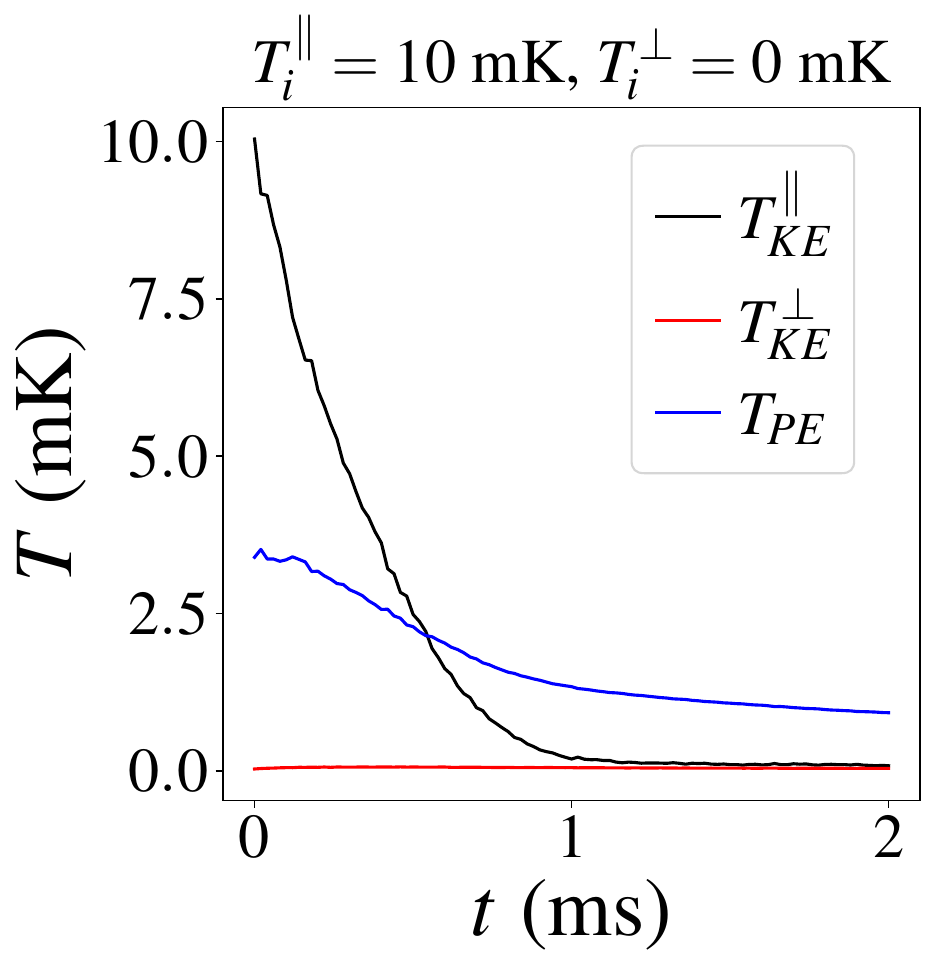}
        %\caption{}
        \label{fig:3D_T10mK_Tplanar0mK}
    \end{subfigure}%
    \hspace*{0.5cm}
    \begin{subfigure}{0.2\textwidth}
        \centering
        %\hspace*{-0.45cm}  % shift image left
        \subcaption{}
        \includegraphics[scale=0.26]{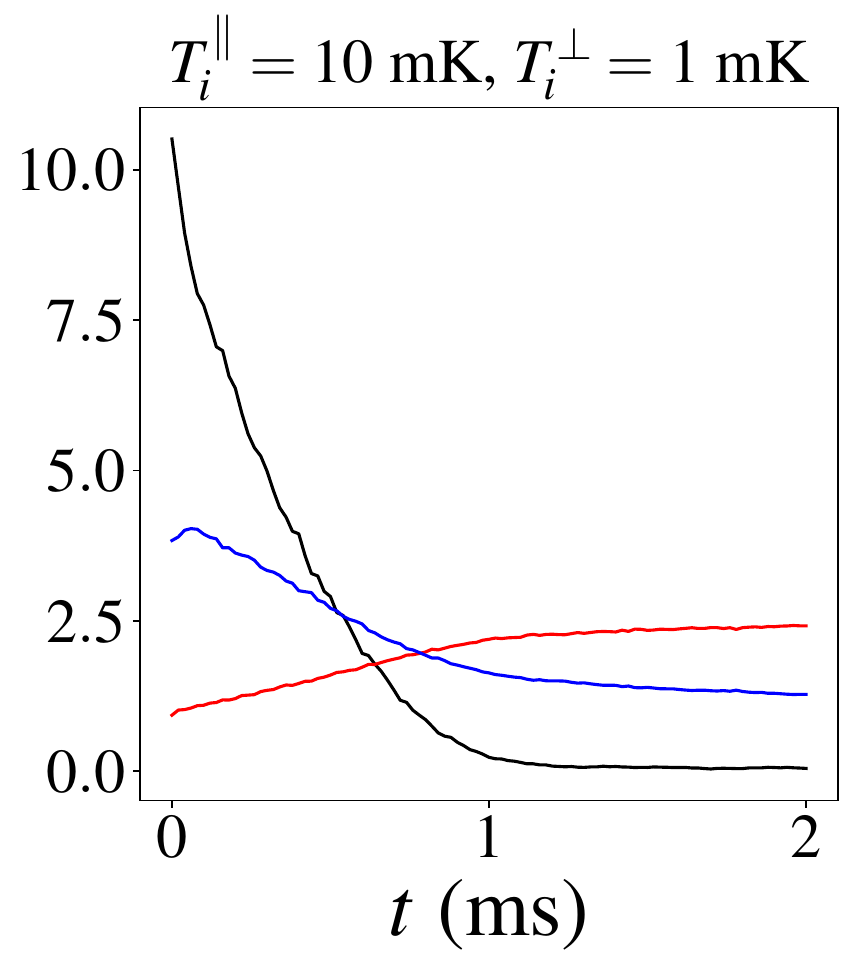}
        %\caption{}
        \label{fig:3D_T10mK_Tplanar1mK}
    \end{subfigure}
    ~ 
    % Row 2
    \begin{subfigure}{0.2\textwidth}
        \centering
        %\hspace*{0.15cm}  % shift image left
        \subcaption{}
        \includegraphics[scale=0.26]{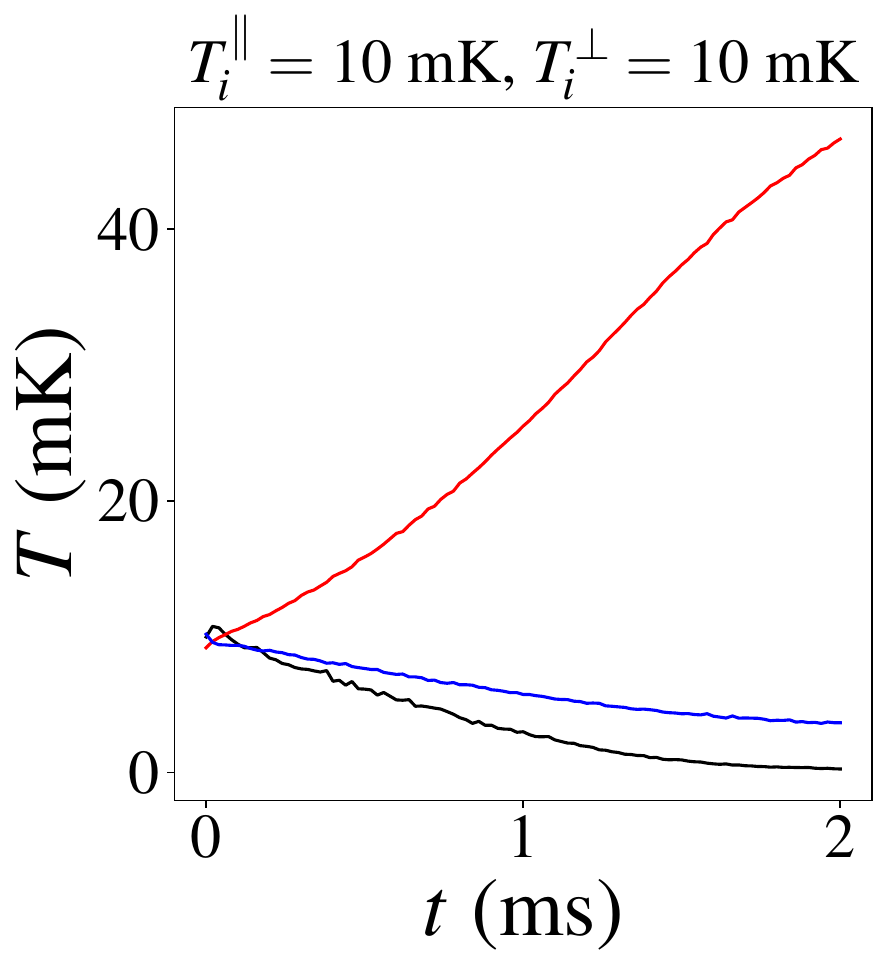}
        %\caption{}
        \label{fig:3D_T10mK_Tplanar10mK}
    \end{subfigure}%
    \hspace*{0.5cm}
    \begin{subfigure}{0.2\textwidth}
        \centering
        %\hspace*{-0.45cm}  % shift image left
        \subcaption{}
        \includegraphics[scale=0.26]{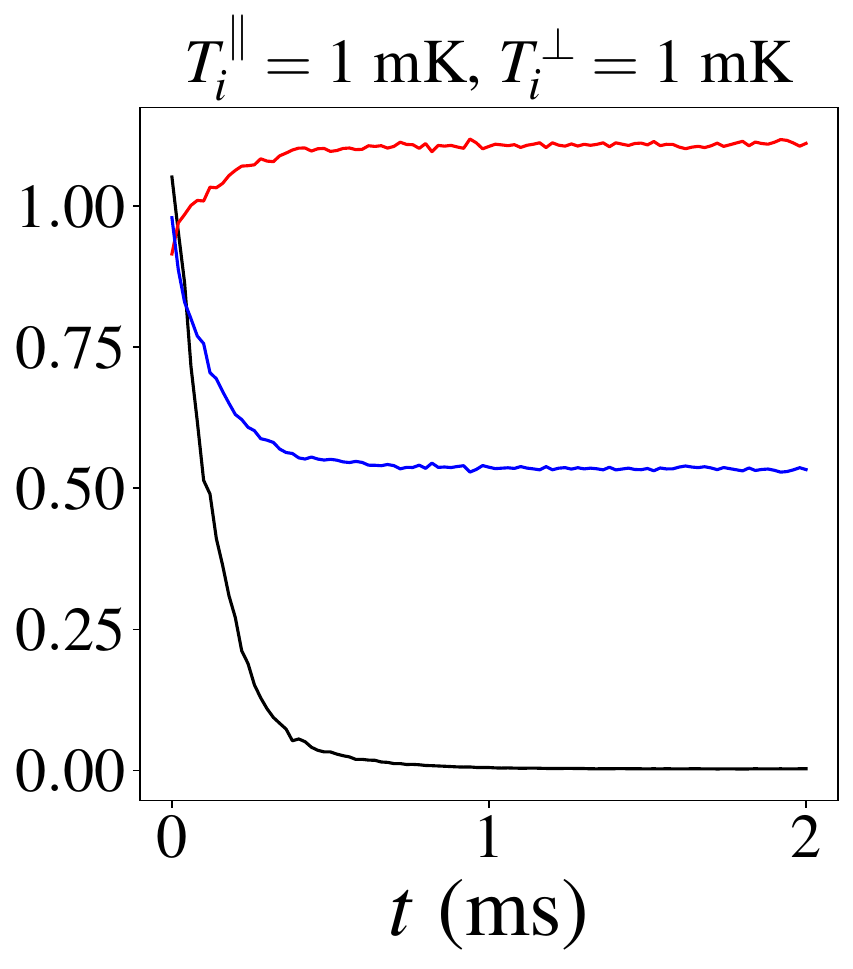}
        %\caption{}
        \label{fig:3D_T1mK_Tplanar1mK}
    \end{subfigure}
    ~
    \caption{The cooling of an $N=100$ ion 3D crystal with $\omega_r=2\pi\times 240$ kHz and $\delta=0.035$ is shown for a variety of initial mode temperatures.  We plot the temperatures associated with the axial kinetic energy ($T_{\text{KE}}^\parallel$), perpendicular kinetic energy ($T_{\text{KE}}^\perp$), and total potential energy ($T_{\text{PE}}$).  The latter is defined according to Eq.~\eqref{PE_3D}. The initial axial and planar crystal temperatures are provided above each plot.  When $T_i^\perp$ is large, the perpendicular kinetic energy tends to increase rapidly during cooling of the axial motion.  On the other hand, the decrease in the potential energy shows that the $\boldsymbol{E}\times\boldsymbol{B}$ modes can be sympathetically cooled via mode coupling.} 
    \label{fig:3D_planar_excited}
\end{figure}

The previous discussion suggests that coupling between the $\boldsymbol{E}\times\boldsymbol{B}$ and axial mode branches can degrade cooling by heating the initially unexcited $\boldsymbol{E}\times\boldsymbol{B}$ modes. However, in realistic settings, the $\boldsymbol{E}\times\boldsymbol{B}$ modes are typically excited to temperatures comparable to or greater than the axial modes. In such cases, it is expected that energy will be transferred in the opposite direction, from the $\boldsymbol{E}\times\boldsymbol{B}$ branch to the axial branch. Since the axial modes are efficiently cooled, this represents a possible mechanism for removing energy from the $\boldsymbol{E}\times\boldsymbol{B}$ modes during EIT cooling.  

In this section, we study how the cooling of an $N=100$ ion 3D crystal depends on the initial temperatures of the mode branches. In particular, we are interested in whether the $\boldsymbol{E}\times\boldsymbol{B}$ mode temperatures are reduced during EIT cooling.  Since the trap parameters from Fig.~\ref{fig:3D_cooling} appeared to produce significant inter-branch coupling in the $N=100$ ion crystal, we use the same values in this section. 

To initialize the crystal's modes, we use the method described in Sec.~IVB. Specifically, the axial and cyclotron modes are individually excited to $T_i^\parallel$ and $T_i^\perp$, respectively, while an MH algorithm is used to initialize the potential energy by perturbing the ion positions in the plane perpendicular to the magnetic field.  While the $\boldsymbol{E}\times\boldsymbol{B}$ modes now include some motion in the axial direction, perpendicular displacements still primarily excite this branch.

Cooling curves for various values of $T_i^\parallel$ and $T_i^\perp$ are shown in Fig.~\ref{fig:3D_planar_excited}.  In these figures, the potential-energy temperature is computed according to 

\begin{equation}
\label{PE_3D}
    T_{\text{PE}} (t)= \frac{2[E_{\text{pot}}(\{\boldsymbol{r}_i(t)\ | 1\leq i \leq N\})-E_{\text{pot}}^{\text{eq}}]}{3Nk_B},
\end{equation}
which accounts for the distribution of energy over three spatial dimensions. The results from the simulations in Fig.~\ref{fig:3D_cooling} are shown explicitly in Fig.~\ref{fig:3D_T10mK_Tplanar0mK}.  In this case, the planar motion is initially unexcited.  Due to the different normalizations in Eqs.~\eqref{PE_1D} and ~\eqref{PE_3D}, the initial potential-energy temperature is approximately $10/3$ mK $\approx 3.3$ mK.  Fig.~\ref{fig:3D_T10mK_Tplanar1mK} shows the cooling results when the initial planar temperature is $1$ mK.  In this case, the perpendicular kinetic temperature increases during cooling of the axial motion, similar to the behavior seen in Fig.~\ref{fig:2D_T10mK_Tplanar1mK}.  This effect is significantly more pronounced in Fig.~\ref{fig:3D_T10mK_Tplanar10mK}, where the planar temperature is initialized at 10 mK.  Here, the perpendicular kinetic temperature increases to above 40 mK after 2 ms.  Comparing this result to Fig.~\ref{fig:3D_T10mK_Tplanar1mK}, one again observes the importance of optimizing the Doppler laser cooling stage.  If the cyclotron modes can be Doppler cooled to a temperature closer to 1 mK, they will experience significantly reduced heating during the EIT cooling stage. On the other hand, when the planar modes are initialized at 10 mK, the total potential-energy temperature decreases significantly during cooling, from 10 mK to under 4 mK.  Since the majority of the potential energy is initially contained in the $\boldsymbol{E}\times\boldsymbol{B}$ branch, this suggests that $\boldsymbol{E}\times\boldsymbol{B}$ modes can be cooled via mode coupling in the EIT setup.  Finally, Fig.~\ref{fig:3D_T1mK_Tplanar1mK} shows the cooling results when the axial and planar modes are initialized at 1 mK.  Here, the perpendicular kinetic energy is heated only very slightly. Since the axial motion is rapidly cooled to near 0 mK in the semiclassical simulation, the system quickly approaches the dark state, which limits the perpendicular heating.  The potential energy decreases to about half of its initial value due to cooling of the axial and $\boldsymbol{E}\times\boldsymbol{B}$ modes.

\section{Conclusion}

EIT cooling has been successfully utilized in experiments to cool the axial modes of 2D ion crystals to near their quantum ground states. It is hoped that similar techniques can be applied to near-future experiments with large 3D crystals.  In this work, we have developed numerical methods to directly simulate the EIT cooling of trapped-ion crystals. The semiclassical description implemented in our MD code accurately models the cooling of crystals over a wide range of energies and is valid, in particular, outside of the LD regime. We have used our code to elucidate the effects of various trap and laser beam parameters on cooling efficacy.  Specifically, we have simulated cooling for various crystal sizes, shapes, and cooling beam waists.

Our results suggest that EIT cooling setups currently used to cool 2D crystals could be applied to large 3D crystals with minimal modifications. Our simulations predict that the axial modes are efficiently cooled in 3D crystals.  Furthermore, there is evidence that the mode structure of 3D crystals can produce sympathetic cooling of $\boldsymbol{E}\times\boldsymbol{B}$ modes during EIT cooling.  This prospect is particularly promising as large 3D crystals are promising platforms for near-future quantum sensing experiments with dramatically improved sensitivity. Cooling the modes of interest to near their quantum ground state in these crystals will be an important experimental step, and the numerical framework developed here will be useful in understanding experimental results and identifying parameter values that optimize cooling.  For instance, we have shown that the initial temperature of the crystal can dramatically affect EIT cooling rates.  This result is seen in simulations of both 2D and 3D crystals and highlights the importance of optimizing Doppler laser cooling protocols \cite{torrisi2016,zaris2026}.

To our knowledge, this is the first MD code to evolve both the motional and internal states of individual ions in a trapped-ion crystal during EIT laser cooling.  Our framework bridges the gap between previous MD simulations, which evolved only the motional states \cite{tang2019,johnson2024,zaris2024}, and recent theoretical studies of EIT cooling in ion crystals \cite{shankar2019,bartolotta2024}, which ignore complexities such as mode coupling.  In the future, our code could be modified to include individual photon scattering events.  This would provide cooling results that are physically accurate on shorter timescales and include heating mechanisms ignored in the current model, such as photon recoil from spontaneous emission.  Furthermore, our code can be modified to simulate related cooling techniques.  Recently, sub-Doppler cooling of a single ion in all three spatial directions was demonstrated by combining dark-state cooling of the axial mode with parametric coupling between the axial and radial modes \cite{mcmahon2026}.  By including such parametric coupling in our MD simulations of large ion crystals, we can investigate whether energy can be transferred from the planar modes into the efficiently cooled axial modes, thereby enabling sub-Doppler cooling of the planar degrees of freedom.

Additionally, it would be useful to simulate the EIT cooling of large 3D crystals, with on the order of $10^6$ ions, in future work.  The sensitivity of relevant quantum sensing protocols scales with $\sqrt{N}$, so large crystals could provide orders-of-magnitude improvements compared to current experiments.  For sufficiently large $N$, the simulation time is dominated by Coulomb force calculations, as the computational cost scales as $N^2$.  However, our code is set up to utilize the fast multipole method (FMM) \cite{greengard1987,greengard1997} to compute Coulomb interactions for large crystals.  The FMM dramatically improves the scaling from $O(N^2)$ to $O(N)$ while maintaining high precision.  We have carefully benchmarked the use of the FMM in our code \cite{zaris2024} and utilized this method to simulate Doppler laser cooling of crystals of up to $N=10^5$ ions \cite{zaris2026}.

\begin{acknowledgments}
J.Z, W.J, and S.E.P. were supported by U.S. Department of Energy grant DE-SC0020393.  A.S. acknowledges support by the Department of Science and Technology, Govt. of India through the INSPIRE Faculty Award (DST/INSPIRE/04/2023/001486), by the Anusandhan National Research Foundation (ANRF), Govt. of India through the Prime Minister’s Early Career Research Grant (PMECRG) (ANRF/ECRG/2024/001160) and by IIT Madras through the New Faculty Initiation Grant (NFIG). A.L.C. and J.J.B acknowledge support from AFOSR grant FA9550-25-1-0080 and from the DOE NQIS Research Center, Quantum Systems Accelerator. M.J.H. acknowledges support from NSF Grant Nos. 2317149 and 2016244 and by the U.S. Department of Energy, Office of Science, National Quantum Information Science Research Center and Quantum Systems Accelerator.
\end{acknowledgments}

\vspace{1em}
\noindent This document has not been peer reviewed but has been cleared by NIST for release.

\appendix

\section{MD Numerical Integrator}
\label{app:integrator}

Here we provide additional details of the numerical integration scheme used to evolve the trapped-ion system and validate the method by testing time-step convergence of the cooling rates.

\subsection{Integration method}
The motion of the ions is propagated using a cyclotronic integrator \cite{patacchini2009}, while the internal density matrices are evolved using a fourth-order Runge--Kutta (RK4) method. The cyclotronic integrator is symplectic and therefore preserves the Hamiltonian phase-space structure of the ion dynamics. In the absence of forces that add or remove energy from the system, this results in a bounded energy error with no secular energy drift.  The cyclotronic integration scheme used in this work is described in \cite{tang2019,zaris2024}; we summarize it here for completeness and explain how the internal-state evolution is incorporated.

The motional integrator separates the dynamics into two parts: (i) exact propagation under the Lorentz force associated with the static magnetic field and (ii) a velocity kick due to all remaining forces. Including the density-matrix evolution, a single time step may be represented schematically as
\begin{widetext}
\begin{equation}
\label{eq:full_integrator}
\begin{pmatrix}
\boldsymbol{r}(t+\Delta t)\\
\boldsymbol{v}(t+\Delta t)\\
\hat{\rho}(t+\Delta t)
\end{pmatrix}
=
U_0\left(\frac{\Delta t}{2}\right)
U_{\rho}^{(2)}
U_{\rm kick}(t;\Delta t)
U_{\rho}^{(1)}
U_0\left(\frac{\Delta t}{2}\right)
\begin{pmatrix}
\boldsymbol{r}(t)\\
\boldsymbol{v}(t)\\
\hat{\rho}(t)
\end{pmatrix}.
\end{equation}
\end{widetext}

Here, $U_0$ propagates the ion motion in the magnetic field, $U_{\rm kick}$ applies the nonmagnetic forces, and $U_{\rho}^{(1,2)}$ denotes the internal-state propagation required to evaluate and update the density matrix during the time step.

The magnetic-field propagator is
\begin{equation}
\label{beta}
U_0(\Delta t) = 
\begin{pmatrix}
    \mathbb{1} &\mathbb{X}_B & \mathbb{0}\\ \mathbb{0} &\mathbb{V}_B & \mathbb{0} \\ \mathbb{0} &\mathbb{0}& \mathbb{1}
\end{pmatrix},
\end{equation}
where
\begin{subequations}
\begin{align}
\mathbb{X}_B
&=
\begin{pmatrix}
\sin(\omega_c\Delta t)/\omega_c &
[\cos(\omega_c\Delta t)-1]/\omega_c &
0\\
-[\cos(\omega_c\Delta t)-1]/\omega_c &
\sin(\omega_c\Delta t)/\omega_c &
0\\
0 & 0 & \Delta t
\end{pmatrix},
\\
\mathbb{V}_B
&=
\begin{pmatrix}
\cos(\omega_c\Delta t) &
-\sin(\omega_c\Delta t) &
0\\
\sin(\omega_c\Delta t) &
\cos(\omega_c\Delta t) &
0\\
0 & 0 & 1
\end{pmatrix}.
\end{align}
\end{subequations}
Thus, $U_0$ exactly propagates the free cyclotron motion in the $x$-$y$ plane.

After the first half-step magnetic propagation, we define the intermediate position and velocity as
\begin{align}
\boldsymbol{r}_{1/2}
&=
\boldsymbol{r}(t)
+
\mathbb{X}_B\left(\frac{\Delta t}{2}\right)
\boldsymbol{v}(t),
\\
\boldsymbol{v}_{1/2}
&=
\mathbb{V}_B\left(\frac{\Delta t}{2}\right)
\boldsymbol{v}(t).
\end{align}

All nonmagnetic forces are then evaluated at the midpoint position $\boldsymbol{r}_{1/2}$. The corresponding velocity kick is
\begin{equation}
\boldsymbol{v}_{1/2}^{+}
=
\boldsymbol{v}_{1/2}
+
\frac{\Delta t}{m}
\boldsymbol{F}_{\rm total}
\left(
\boldsymbol{r}_{1/2},
\hat{\rho}_{1/2},
t+\frac{\Delta t}{2}
\right).
\end{equation}

The total force contains contributions from the axial confinement (Eq.~\eqref{pot_trap}), rotating wall (Eq.~\eqref{pot_wall}), Coulomb interactions, and the EIT lasers (Eq.~\eqref{eq:laser_force}). The Lorentz force is excluded from $\boldsymbol{F}_{\rm total}$, since its effect is already incorporated exactly through $U_0$. Using the force expression from Eq.~\eqref{eq:total_force}, one can write $\boldsymbol{F}_{\text{total}}^j = \boldsymbol{F}_j - q\dot{\boldsymbol{r}}_j\times B_z\hat{\boldsymbol{z}}$. Because the rotating-wall force depends explicitly on time, it is evaluated at $t+\Delta t/2$.

The EIT laser force depends on both the ion coordinates and its internal density matrix,
\begin{equation}
\boldsymbol{F}_{\rm laser}
=
\boldsymbol{F}_{\rm laser}
\left(
\boldsymbol{r},
\hat{\rho}
\right).
\end{equation}
Consequently, the density matrix must first be propagated to the temporal midpoint before the laser force is evaluated. During this propagation, the ion coordinates are held fixed at the midpoint values $\boldsymbol{r}_{1/2}$.

For ion $j$, the first RK4 propagation advances the density matrix through a half time step. Defining
\begin{subequations}
\begin{align}
c_1^j
&=
\dot{\hat{\rho}}_j
\left(
\hat{\rho}_j(t);
\boldsymbol{r}_{1/2}
\right),
\\
c_2^j
&=
\dot{\hat{\rho}}_j
\left(
\hat{\rho}_j(t)+\frac{\Delta t}{4}c_1^j;
\boldsymbol{r}_{1/2}
\right),
\\
c_3^j
&=
\dot{\hat{\rho}}_j
\left(
\hat{\rho}_j(t)+\frac{\Delta t}{4}c_2^j;
\boldsymbol{r}_{1/2}
\right),
\\
c_4^j
&=
\dot{\hat{\rho}}_j
\left(
\hat{\rho}_j(t)+\frac{\Delta t}{2}c_3^j;
\boldsymbol{r}_{1/2}
\right).
\end{align}
\end{subequations}

The midpoint density matrix is then
\begin{equation}
\hat{\rho}_j\left(t+\frac{\Delta t}{2}\right)
=
\hat{\rho}_j(t)
+
\frac{\Delta t}{12}
\left(
c_1^j+2c_2^j+2c_3^j+c_4^j
\right).
\end{equation}
This midpoint density matrix is used to evaluate the laser force entering the velocity kick.

Following the kick, the density matrix is propagated to the end of the time step while the coordinates remain fixed at $\boldsymbol{r}_{1/2}$. In the implementation used here, the final density matrix is obtained by performing a full RK4 propagation of duration $\Delta t$ from $\hat{\rho}_j(t)$, rather than by performing two successive RK4 half-steps. Defining
\begin{subequations}
\begin{align}
d_1^j
&=
\dot{\hat{\rho}}_j
\left(
\hat{\rho}_j(t);
\boldsymbol{r}_{1/2}
\right),
\\
d_2^j
&=
\dot{\hat{\rho}}_j
\left(
\hat{\rho}_j(t)+\frac{\Delta t}{2}d_1^j;
\boldsymbol{r}_{1/2}
\right),
\\
d_3^j
&=
\dot{\hat{\rho}}_j
\left(
\hat{\rho}_j(t)+\frac{\Delta t}{2}d_2^j;
\boldsymbol{r}_{1/2}
\right),
\\
d_4^j
&=
\dot{\hat{\rho}}_j
\left(
\hat{\rho}_j(t)+\Delta t\,d_3^j;
\boldsymbol{r}_{1/2}
\right),
\end{align}
\end{subequations}
the final density matrix is then
\begin{equation}
\hat{\rho}_j(t+\Delta t)
=
\hat{\rho}_j(t)
+
\frac{\Delta t}{6}
\left(
d_1^j+2d_2^j+2d_3^j+d_4^j
\right).
\end{equation}

Finally, the second $U_0(\Delta t/2)$ operation advances the ions from the midpoint coordinates to their positions and velocities at $t+\Delta t$. 

\subsection{Time-step convergence}

\begin{figure}
    \centering
    \hspace*{-1.5cm}
    \begin{subfigure}{0.2\textwidth}
        \centering
        %\hspace*{0.15cm}  % shift image left
        \subcaption{}
        \includegraphics[scale=0.33]{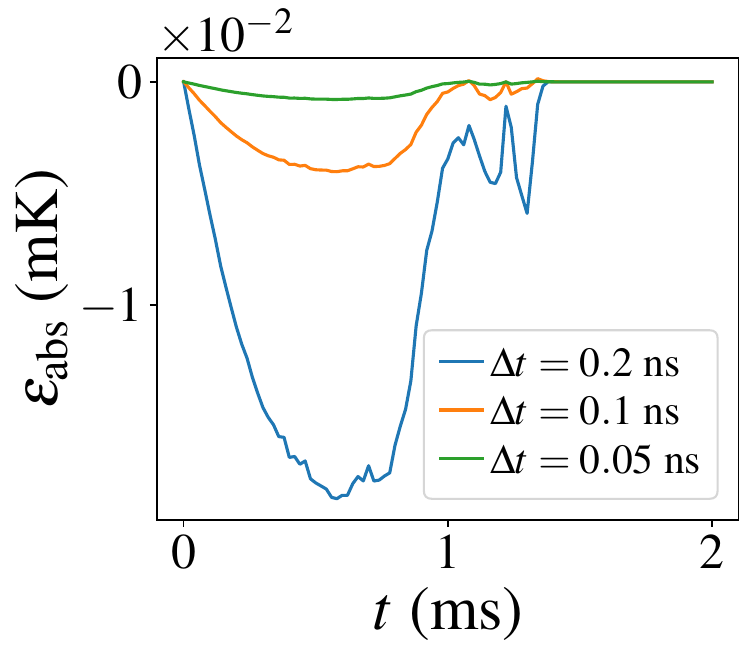}
        %\caption{}
        \label{fig:crystal_time_conv_a}
    \end{subfigure}%
    \hspace*{0.55cm}
    \begin{subfigure}{0.2\textwidth}
        \centering
        %\hspace*{-0.45cm}  % shift image left
        \subcaption{}
        \includegraphics[scale=0.33]{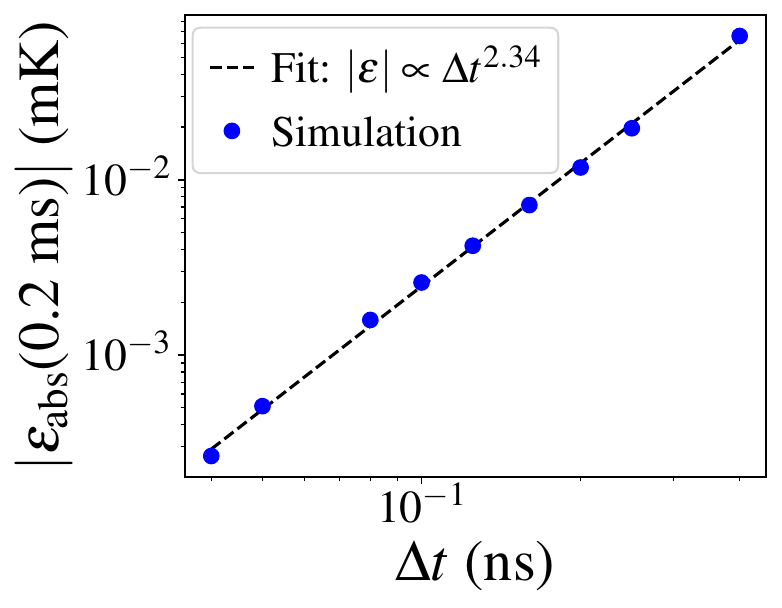}
        %\caption{}
        \label{fig:crystal_time_conv_b}
    \end{subfigure}
    ~ 
    \hspace*{-1.5cm}
    % Row 2
    \begin{subfigure}{0.2\textwidth}
        \centering
        %\hspace*{0.15cm}  % shift image left
        \subcaption{}
        \includegraphics[scale=0.33]{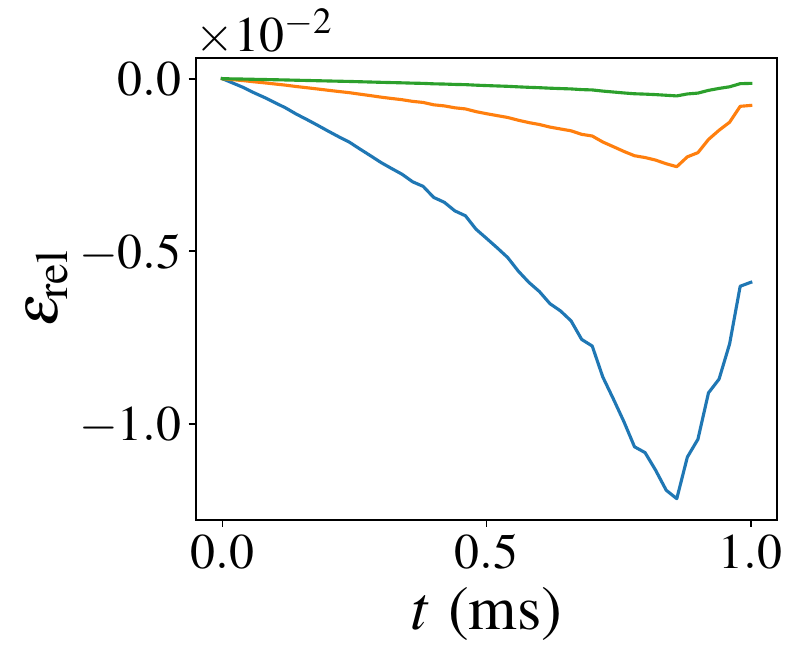}
        %\caption{}
        \label{fig:crystal_time_conv_c}
    \end{subfigure}%
    \hspace*{0.7cm}
    \begin{subfigure}{0.2\textwidth}
        \centering
        %\hspace*{-0.45cm}  % shift image left
        \subcaption{}
        \includegraphics[scale=0.33]{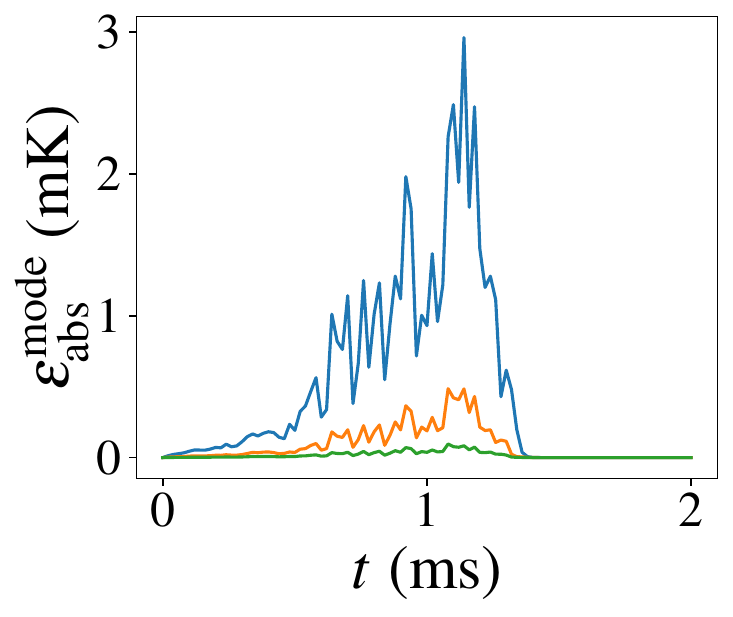}
        %\caption{}
        \label{fig:crystal_time_conv_d}
    \end{subfigure}
    \caption{Cooling simulations of the $N=100$ ion crystal from Fig.~\ref{fig:2D_crystal} demonstrate convergence with respect to the time step.  (a) The absolute error in the mean axial mode temperature is plotted versus time. Using a time step of $\Delta t=0.1$ ns, the maximum temperature error during cooling is less than $0.01$ mK. (b) The absolute value of the temperature error after $0.2$ ms of cooling is plotted as a function of the time step.  A polynomial fit to the simulation data shows that the error scales approximately quadratically with the time step ($|\varepsilon| \propto \Delta t^{2.34}$)   (c)  The relative error in the mean axial mode temperature is plotted for the first millisecond of cooling. (d)  The error in the temperature of individual modes can be analyzed, as well.  At each time step, the maximum absolute axial mode temperature error is plotted, establishing that individual mode temperatures are convergent with respect to the time step.}
    \label{fig:crystal_time_conv}
\end{figure}

When evolving the density matrix, we assume that the ion position $\boldsymbol{r}_j$ is constant during each time step. This implies that the laser phase $\boldsymbol{k}_{\mu}\cdot\boldsymbol{r}_j$ experienced by ion $j$ is also constant. However, since the ion crystal rotates at frequency $\omega_r$ about the center of the trap, individual ions rapidly traverse distances comparable to the laser wavelength.  Therefore, it important to validate the constant phase assumption.  To this end, we have studied the convergence of crystal cooling rates with respect to the time step.  Note that the single-ion convergence tests performed in Section \ref{sec:single_ion} do not capture the effect described here, since the ion is located at the center of the trap.  Since the velocity associated with crystal's rotation is proportional to the radial distance, it is desirable to test convergence using a crystal with a large planar extent.  As such, we use the case of the 100 ion 2D crystal shown in Fig.~\ref{fig:2D_crystal}.  We note that the largest crystal studied in this work has 500 ions, but that simulations with large crystals become expensive when the time step is decreased.  Furthermore, the planar extent of the 500 ion crystal is only about twice that of the 100 ion crystal, so convergence results are not expected to differ significantly between the two cases.

Results of the convergence tests are shown in Fig.~\ref{fig:crystal_time_conv}.  We first analyze the growth of the error in the axial temperature.  As before, the axial temperature is found by computing the mean temperature of all $N$ axial modes.  The absolute error ($\varepsilon_{\text{abs}}$), shown in Fig.~\ref{fig:crystal_time_conv_a}, is then calculated by finding the difference in temperature with respect to the simulation with the smallest time step ($\Delta t=0.025$ ns).  Explicitly, 

\begin{equation}
\label{eq:eps_abs}
\varepsilon_{\text{abs}}(t) = \frac{1}{N}\sum_{j=1}^N \left[T_{j,\Delta t}(t) - T_{j,\text{ref}}(t)\right],
\end{equation}
where $T_{j,\Delta t}(t)$ is the temperature of axial mode $j$ at time $t$ in the simulation with time step $\Delta t$, and `ref' refers to $\Delta t = 0.025$ ns. 

In Fig.~\ref{fig:crystal_time_conv_b}, we plot the temperature error versus time step.  Specifically, the magnitude of the temperature error $|\varepsilon_{\text{abs}}|$, after $0.2$ ms of cooling, is shown for various time steps.  A polynomial fit to the simulation data shows that $|\varepsilon_{\text{abs}}| \propto \Delta t^{2.34}$, so the time-step convergence is approximately quadratic.  

In Fig.~\ref{fig:crystal_time_conv_c}, we plot the relative error in the axial temperature, given by

\begin{equation}
    \varepsilon_{\text{rel}}(t) = \frac{\sum_{j=1}^N \left[T_{j,\Delta t}(t) - T_{j,\text{ref}}(t)\right]}{\sum_{j=1}^N T_{j,\text{ref}}(t)}
\end{equation}
The relative error associated with a time step of $0.1$ ns remains small during cooling, justifying our use of this time step throughout this work.  

The error in the temperatures of individual modes also shows convergence.  In Fig.~\ref{fig:crystal_time_conv_d}, we plot the maximum absolute mode temperature error versus time,
\begin{equation}
    \varepsilon_{\mathrm{abs}}^{\mathrm{mode}}(t)
    =
    \max_j
    \left|
        T_{j,\Delta t}(t)
        -
        T_{j,\text{ref}}(t)
    \right|.
\end{equation}
Assuming $\Delta t=0.1$ ns, all individual mode temperatures are valid to within a fraction of a millikelvin during cooling. 

\section{Fully quantum code details}
\label{app:FQ}

Details of the fully quantum code are presented here.  We first describe the numerical integration scheme.  Then, we discuss a method for estimating the cooling rates of PAC-initialized states from our simulations, which initialize the ion in a Fock state.

\subsection{Numerical integration}

As described in Section \ref{sec:fq}, the state $\ket{\psi(t)}$ is updated at each time step according to $\ket{\psi(t+\Delta t)} = e^{-i\hat{H}_{\text{eff}}\Delta t/\hbar}\ket{\psi(t)}$. To implement this update, we utilize a split-operator Fourier transform (SOFT) technique. The terms in the Hamiltonian $H_{\text{eff}}$ can be divided into those which depend only on the position operator $\hat{z}$ and those which depend only on momentum $\hat{p}$.  The Hamiltonian can then be written as $\hat H_{\text{eff}} = \hat H_{\text{eff}}^{(z)} + \hat H_{\text{eff}}^{(p)}$, where

\begin{subequations}
\label{eq:split_H}
\begin{align}
\hat H_{\text{eff}}^{(z)}
&= \frac{1}{2}m\omega_z^2\hat{z}^2
+ \hbar\sum_{\mu} \bigg(
\frac{\Omega_{\mu}}{2}\hat{\sigma}_{e g_\mu}
e^{i k_{z,\mu}\hat{z}}
+ \text{h.c.}\bigg)
\nonumber\\
&\quad -\frac{i\hbar}{2}\sum_\mu L_\mu^\dagger L_\mu,
\\
\hat H_{\text{eff}}^{(p)}
&= \frac{\hat{p}^2}{2m}.
\end{align}
\end{subequations}
Using the fast Fourier transform (FFT), the state $\ket{\psi(t)}$ can be transformed from the position basis to the momentum basis, and vice versa.  Therefore,  $\hat H_{\text{eff}}^{(z)}$ and $\hat H_{\text{eff}}^{(p)}$ can be applied in the position and momentum bases, respectively.  To this end, the update can be decomposed, using a second-order Trotter formula \cite{janke1992}, into

\begin{equation}
\begin{split}
    \ket{\psi(t + \Delta t)}
    &= e^{-\frac{i}{\hbar}(\hat H_{\text{eff}}^{(z)} + \hat H_{\text{eff}}^{(p)})\Delta t} \ket{\psi(t)} \\
    &\approx e^{-\frac{i}{\hbar} \hat H_{\text{eff}}^{(p)} \Delta t/2} e^{-\frac{i}{\hbar} \hat H_{\text{eff}}^{(z)} \Delta t}  e^{-\frac{i}{\hbar} \hat H_{\text{eff}}^{(p)} \Delta t/2} \ket{\psi(t)}  \\
    &\;\;\;\;\;\;\;\;+ O(\Delta t^3) .
\end{split}
\end{equation}
Here, it is understood that the state undergoes an inverse Fourier transform before application of the second operator, which is position-dependent.  Similarly, the state is transformed back to the momentum basis after the second operator is applied.

\subsection{PAC cooling rates}
In the FQ simulations, the ion is initialized in a Fock state of the Hamiltonian $\hat{H}_{\text{trap}} = \hat{p}^2/2m + m\omega_z^2\hat{z}^2/2$.  However, a more consistent comparison to the semiclassical results is obtained using a PAC-state initialization \cite{bartolotta2024}.  The form of a PAC state is given in Eq.~\eqref{eq:pac}.  Here, we explain how to estimate the cooling rate of a PAC state using the results of the Fock state simulations.

If the ion is initialized in the Fock state $\ket{n_0}$, then the ensemble average of the occupation number as a function of time, $\braket{n_{n_0}(t)}$, can be obtained by averaging over $N_t$ quantum trajectories, using Eq.~\eqref{FQ_exp_value} with $\hat{O} = \hat{n}$ (where $\hat{n}$ is the number operator). While the exact number of trajectories which we use varies with $n_0$, $N_t \geq 2000$ in all cases, which produces numerical convergence of the cooling rates. The Fock-state cooling rate at occupation number $n_0$ is found using 

\begin{equation}
\label{eq:fock_cooling_rate}
R_{n_0}^{F} = \frac{1}{T_{\text{av}}}\int_0^{T_{\text{av}}} \frac{d\braket{n_{n_0}(t)}}{dt}dt.
\end{equation}
In the FQ simulations, we use an averaging time of $10\;\mu\text{s} \leq T_{\text{av}}\leq  50 \;\mu\text{s}$. Since a PAC state is a linear combination of the Fock states (Eq.~\eqref{eq:pac}), the cooling rate of a PAC state with mean occupation number $\bar n$ can be computed using 
\begin{equation}
\label{eq:pac_cooling_rate}
R_{\bar n}^{PAC} = \sum_{m=0}^\infty P_m(\bar n)R_m^F,
\end{equation}
where $P_m(\bar n) = e^{-\bar n}[\bar n^m/(m!)]$.  In practice, we run FQ simulations only for a selected set of Fock states \(\{\lvert n'_j\rangle\}_{j=1}^{N'}\).  Therefore, the sum in Eq.~\eqref{eq:pac_cooling_rate} cannot be evaluated directly to compute the PAC state cooling rate.  Instead, a binning procedure is utilized to assign approximate weights, $w_{j,\bar n}$, to each of the Fock state cooling rates and the PAC-state cooling rate is then estimated using

\begin{equation}
\label{eq:pac_cooling_rate_approx}
R_{\bar n}^{\text{PAC}} \approx \sum_{j=1}^{N'} w_{j,\bar n}R_{n_j'}^F.
\end{equation}
The weights are obtained by assigning the probability $P_m(\bar n)$ to the Fock state which minimizes $|n_j'-m|$.

\section{Ion Crystal Normal Mode Analysis}
\label{app:normal_modes}

In this section we derive the normal modes of a trapped-ion crystal and present tools used to analyze ion crystals dynamics in the normal-mode basis.  First, we form the Lagrangian and solve for the equations of motion of an ion crystal.  Next, we diagonalize the system of equations to find the eigenmodes.  We then explain the mode initialization procedure and the analysis of simulation data in the normal-mode basis.

\subsection{Ion crystal Lagrangian}

The motion of ion crystals is typically analyzed in a frame which rotates at frequency $\omega_r$, where the crystal is stationary at equilibrium.  In this frame, the Lagrangian of a Penning trap ion crystal can be expressed as
\begin{equation}
\begin{split}
\label{beta}
\mathcal{L} = \sum_{i=1}^N \Big[ & \frac{1}{2}m|\dot{\boldsymbol{r}}_{r,i}|^2 -\frac{1}{2}qB_{\text{eff}}(\dot{x}_{r,i}y_{r,i} - x_{r,i}\dot{y}_{r,i}) \\ & + \frac{1}{2}m\left( \omega_r^2 - \omega_c\omega_r + \frac{1}{2}\omega_z^2\right)(x_{r,i}^2+y_{r,i}^2) \\ & - \frac{1}{4}q k_z\delta(x_{r,i}^2-y_{r,i}^2) - \frac{1}{2}m\omega_z^2z_{r,i}^2 \\& -\frac{q^2}{8\pi\epsilon_0}\sum_{j\neq i} \frac{1}{|\boldsymbol{r}_{r,j}-\boldsymbol{r}_{r,i}|}\Big].
\end{split}
\end{equation}
Here, the subscript `r' denotes that the ion coordinates are taken to be in the rotating frame. We will drop this subscript in the remainder of this section for brevity. The first term in the Lagrangian describes the ions' kinetic energy. The second term describes the Lorentz force in the rotating frame, where the effective magnetic field is $B_\text{eff} = B_z - 2m\omega_r/q$.  The remaining terms correspond to the potential energy of the ion crystal.  

To compute the normal modes, the Lagrangian is expanded about the equilibrium ion  configuration $\{\boldsymbol{r}_i^0 = (x_i^0,y_i^0,z_i^0)|1 \leq i \leq N\}$, giving $\mathcal{L} = \mathcal{L}_{\text{eq}} + \mathcal{L}_{\text{nm}}$, where $\mathcal{L}_{\text{nm}}$ determines the normal modes.  Explicitly,
\begin{equation}
\begin{split}
\label{eq:nm_lagrange}
\mathcal{L}_{\text{nm}} = \frac{1}{2}\sum_{i,j}\sum_{\alpha,\beta}\Bigg[& \frac{\partial^2\mathcal{L}}{\partial\alpha_i\partial\beta_j}\Bigg|_{eq}\delta\alpha_i\delta\beta_j + \frac{\partial^2\mathcal{L}}{\partial\dot{\alpha}_i\partial\dot\beta_j}\Bigg|_{eq}\delta\dot\alpha_i\delta\dot\beta_j\\ & + 2\frac{\partial^2\mathcal{L}}{\partial\dot{\alpha}_i\partial \beta_j}\Bigg|_{eq}\delta\dot\alpha_i\delta\beta_j\Bigg],
\end{split}
\end{equation}
where $i$ and $j$ denote ion indices, while $\alpha$ and $\beta$ stand for Cartesian coordinates. To simplify notation, we define the $N\times N$ matrices $\mathbb{K}$, $\mathbb{M}$, and $\mathbb{W}$ whose elements are computed according to

\begin{widetext}
\begin{align}
    K_{ij}^{\alpha\beta} &=  -\frac{\partial^2\mathcal{L}}{\partial\alpha_i\partial\beta_j}\Bigg|_{\text{eq}}  = \begin{cases}-m(\omega_r^2-\omega_c\omega_r+\frac{1}{2}\omega_z^2) + \frac{1}{2}qk_z\delta + \frac{q^2}{4\pi\epsilon_0}\left[\sum_{k\neq i}\frac{-(r_{ik}^0)^2 + 3(x_{ik}^0)^2}{(r_{ik}^0)^5}\right] & \alpha=\beta=x, i=j\\\frac{q^2}{4\pi\epsilon_0}\frac{(r_{ij}^0)^2-3(x_{ij}^0)^2}{(r_{ij}^0)^5} & \alpha=\beta=x, i\neq j \\ -m(\omega_r^2-\omega_c\omega_r+\frac{1}{2}\omega_z^2) - \frac{1}{2}qk_z\delta + \frac{q^2}{4\pi\epsilon_0}\left[\sum_{k\neq i}\frac{-(r_{ik}^0)^2 + 3(y_{ik}^0)^2}{(r_{ik}^0)^5}\right] & \alpha=\beta=y, i=j\\\frac{q^2}{4\pi\epsilon_0}\frac{(r_{ij}^0)^2-3(y_{ij}^0)^2}{(r_{ij}^0)^5} & \alpha=\beta=y, i\neq j \\ m\omega_z^2 + \frac{q^2}{4\pi\epsilon_0}\left[\sum_{k\neq i}\frac{-(r_{ik}^0)^2 + 3(z_{ik}^0)^2}{(r_{ik}^0)^5}\right] & \alpha=\beta=z, i=j\\\frac{q^2}{4\pi\epsilon_0}\frac{(r_{ij}^0)^2-3(z_{ij}^0)^2}{(r_{ij}^0)^5} & \alpha=\beta=z, i\neq j \\  \frac{q^2}{4\pi\epsilon_0}\left[\sum_{k\neq i}\frac{3(\alpha_{ik}^0)(\beta_{ik}^0)}{(r_{ik}^0)^5}\right] & \alpha\neq \beta, i=j\\-\frac{q^2}{4\pi\epsilon_0}\frac{3(\alpha_{ij}^0)(\beta_{ij}^0)}{(r_{ij}^0)^5} & \alpha\neq\beta, i\neq j\end{cases}\\
    M_{ij}^{\alpha\beta} &=  \frac{\partial^2\mathcal{L}}{\partial\dot\alpha_i\partial\dot\beta_j}\Bigg|_{\text{eq}} = m\delta_{ij}\delta_{\alpha\beta}\\
    W_{ij}^{\alpha\beta} &=  -\frac{\partial^2\mathcal{L}}{\partial\dot\alpha_i\partial\beta_j}\Bigg|_{\text{eq}} = \begin{cases} \frac{1}{2}\delta_{ij}qB_{\text{eff}} & \alpha=x,\beta=y \\ -\frac{1}{2}\delta_{ij}qB_{\text{eff}} & \alpha=y,\beta=x\end{cases}
\end{align}
\end{widetext}
Here, $\alpha_{ik} = \alpha_i-\alpha_k$ and $r_{ik} = |\boldsymbol{r}_i -\boldsymbol{r}_k|$. The matrix $\mathbb{K}$ is known as the stiffness matrix, and is real and symmetric.  The matrix $\mathbb{W}$ accounts for the Lorentz force and is antisymmetric and block diagonal. We use typical sign conventions for the matrices.  Having computed the matrix elements, we construct the matrices explicitly as

\begin{align}
    &\mathbb{K} = \begin{bmatrix}
        \mathbb{K}^{xx} & \mathbb{K}^{xy} & \mathbb{K}^{xz}\\
        \mathbb{K}^{yx} & \mathbb{K}^{yy} & \mathbb{K}^{yz}\\
        \mathbb{K}^{zx} & \mathbb{K}^{zy} & \mathbb{K}^{zz}\\
        \end{bmatrix}\\
    &\mathbb{M} =\begin{bmatrix}
        \mathbb{M}^{xx} & \mathbb{0}& \mathbb{0}\\
        \mathbb{0} & \mathbb{M}^{yy} & \mathbb{0}\\
        \mathbb{0} & \mathbb{0} & \mathbb{M}^{zz}\\
        \end{bmatrix} = \begin{bmatrix}
        m\mathbb{I}_{N} & \mathbb{0}& \mathbb{0}\\
        \mathbb{0} & m\mathbb{I}_{N} & \mathbb{0}\\
        \mathbb{0} & \mathbb{0} & m\mathbb{I}_{N}\\
        \end{bmatrix}\\
    &\mathbb{W} =\begin{bmatrix}
        \mathbb{0}& \mathbb{W}^{xy}& \mathbb{0}\\
        \mathbb{W}^{yx} & \mathbb{0} & \mathbb{0}\\
        \mathbb{0} & \mathbb{0} & \mathbb{0}\\
        \end{bmatrix}= \begin{bmatrix}
        \mathbb{0} & \frac{1}{2}qB_{\text{eff}}\mathbb{I}_N & \mathbb{0}\\
        -\frac{1}{2}qB_{\text{eff}}\mathbb{I}_N  & \mathbb{0} & \mathbb{0}\\
        \mathbb{0} & \mathbb{0} & \mathbb{0}\\
        \end{bmatrix},
\end{align}
where $\mathbb{I}_N$ is the $N\times N$ identity matrix.  The above matrices are $3N\times 3N$ dimensional.  From now on we will use, for instance, the notation $K_{ij}$ to refer to the element in the $i^{th}$ row and $j^{th}$ column of matrix $\mathbb{K}$ (i.e. $i$ and $j$ are no longer ion indices). 

\subsection{Normal modes}

From the Lagrangian, one can compute the Euler-Lagrange equations of motion.  It is useful to first introduce the $3N$-dimensional vectors $\boldsymbol{\delta r} = (\delta x_1, ..., \delta x_N,\delta y_1, ..., \delta y_N,\delta z_1, ..., \delta z_N)^T$ and $\boldsymbol{v} = (\delta \dot{x}_1, ..., \delta \dot{x}_N,\delta \dot{y}_1, ..., \delta \dot{y}_N,\delta \dot{z}_1, ..., \delta \dot{z}_N)^T$, which tabulate the ions' displacements from equilibrium and velocities, respectively.  Then Eq.~\eqref{eq:nm_lagrange} can be rewritten as

\begin{equation}
\begin{split}
\label{beta}
\mathcal{L}_{\text{nm}} = \frac{1}{2}\sum_{i,j=1}^{3N}\Big[& -K_{ij}\delta r_i\delta r_j + M_{ij} v_i v_j- 2W_{ij} v_i\delta r_j\Big].
\end{split}
\end{equation}
The required partial derivatives are
\begin{subequations}
\label{beta}
\begin{align}
&\frac{\partial\mathcal{L}_{\text{nm}}}{\partial v_k} = M_{kk} v_k - \sum_{j=1}^{3N} W_{kj}\delta r_j,\\
&\frac{\partial\mathcal{L}_{\text{nm}}}{\partial r_k} = -\sum_{j=1}^{3N}K_{kj}\delta r_j + \sum_{j=1}^{3N} W_{kj} v_j.
\end{align}
\end{subequations}
Here, we have used the symmetry properties of the matrices to simplify the expressions.  The Euler-Lagrange equations are then given by
\begin{equation}
\label{beta}
m\dot{\boldsymbol{v}}=-\mathbb{K}\boldsymbol{\delta r} + 2\mathbb{W}\boldsymbol{v}.
\end{equation}

We define the composite vector $\ket{q}$ of positions and velocities as $\ket{q} = \ket{\boldsymbol{\delta r}, \boldsymbol{v}}$.  We use Dirac notation here, which will be convenient when dealing with the complex eigenmodes of the system. The crystal dynamics are then completely described by the matrix equation
\begin{equation}
\begin{split}
\label{eq:nm_eom}
\frac{d\ket{q}}{dt} &= \begin{bmatrix}
        \mathbb{0}_{3N}& \mathbb{I}_{3N}\\
        -\mathbb{K}/m & 2\mathbb{W}/m \\
        \end{bmatrix}\ket{q} \equiv \mathbb{D}\ket{q}.
\end{split}
\end{equation}
To determine the normal modes, we assume a harmonic solution $\ket{q} \propto \ket{u_j}e^{-i\omega_j t}$.  Plugging this ansatz into Eq.~\eqref{eq:nm_eom} produces
\begin{equation}
\begin{split}
\label{eq:nm_eigenvalue_eq}
\mathbb{D}\ket{u_j} = -i\omega_j\ket{u_j}.
\end{split}
\end{equation}
Therefore, the normal modes $\ket{u_j}$ and associated frequencies $\omega_j$ are found by solving the eigenvalue problem for $\mathbb{D}$. Complex conjugating Eq.~\eqref{eq:nm_eigenvalue_eq}, one finds that $\ket{u_j^*}$ is also an eigenvector of $\mathbb{D}$ and its eigenvalue is $i\omega_j$.  Thus, we can label the $6N$ eigenmodes such that $\ket{u_{j+3N}} = \ket{u_j^*}$ and $\omega_{j+3N} = -\omega_j$ for $1\leq j \leq3N$.

As we will see later, it is helpful to normalize the eigenvectors by their energies.  We define the $6N\times 6N$ energy matrix
\begin{equation}
\begin{split}
\label{beta}
\mathbb{E} &= \begin{bmatrix}
        \mathbb{K}& \mathbb{0}\\
        \mathbb{0} & m\mathbb{I}_{3N} \\
        \end{bmatrix}.
\end{split}
\end{equation}
The energy of the crystal above its equilibrium value can be expressed in terms of the matrix $\mathbb{E}$ as  

\begin{equation}
\begin{split}
\label{beta}
H_{\text{cry}} &= \frac{1}{2}\braket{q|\mathbb{E}|q} = \frac{1}{2}\boldsymbol{\delta r}^T\mathbb{K}\boldsymbol{\delta r} + \frac{1}{2}m\boldsymbol{v}^T\boldsymbol{v}.
\end{split}
\end{equation}
If the unnormalized eigenvectors are denoted $\ket{u'_n}$, then the energy-normalized eigenvectors $\ket{u_n}$ are computed using
\begin{equation}
\begin{split}
\label{beta}
\ket{u_n} &= \frac{\ket{u'_n}}{\sqrt{\braket{u_n'|\mathbb{E}|u_n'}}}.
\end{split}
\end{equation}
This relation guarantees $\braket{u_j|\mathbb{E}|u_j} = 1$.  It can also be shown that $\braket{u_j|\mathbb{E}|u_k} = 0$ when $j\neq k$ \cite{shankar2020}. This is done by multiplying Eq. \eqref{eq:nm_eigenvalue_eq} by $i\mathbb{E}$, which results in $i\mathbb{E}\mathbb{D}\ket{u_j} = \omega_j\mathbb{E}\ket{u_j}$.  This is a generalized eigenvalue problem with Hermitian matrices $i\mathbb{E}\mathbb{D}$ and $\mathbb{E}$, implying $\braket{u_j|\mathbb{E}|u_k} = \delta_{jk}$.  In particular, choosing $k=j+3N$, we find $\braket{u_j|\mathbb{E}|u_j^*} = \delta_{jk}$. Since the eigenvectors form a complete basis, a general composite vector can be expressed as a linear combination of the eigenvectors according to 

\begin{equation}
\begin{split}
\label{eq:phase_space_vec_general}
\ket{q} &= \sum_{j=1}^{6N} A_j e^{-i\omega_j t}\ket{u_j}\\ &= \sum_{j=1}^{3N}\left( A_j e^{-i\omega_j t}\ket{u_j}+A_j^*e^{i\omega_j t}\ket{u_j^*}\right).
\end{split}
\end{equation}
Here, $A_j$ is the complex mode amplitude.  The state vector $\ket{q}$ must be real, which constrains $A_{j+3N} = A_j^*$.

\subsection{Initialization and data analysis in the normal mode basis}

We generally initialize certain normal modes at a given temperature before simulating EIT cooling. Afterwards, the raw simulation data is analyzed in the normal mode basis.  Here, we describe this simulation pipeline in detail.

Typically, all modes within a given mode branch are initialized at the same temperature.  Consider initializing all axial modes (assumed to be labeled by mode numbers $N+1\leq j \leq 2N$) with real amplitude $A_0$.  The composite phase-space vector is found using Eq.~\eqref{eq:phase_space_vec_general}, resulting in
\begin{equation}
\begin{split}
\label{beta}
\ket{q(t=0)} &= \sum_{j=N+1}^{2N}\left( A_0 e^{i\varphi_j }\ket{u_j}+A_0e^{-i\varphi_j}\ket{u_j^*}\right)\\
 &= 2\sum_{j=N+1}^{2N}\text{Re}\left( A_0 e^{i\varphi_j }\ket{u_j}\right).
\end{split}
\end{equation}
Here we randomize the phase of mode $j$ by randomly sampling $\varphi_j$ from the distribution $U(0,2\pi)$. To determine the relationship between the amplitude $A_0$ and the initial mode temperature $T_0$, we note that the energy in mode $j$ is

\begin{equation}
\begin{split}
\label{beta}
E_j &= \frac{A_0^2}{2}\left( \bra{u_j}e^{-i\varphi_j} + \bra{u_j^*}e^{i\varphi_j} \right)\mathbb{E} \left(e^{i\varphi_j}\ket{u_j} + e^{-i\varphi_j}\ket{u_j^*} \right) \\
&= A_0^2.
\end{split}
\end{equation}
Here we have used the orthonormality condition.  Setting $E_j(t=0) = k_BT_0$, we find $A_0 = \sqrt{k_BT_0}$.  This relation allows us to choose $A_0$ associated with the desired initial temperature of the crystal.  The cyclotron mode branch can be excited in the same manner described above.

After our MD simulations, we analyze the cooling of the crystal in the normal mode basis.  In particular, we compute the mode energies at each recorded time step.  Suppose that at time $t$, the composite phase-space vector is given by
\begin{equation}
\begin{split}
\label{beta}
\ket{q(t)} = \sum_{j=1}^{3N}\left( A_j\ket{u_j}+A_j^*\ket{u_j^*}\right) = 2\sum_{j=1}^{3N}\text{Re}\left( A_j\ket{u_j}\right),
\end{split}
\end{equation}
where we have absorbed the complex phase into $A_j$, which is now a complex number.  To extract $A_j$ from the positions and velocities of the ions at time $t$, we use $\mathbb{E}$-orthonormality of the eigenvectors to find
\begin{equation}
\begin{split}
\label{beta}
\braket{u_j|\mathbb{E}|q(t)} = A_j.
\end{split}
\end{equation}
Therefore, $|A_j| = |\braket{u_j|\mathbb{E}|q(t)}|$ and the temperature is found using $T_j(t) = |A_j|^2/k_B$.

\section{Mode branch coupling}
\label{app:mode_coupling}

\begin{figure}
    \centering
    \hspace*{-1.5cm}
    \begin{subfigure}{0.2\textwidth}
        \centering
        %\hspace*{0.15cm}  % shift image left
        \subcaption{}
        \includegraphics[scale=0.28]{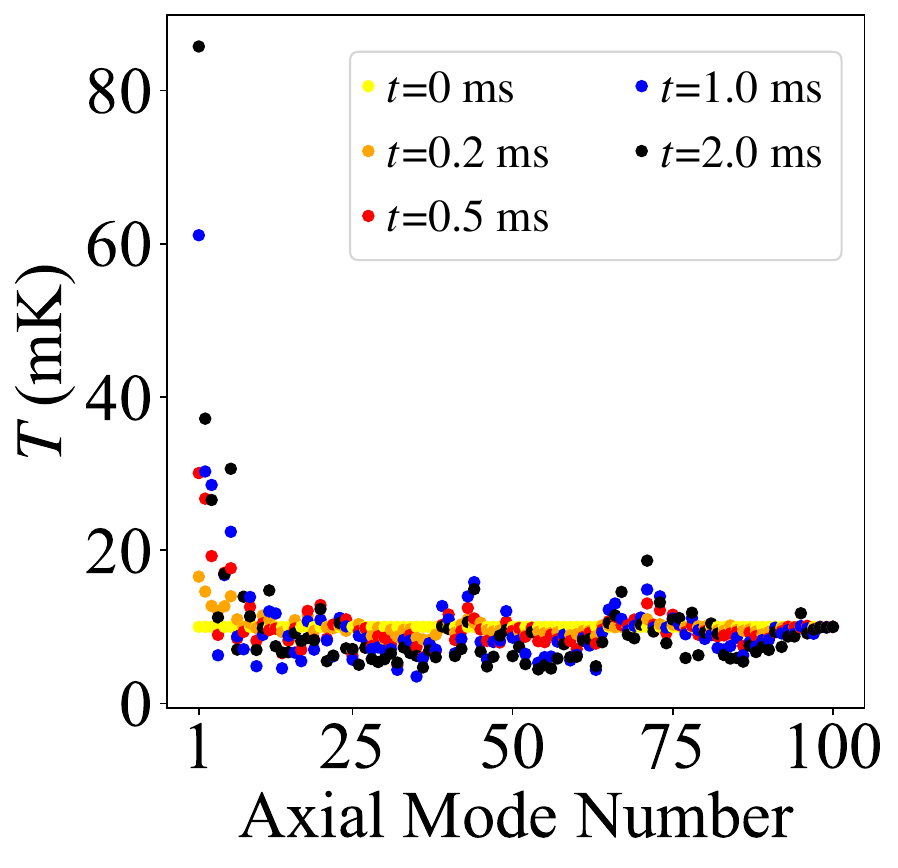}
        %\caption{}
        \label{fig:free_ev_T10mK_mode_resolved}
    \end{subfigure}%
    \hspace*{0.55cm}
    \begin{subfigure}{0.2\textwidth}
        \centering
        %\hspace*{-0.45cm}  % shift image left
        \subcaption{}
        \includegraphics[scale=0.28]{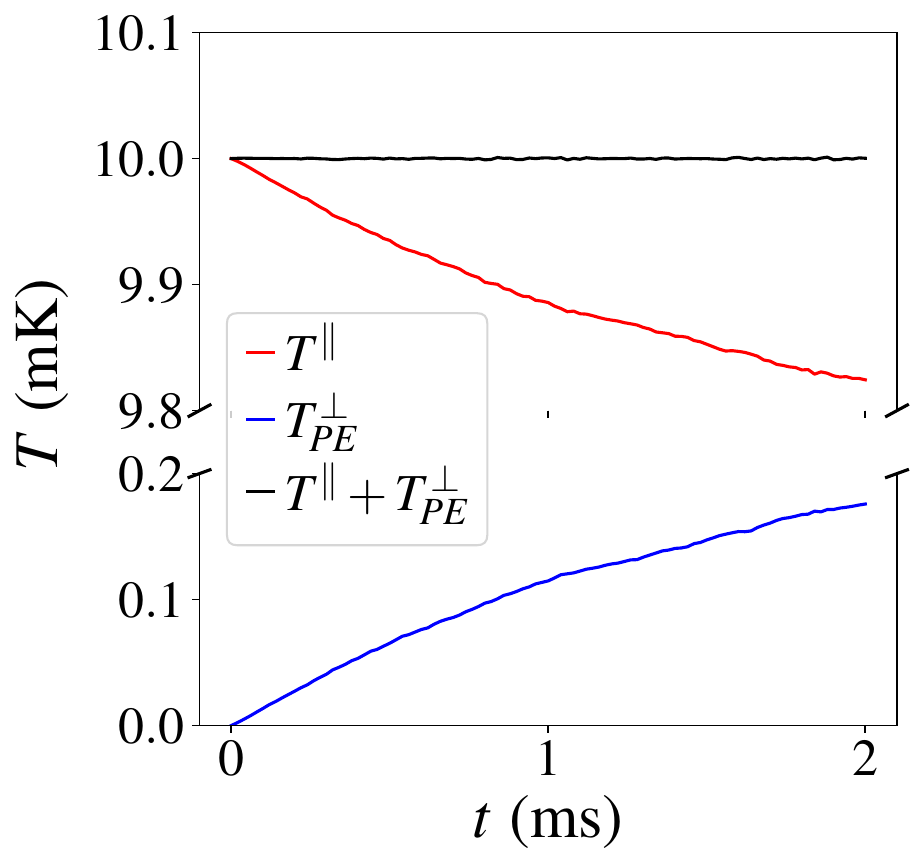}
        %\caption{}
        \label{fig:free_ev_T10mK_coupling}
    \end{subfigure}
    ~ 
    \hspace*{-1.5cm}
    % Row 2
    \begin{subfigure}{0.2\textwidth}
        \centering
        %\hspace*{0.15cm}  % shift image left
        \subcaption{}
        \includegraphics[scale=0.267]{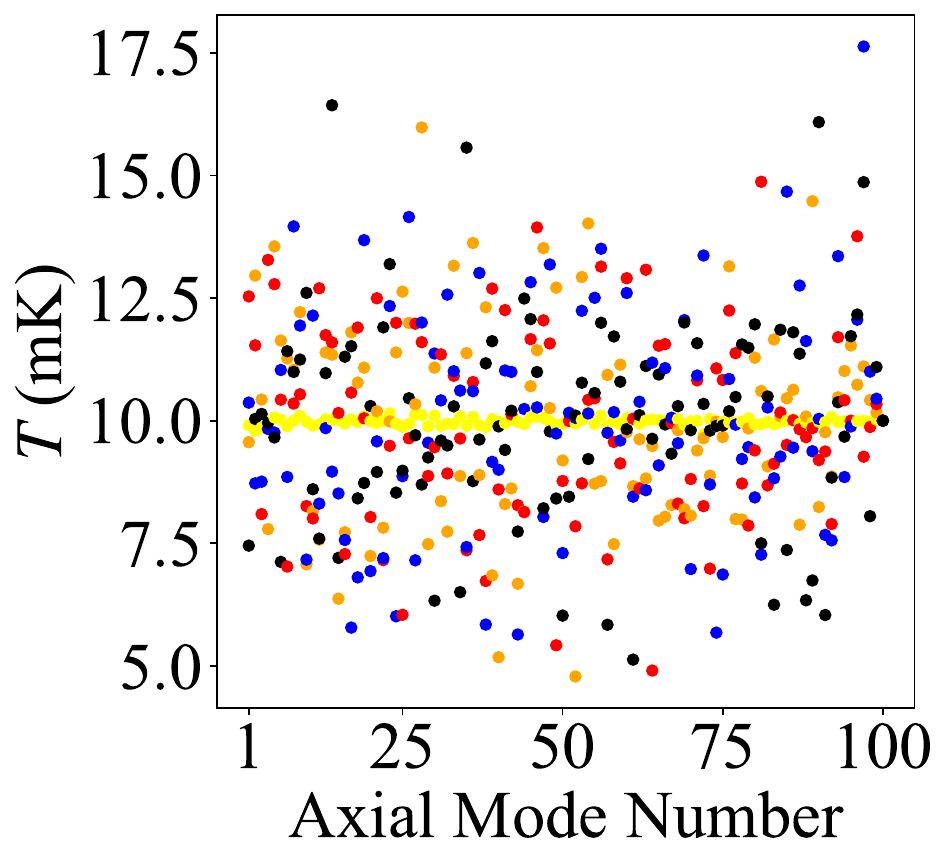}
        %\caption{}
        \label{fig:free_ev_T10mK_Tplanar10mK_mode_resolved}
    \end{subfigure}%
    \hspace*{0.6cm}
    \begin{subfigure}{0.2\textwidth}
        \centering
        %\hspace*{-0.45cm}  % shift image left
        \subcaption{}
        \includegraphics[scale=0.27]{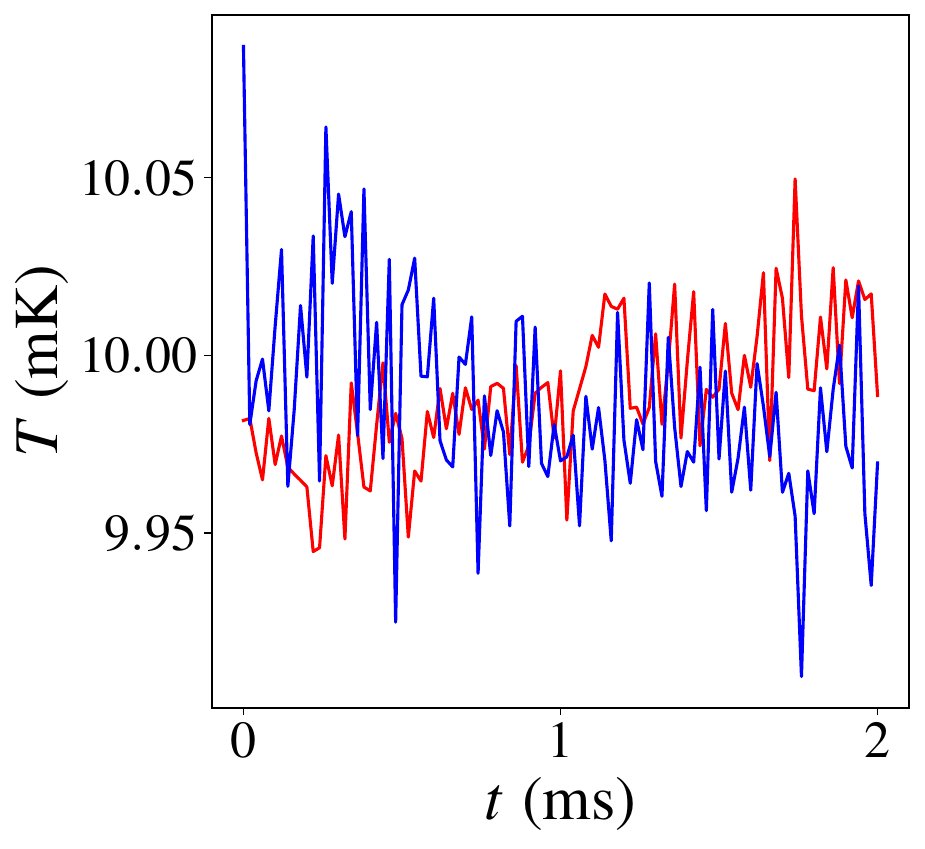}
        %\caption{}
        \label{fig:free_ev_T10mK_Tplanar10mK_coupling}
    \end{subfigure}
    \caption{(a) The axial mode temperatures at different times during free evolution are shown for an $N=100$ ion 2D crystal in which only the axial modes are initially excited ($T_i^{\parallel}=10$ mK).  Mode temperatures are averaged over 20 independent runs.  The lowest frequency axial modes are heated significantly during free-evolution. (b) The axial temperature and planar potential-energy temperature are plotted versus time in red and blue, respectively.  The latter approximates the temperature of the $\boldsymbol{E} \times \boldsymbol{B}$ branch.  Coupling between the axial and $\boldsymbol{E} \times \boldsymbol{B}$ branches causes energy to be transferred from the former to the latter.  This process explains the increase in temperature of the low-frequency axial modes seen in (a), as explained in the text. (c)  When all mode branches are initialized at $T=10$ mK, heating of the low-frequency axial modes does not occur.  (d)  In this case, the axial and planar potential-energy temperatures are approximately constant in time, as expected.}
    \label{fig:free_ev_T10mK}
\end{figure}

In this section, we present results from free-evolution simulations of trapped-ion crystal dynamics.  In these simulations, the EIT laser beams are turned off and the ions' motion is evolved in time according to $\boldsymbol{F}^{(j)}_{\text{free}} = \boldsymbol{F}_j - \boldsymbol{F}^{(j)}_{\text{laser}}$. The purpose of these simulations is to characterize the coupling between the $\boldsymbol{E} \times \boldsymbol{B}$ and axial mode branches, which influences results presented in the main text.

The first set of free-evolution simulations uses the same crystal initializations as the $N=100$ ion simulations from Fig.~\ref{fig:2D_crystal}.  The axial modes are each initialized at $T_i = 10$ mK in 20 independent realizations with random initial phases.  The temperatures of the axial modes over time during free evolution (averaged over 20 runs) are shown in Fig.~\ref{fig:free_ev_T10mK_mode_resolved}. The lowest temperature axial modes are heated significantly.  This phenomenon was first observed in Fig.~\ref{fig:2D_100i_moed_resolved_cooling}.  This mode-specific heating is a result of resonant mode coupling.  In Fig.~\ref{fig:free_ev_T10mK_coupling}, we plot the crystal's axial temperature $T^{\parallel}$ and planar potential-energy temperature, $T_{\text{PE}}^\perp$.  The latter is computed as described in Sec.~\ref{sec:2D_excited_planar_modes} and is approximately equal to the temperature of the $\boldsymbol{E} \times \boldsymbol{B}$ modes.  The sum $T^{\parallel}+T_{\text{PE}}^\perp$ is approximately constant in time, but a fraction of the axial kinetic energy is converted into planar potential energy. This supports the explanation that axial mode excitations decay into $\boldsymbol{E} \times \boldsymbol{B}$ mode excitations.  Specifically, if there exist two axial modes with frequencies $\omega^\parallel_1$ and $\omega_2^\parallel$ and an $\boldsymbol{E} \times \boldsymbol{B}$ mode with frequency $\omega^\perp$, such that $\omega_1^\parallel \approx \omega_2^\parallel+\omega^\perp$, then the higher frequency axial mode can decay into the sum of the lower frequency axial mode and the $\boldsymbol{E} \times \boldsymbol{B}$ mode \cite{marquet2003,johnson2026}.  This also explains why the lower frequency axial modes are preferentially heated in Fig.~\ref{fig:free_ev_T10mK_mode_resolved}.

Next, we run free-evolution simulations for the case where the axial and planar temperatures are initialized at 10 mK ($T_i^\parallel = T_i^\perp =10$ mK).  For these simulations we use the same initial crystals as in the simulations from Fig.~\ref{fig:2D_T10mK_Tplanar10mK}.  The time evolution of the axial mode temperatures are illustrated in Fig,~\ref{fig:free_ev_T10mK_Tplanar10mK_mode_resolved}.  In this case, while there are fluctuations in the mode temperatures, there is no obvious frequency-dependent mode heating.  In Fig.~\ref{fig:free_ev_T10mK_Tplanar10mK_coupling}, we plot the crystal's axial temperature and planar potential-energy temperature.  Since the axial and $\boldsymbol{E} \times \boldsymbol{B}$ mode branches are initialized approximately in thermal equilibrium, the temperatures are approximately constant in time.  

\begin{figure}
    \centering
    \includegraphics[scale=0.5]{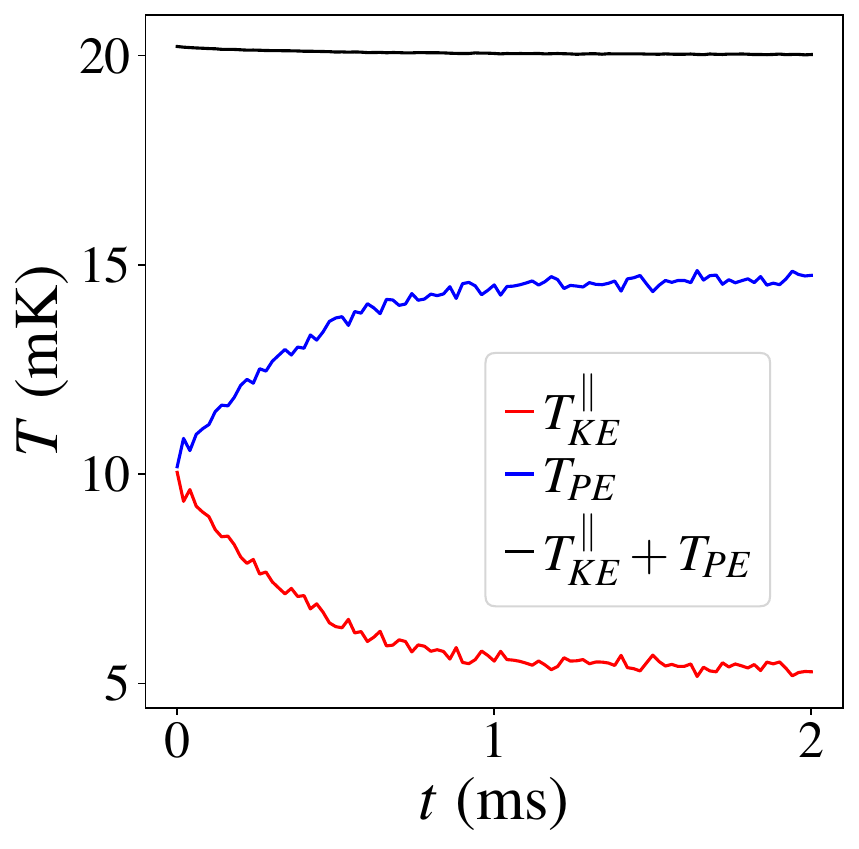}
    ~ 
    \caption{The axial kinetic temperature and total potential-energy temperature are plotted during free-evolution simulations of a 3D ion crystal.  Initially, the axial modes are excited to $T_i=10$ mK and the other mode branches are unexcited.  The total potential energy is computed using Eq.~\eqref{PE_1D}.  The rate of energy transfer from the axial mode branch to the $\boldsymbol{E} \times \boldsymbol{B}$ mode branch is fast compared to the 2D crystal case due to the reduced frequency gap between the branches (see Fig.~\ref{fig:3D_spectrum}).}
    \label{fig:free_ev_3D}
\end{figure}

Lastly, we study the free evolution of the $N=100$ ion 3D crystal from Fig.~\ref{fig:3D_cooling} ($\omega_r=2\pi\times 240$ kHz and $\delta=0.035$). The axial modes are initialized at $T_i=10$ mK, while the $\boldsymbol{E} \times \boldsymbol{B}$ and cyclotron modes are unexcited.  Plots of the axial kinetic temperature and the total potential-energy temperature of the crystal (computed using Eq.~\eqref{PE_1D}) are shown in Fig.~\ref{fig:free_ev_3D}. A decrease in the axial kinetic energy and an accompanying increase in the potential energy shows that energy is transferred from the axial modes to the $\boldsymbol{E} \times \boldsymbol{B}$ modes. The rate of energy transfer between the mode branches is significantly higher than in the 2D crystal case, due to the decreased frequency gap between the branches \cite{johnson2024}.

%\nocite{*}

\bibliography{apssamp}% Produces the bibliography via BibTeX.

\end{document}